\documentclass[aps,pra,superscriptaddress,longbibliography]{revtex4-1}
\usepackage{graphicx}
\usepackage{amsfonts}
\usepackage{amsmath}
\usepackage[usenames,dvipsnames]{color}
\usepackage{bm}
\usepackage{amssymb}
\usepackage{braket}
\usepackage{multirow}
\usepackage{epstopdf}
\usepackage{natbib}
\usepackage{hyperref}
\hypersetup{
colorlinks=true,
linkcolor=blue,
citecolor=blue,
filecolor=green,
urlcolor=blue,
}

\usepackage[usenames,dvipsnames]{color}

\usepackage[
monochrome
]{xcolor}

\usepackage{hyperref}
\hypersetup{
colorlinks=true,
linkcolor=blue,
citecolor=blue,
filecolor=green,
urlcolor=blue,
}

\begin{document}
\title{Long-lived quantum coherent dynamics of a $\Lambda$-system driven by a thermal environment}
%: Enhancing coherence lifetimes via incoherent driving
\author{Suyesh Koyu and Timur V. Tscherbul\footnote{Corresponding author. Electronic address: {\tt ttscherbul@unr.edu}} }
\noaffiliation
\affiliation{Department of Physics, University of Nevada, Reno, NV, 89557, USA}

\begin{abstract}
We present a theoretical study of quantum coherent dynamics of a three-level $\Lambda$ system driven by a thermal environment (such as blackbody radiation), which serves as an essential building block of photosynthetic light-harvesting models and quantum heat engines.  By solving the nonsecular Bloch-Redfield master equations, we obtain analytical results for  the ground-state population and coherence dynamics and classify the dynamical regimes of the incoherently driven  $\Lambda$-system as underdamped and overdamped  depending on whether the ratio $\Delta/[r f(p)]$ is greater or less than one, where $\Delta$ is the ground-state energy splitting,  $r$ is the incoherent pumping rate, and $f(p)$ is a function of the transition dipole alignment parameter $p$.
 In the underdamped regime, we observe long-lived coherent dynamics that lasts for $\tau_c\simeq 1/r$, even though the initial state of the $\Lambda$-system contains no coherences in the energy basis.
In the overdamped regime for $p = 1$, we observe the emergence of coherent quasi-steady states with the lifetime $\tau_{c} = 1.34 (r/\Delta^{2})$, which have low von Neumann entropy  compared to the conventional thermal states. We propose an experimental scenario for observing noise-induced coherent dynamics  in metastable He$^*$ atoms driven by x-polarized incoherent light.
 Our results suggest that thermal excitations can generate experimentally observable long-lived quantum coherent dynamics in the ground-state subspace of atomic and molecular $\Lambda$ systems in the absence of coherent driving.
\end{abstract}

%generate low-entropy quantum states and

\date{\today}
\maketitle

\newpage

\section{Introduction}

The quantum dynamics of multilevel atoms and molecules interacting with noisy electromagnetic fields [most notably, blackbody radiation (BBR)] plays a central role in  photosynthetic energy transfer in biological systems \cite{Brumer:18,Cao:20,Dodin:21,Tscherbul:14,Olsina:14}, quantum thermodynamics \cite{Kosloff:14,Scully:11,Dorfman:13} and precision measurement  \cite{Beloy:06,Safronova:12,Ovsiannikov:11,Lisdat:21}.   
In particular, the  essential first steps in photosynthetic energy transfer \cite{Kassal:13,Leon-Montiel:14,Brumer:18,Tscherbul:18,Yang:20,Chuang:20} and in vision \cite{Polli:10,Schulten:14,Tscherbul:14b,Tscherbul:15,Chuang:22} involve photoexcitation of biological chromophore molecules by incoherent solar light, which can be approximated as a  BBR source \cite{Kassal:13,Leon-Montiel:14}.  The role of quantum coherence in these primary biological processes remains incompletely understood, and continues to attract significant interest \cite{Brumer:18,Cao:20,Dodin:21}. 
In addition, BBR shifts atomic and molecular energy levels and causes incoherent transitions between them, limiting the precision of highly accurate spectroscopic measurements \cite{Beloy:06,Safronova:12,Ovsiannikov:11,Lisdat:21}.

The standard theoretical treatment of quantum dynamics of multilevel atoms and molecules driven by the blackbody radiation is based on solving the secular (Pauli) rate equations \cite{Cohen-Tannoudji:04,Blum:11} for the time evolution of state populations.  While adequate in many cases of practical interest, the rate equations rely on the secular approximation, which neglects the coherences between the atomic and/or molecular energy levels \cite{Cohen-Tannoudji:04,Blum:11}. The secular approximation cannot be justified in the presence of nearly degenerate energy levels (also known as Liouvillian degeneracies \cite{Wilhelm:07}), in which case the more sophisticated Bloch-Redfield (BR) master equations, which account for the population-to-coherence coupling terms, should be used \cite{Tscherbul:14,Jeske:15,Dodin:16,Dodin:18,Koyu:18,Tscherbul:15b,Eastham:16,Wang:19,Liao:21,Merkli:15,Trushechkin:21}.
The population-to-coherence couplings lead to interesting and underexplored physical effects,  such as the  generation of noise-induced  quantum coherences by thermal driving alone (i.e. in the absence of coherent driving) starting  from a coherence-free initial state \cite{Fleischhauer:92,Hegerfeldt:93,Kozlov:06,Ou:08,Scully:11,Dorfman:13,Tscherbul:14,Dodin:16,Koyu:18,Dodin:18,Koyu:21}. These noise-induced Fano coherences arise due to the interference of different incoherent transition pathways \cite{Tscherbul:14,Tscherbul:15b,Dodin:16,Dodin:21} caused by the cross-coupling terms in the light-matter interaction Hamiltonian of the form $\bm{\mu}_{ik}\cdot \bm{\mu}_{jk}$ \cite{Ficek:05}, where  $\bm{\mu}_{ij}=\langle i | \bm{\mu} | j\rangle$ are the transition dipole matrix elements in the system eigenstate basis.
Physically, these cross couplings arise from the transitions $k\leftrightarrow i$ and $k\leftrightarrow j$ being driven by the same mode of the incoherent radiation field \cite{Dodin:21}.  Importantly, these couplings do not average to zero under isotropic incoherent excitation if the corresponding transition dipole moments are non-orthogonal ($\bm{\mu}_{ik}\cdot \bm{\mu}_{jk}\ne0$) \cite{Patnaik:99,Tscherbul:14}.

% and they defy conventional wisdom, which sees thermal noise as leading to an irreversible decay of quantum coherent dynamics. 

  Early theoretical studies of Fano coherences focused on coherent population trapping and resonance fluorescence of trapped ions \cite{Fleischhauer:92,Hegerfeldt:93}. Ref.~\cite{Fleischhauer:92} considered a four-level system driven by polarized incoherent radiation in addition to a coherent pump field, and proposed it for lasing without inversion.
More recent theoretical work investigated the role of Fano coherences in suppressing spontaneous emission from three-level atoms \cite{Kapale:03}, in enhancing the efficiency of quantum heat engines \cite{Scully:11,Dorfman:13}, in biological processes induced by solar light \cite{Brumer:18,Tscherbul:18,Jung:20,Yang:20,Jankovic:20,Tomasi:21} and in negative entropy production \cite{Latune:20}.
We have explored the dynamical evolution of Fano coherences in a model three-level V-system excited by isotropic  \cite{Tscherbul:14,Dodin:16,Koyu:18} and polarized  \cite{Dodin:18,Koyu:21} incoherent light. In the latter  case, we explored the properties of coherent quasi-steady states \cite{Agarwal:01}, which form in the long-time limit and differ substantially from the conventional thermal states predicted by the secular rate equations.
Recently, closely related vacuum-induced Fano coherences have been detected experimentally in a cold ensemble of Rb atoms \cite{Han:21}.

A three-level $\Lambda$-system consisting of two nearly degenerate ground levels radiatively coupled to a single excited level  [see Fig. \ref{fig:Lambda_sys}(a)] is a paradigmatic quantum multilevel system, which serves as a fundamental building block of  photosynthetic light-harvesting complexes \cite{Brumer:18,Tscherbul:18,Yang:20,Chuang:20}, quantum heat engines \cite{Scully:11,Dorfman:13}, and quantum optical systems \cite{Fleischhauer:05}.
However, despite its profound significance, the thermally driven $\Lambda$-system has only been explored in the regime of degenerate ground levels ($\Delta=0$) \cite{Ou:08}, which is a theoretical idealization due to the presence of  degeneracy-lifting Stark and Lamb shifts \cite{Ficek:05,Altenmuller:95}. This leaves open the question of whether thermal environments can induce and sustain coherent ground-state {dynamics} in realistic atomic and/or molecular $\Lambda$-systems.
% with $\Delta>0$.

Here, we study the quantum dynamics of the $\Lambda$-system driven by a thermal environment such as BBR. By solving nonsecular Bloch-Redfield quantum master equations, we obtain analytic results for the time evolution of noise-induced Fano coherences between the ground levels of the $\Lambda$-system and establish the existence of two distinct dynamical regimes, where the coherences exhibit either long-lived underdamped oscillations or quasi-steady states with low entropy compared to the conventional thermal states. Our results show that it is possible to generate long-lived quantum coherent dynamics between the ground states of the $\Lambda$-system using solely BBR driving, and we suggest an experimental scenario for observing this dynamics (along with the resultant coherent steady states)  in metastable He$^*$ atoms in an external magnetic field. 

The rest of this paper is structured as follows. In Sec. IIA we formulate the theory of incoherent excitation of the $\Lambda$ system using the Bloch-Redfied quantum master equations, and present their analytical solutions. In Sec. IIB we discuss two primary regimes of population and coherence dynamics (underdamped and overdamped). In Sec. III we suggest an experimental scenario for observing noise-induced Fano coherences in the steady state using metastable He$^*$ atoms driven by polarized incoherent BBR. Section IV concludes by summarizing the main results of this work.

\section{Noise-induced coherent dynamics} 

\subsection{Theory: Bloch-Redfield master equations and their analytical solution}

The energy level structure of the $\Lambda$-system, shown in Fig~1(a), consists of two nearly degenerate ground states $|g_1\rangle$ and $|g_2\rangle$ coupled to a single excited state $|e\rangle$ by the incoherent radiation field. As pointed out in the Introduction, the $\Lambda$ level structure can be regarded as a minimal model of a multilevel quantum system interacting with incoherent BBR. (Another widely used minimal model is the three-level V-system considered elsewhere \cite{Tscherbul:14,Dodin:16}).  
%The $\Lambda$-system is a  fundamental building block of more complex multilevel systems used as models of quantum heat engines \cite{Scully:11,Scully:13} and photosynthetic light-harvesting complexes \cite{Tscherbul:18,Kassal:14,Cao_dimer_JPCL}.
%it is also the central system to quantum optics, with the upper level of the $\Lambda$-system commonly used for coherent state transfer and manipulation of the ground-state doublet \cite{EIT,others}.  
The quantum dynamics of the $\Lambda$-system driven by isotropic incoherent radiation
%(a prototypical thermal environment \cite{})
is described by the Bloch-Redfield (BR) quantum master equations for the density matrix in the eigenstate basis  \cite{Ou:08,Tscherbul:15b}
\begin{align}\notag
\dot\rho_{g_{i}g_{i}} &= -r_{i} \rho_{g_{i}g_{i}} + (r_{i}+\gamma_{i}) \rho_{ee} - p \sqrt{r_{1} r_{2}} \rho^R_{g_{1}g_{2}} 
%\label{BR_Eqns01_Lambda} 
\\
\begin{split}
\dot\rho_{g_{1}g_{2}} &= - i\rho_{g_{1}g_{2}} \Delta - \frac{1}{2}(r_{1}+r_{2}) \rho_{g_{1}g_{2}} \\
&+ p (\sqrt{r_{1} r_{2}}  + \sqrt{\gamma_{1} \gamma_{2}}) \rho_{ee}-\frac{p}{2}\sqrt{r_{1} r_{2}}(\rho_{g_{1}g_{1}}+\rho_{g_{2}g_{2}})  \label{BR_Eqns_Lambda}
\end{split} 
\end{align}
where $\rho_{g_i,g_i}$ are the ground-state populations, $\rho_{g_{1}g_{2}} = \rho^{R}_{g_{1}g_{2}} + i \rho^{I}_{g_{1}g_{2}}$ is the coherence between the ground states  $\ket{g_{1}}$ and $\ket{g_{2}}$ [see Fig. 1 (a)] with the real and imaginary parts $\rho^{R}_{g_{1}g_{2}}$ and $\rho^{I}_{g_{1}g_{2}}$, $\gamma_{i}$ is the radiative decay rate of the excited state $\ket{e}$ into the ground state $\ket{i}$, $r_{i} = \bar{n} \gamma_{i}$ is the incoherent pumping rate, $\bar{n}$ is the average occupation number of the thermal field \textcolor{blue}{($\bar{n}\simeq 10^{-9}$ for typical sunlight-harvesting conditions \cite{Hoki:11,Tscherbul:14})}, $p= (\bm{\mu}_{g_{1}e} \cdot \bm{\mu}_{g_{2}e})/ \mu_{g_{i}e}\mu_{g_{2}e}$ is the transition dipole alignment factor, and $\bm{\mu}_{ij}$ is the transition dipole matrix element between the states $i$ and $j$.

We consider a symmetric $\Lambda$-system ($r_1=r_2=r$) driven by a suddenly turned on incoherent light. This restriction drastically simplifies the solution of the BR equations without losing the essential physics \cite{Tscherbul:14,Dodin:16}. Equations (\ref{BR_Eqns_Lambda}) rely on the Born-Markov approximation, which is known to be very accurate for quantum optical systems \cite{Breuer:06}.
 %Both of these approximations are valid for quantum optical systems interacting with sufficiently broadband incoherent radiation fields \cite{PetruccioneBook}. 
Significantly, we do not assume the validity of the secular approximation, which cannot be justified for nearly degenerate energy levels \cite{Tscherbul:14,Jeske:15,Dodin:16,Dodin:18,Koyu:18,Tscherbul:15b,Eastham:16,Wang:19,Liao:21,Merkli:15,Trushechkin:21}.
This approximation is equivalent to setting $p=0$ in Eqs.~(\ref{BR_Eqns_Lambda}), which eliminates the population-to-coherence coupling terms and hence Fano coherences (see below). Several authors have shown that nonsecular BR equations  \cite{Tscherbul:14,Tscherbul:15b,Eastham:16,Wang:19,Liao:21,Merkli:15,Trushechkin:21} and related Lindblad-form master equations \cite{McCauley:20}  generally provide a more accurate description of open quantum system dynamics than secular rate equations.
    %The resulting Pauli rate equations for population dynamics plus a decoupled set of equations for the coherences \cite{Cohen-Tannoudji:04,Blum:11} are fundamentally un The use of nonsecular  BR equations is essential to properly describe noise-induced Fano coherences.
  %   can exist  This approximation \cite{Tscherbul:14,Dodin:16,Dodin:18}. In the limit of orthogonal  transition dipoles ($p=0$) . Thus, no Fano coherences can be generated from a coherence-free initial state in this limit.
%\textcolor{magenta}{Suyesh, should we add a comment here on the Lindblad form of the BR equations?}

 \begin{figure}[t!]
	\centering
	\includegraphics[width=0.7\columnwidth, trim = 40 350 0 80]{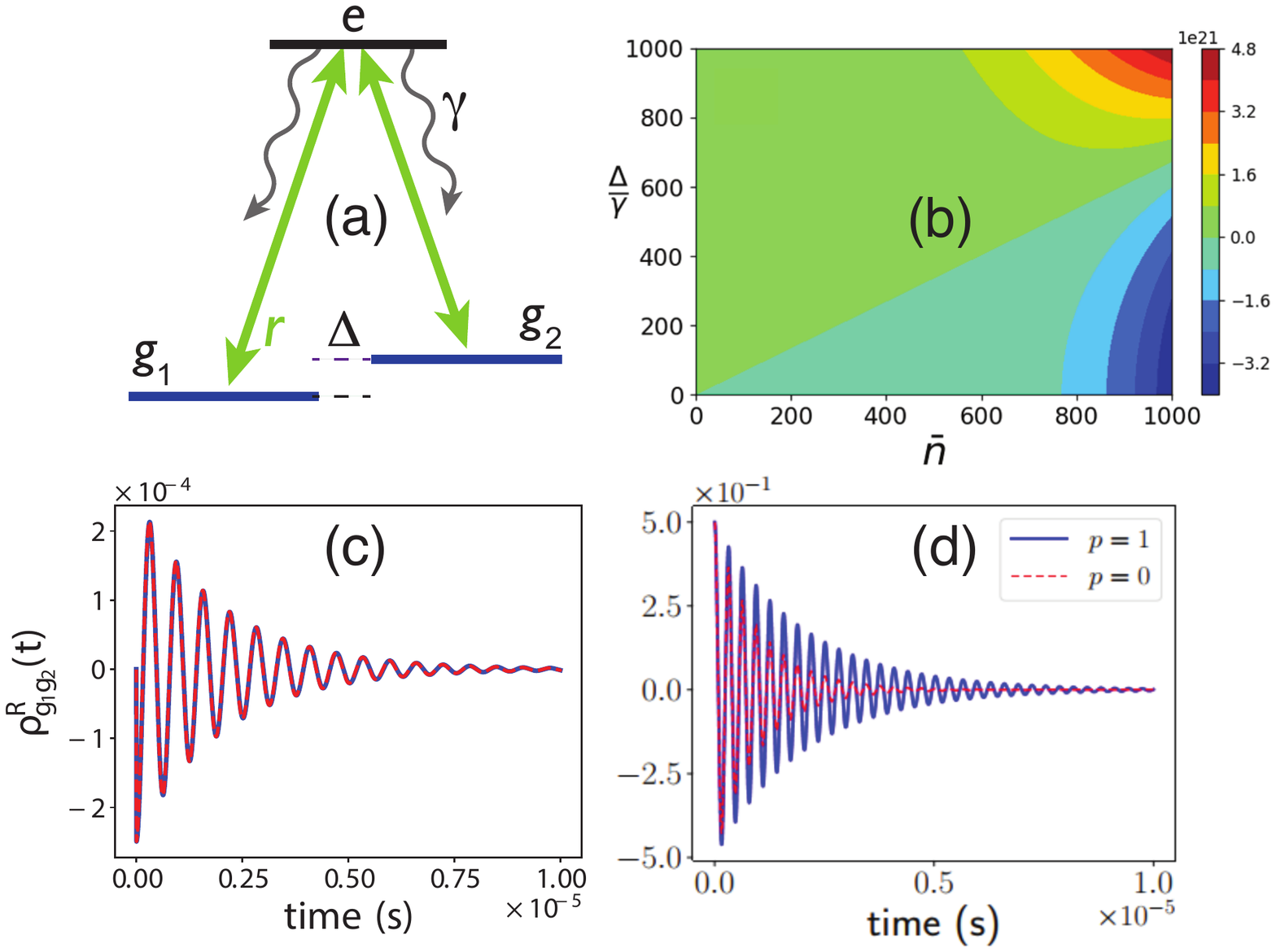}
	\caption{(a) Schematic diagram of the $\Lambda$ system with two ground states $|g_1\rangle$ and $\ket{g_2}$ and a single excited state $|e\rangle$. (b) The discriminant  $D$ as a function of the average photon number $\bar{n}$ and the reduced splitting $\Delta/\gamma$ showing the different dynamical regimes of the incoherently driven $\Lambda$-system. Regions of positive $D$ correspond to the underdamped regime; those of negative $D$ to the overdamped regime. (c) Coherence dynamics of the $\Lambda$ system initially in the fully mixed thermal state \textcolor{blue}{[$\rho_{g_{i}g_{i}}(0)=1/2$, $\rho_{g_1 g_2}(0)=0$]} for $\Delta/\gamma= 10^{-2}$, $\gamma=10^9$~s$^{-1}$, $\bar{n}=10^{-3}$ and $p = 1$. Full line -- numerically exact  results, dashed line -- analytical results. \textcolor{blue}{(d)} Same as (c) for the initial coherent superposition [$\rho_{g_{1}g_{2}}(0) = 1/2$]  \textcolor{blue}{with $p=1$ (full line) and $p = 0$ (dashed line)} for $\Delta/\gamma= 2 \times 10^{-2}$ and $\bar{n}=10^{-3}$.}
	\label{fig:Lambda_sys}
\end{figure}
 
To solve the BR equations (\ref{BR_Eqns_Lambda}) we recast them in matrix form $\dot{\textbf{x}}(t) = \textbf{A} \textbf{x}(t)+ \textbf{d}$, where $ \textbf{x}(t)= ( \rho_{g_{1}g_{1}},  \rho_{g_{1}g_{2}}^{R},  \rho_{g_{1}g_{2}}^{I} )^T $ is the state vector in the Liouville representation, $\mathbf{A}$ is the matrix  of coefficients on the right-hand side of Eq.~(\ref{BR_Eqns_Lambda}), $\textbf{d}$ is a \textcolor{blue}{constant} driving vector, and  $\textbf{x}_{0}$ defines the initial conditions for the density matrix   \cite{SM}.
The solutions of the matrix BR equation $\textbf{x}(t) = e^{\textbf{A}t} \textbf{x}_{0} + \int_{0}^{\infty} ds e^{\textbf{A}(t-s)} \textbf{d}$ may be expressed in terms of the eigenvalues of  $\mathbf{A}$, which determine the decay timescales of the different eigenmodes of the system \cite{SM}. The general features of the solutions can be understood  without  finding the eigenvalues  by considering the discriminant of the characteristic equation \cite{SM}
\begin{equation}\label{Discriminant_Lambda}
D= B^{3} + [C-\frac{3}{2}A(B+A^{2})]^{2} 
\end{equation}
where $A =\frac{1}{3}(5r+2\gamma)$, $B=\frac{1}{3}\big[\Delta^{2}+r^{2}+(2-p^{2})r(3r+2\gamma)\big] - A^{2}$, and 
$C =\frac{1}{2}(3r+2\gamma)\big[\Delta^{2}+(1-p^{2})r^{2}\big] + A^{3}$. 
Three dynamical regimes  can be distinguished depending on the sign of $D$ using the analogy with the damped harmonic oscillator \cite{Dodin:16}. In the underdamped regime $(D>0)$, $\mathbf{A}$ has one \textcolor{blue}{negative} real  and two complex conjugate eigenvalues, giving rise to an exponentially decaying and two oscillating  eigenmodes. 
In the overdamped regime ($D<0$) all of the eigenvalues are real and negative and thus the eigenmodes decay exponentially. In the critical regime  ($D=0$) all eigenvalues are real, negative, \textcolor{blue}{and at least two of them are equal.}

As shown in the Supplementary Material \cite{SM},  $D = \frac{\gamma^{6}}{2916} \sum_{k=0}^{6} d_{k}(p,\frac{\Delta}{\gamma}) \bar{n}^{k}$.
%, where $\bar{n}={r}/{\gamma}$ is the average occupation number of thermal photons.
Figure~\ref{fig:Lambda_sys}(b) illustrates the different dynamical regimes of the incoherently driven $\Lambda$-system obtained by solving the equation $D=0$. 
% for the maximally aligned transition dipoles $(p=1)$. 
We observe that the regions of positive $D$ are separated from those of negative $D$ by the critical line $\Delta/\gamma=f(p)\bar{n}$, {where $f(p)$ is a universal function of $p$ \cite{SM}, which gives the slope of the line.} Thus, {\it the  overdamped regime is realized for  $\Delta/\gamma < f(p)\bar{n}$ and the underdamped regime for  $\Delta/\gamma > f(p) \bar{n}$.}

In the limit $\bar{n}\to 0$ we obtain $D= \frac{\gamma^{6}}{27}(\frac{\Delta}{\gamma})^{2}[(\frac{\Delta}{\gamma})^{2}+4]^{2} >0$. Thus, as shown below, the $\Lambda$-system predomonantly exhibits underdamped oscillatory dynamics under weak thermal driving, in marked contrast with the weakly driven V-system, where the underdamped regime is realized only for $\Delta/\gamma > 1$ \cite{Tscherbul:14, Dodin:16,Koyu:18}. This is because the ground states of the $\Lambda$-system are not subject to spontaneous decay, unlike the excited states of the V-system.  As a result,  the two-photon coherence lifetime of the V-system scales as $ 1/\gamma$, whereas that of the $\Lambda$-system as $1/r$ (as follows from Eq.~(\ref{BR_Eqns_Lambda}) for $p=0$). We note that while the overdamped regime does exist in the weakly driven $\Lambda$-system when $\Delta/r < p^{2}/2$ \cite{SM},  it is a highly restrictive condition, so we limit our discussion below to underdamped dynamics.

The eigenvalues $\lambda_{j}$ of $\mathbf{A}$ give the  decay rates of the corresponding eigenmodes \cite{SM}
% $\textbf{V}_{j}$  
%the We find the eigenvalues $\lambda_{j}$ of the coefficient matrix $\textbf{A}$ (\ref{Amatrix_Lambda}) using SymPy   
\begin{equation}\label{Eigval_Gen}
\lambda_{j} = -A + \alpha_{j} \frac{B}{\mathcal{T}} - \beta_{j} \mathcal{T} \end{equation}
where $ \mathcal{T} = \sqrt[3]{E + \sqrt{D}}$, $ E = [C - \frac{3}{2}A(B+A^{2}) ]$,  $\omega = \frac{-1+i\sqrt{3}}{2}$,
and the quantities  $A$, $B$, and $C$ are defined below Eq. (\ref{Discriminant_Lambda}) and $\alpha_{j}$, $\beta_{j}$ are the cube roots of unity [$(\alpha_{1},\beta_{1}) = (1, 1)$, $(\alpha_{2},\beta_{2}) = (\omega^{2}, \omega)$, and $(\alpha_{3},\beta_{3}) = (\omega, \omega^{2})$]. We next consider the various limits of coherence dynamics defined by the sign of $D$.

In the {\it  weakly driven regime} ($r/\gamma \ll 1$) the general expressions (\ref{Eigval_Gen}) can be simplified
to give $\lambda_{1} = -(3r+2\gamma) = -2\gamma$ and $\lambda_{2,3} = -r Q(\frac{\Delta}{\gamma}) \pm i \Delta$, where $Q(x)\in[1/2,1]$ is a smooth function, which increases monotonically with $x=\Delta/\gamma$ \cite{SM}.
%, which increases from 1/2 to 1 as $x$ in increased.
%. which increases steadily from 1/2 to 1 as $x$ is varied between zero and infinity.
%In the regime $\Delta/\gamma \gg1$ 
 % which increases monotonously from 1/2 ($x\to 0$) to 1  ($x\gg 1$).
 % This decoupling approximation gives accurate results if the ground-state splitting is large compared to the spontaneous decay rate ($\Delta/\gamma > 1$). In the opposite limit of small level spacing, we obtain, in the weak pumping limit, $\lambda_{2,3} = -r (p^2-2)/2$.
The general analytical solutions of the BR equations in the underdamped regime  may be written as (see the Supplementary Material  \cite{SM} for a derivation)
\color{blue}
\begin{align}
\begin{split}
    \rho_{g_{1}g_{2}}^{R}(t) &= \big[\rho^{R}_{g_{1} g_{2}}(0) \cos\Delta t + \rho^{I}_{g_{1} g_{2}}(0)  \sin\Delta t)\big] {e}^{-r Q(\frac{\Delta}{\gamma})t} \\ & + \frac{pr}{2[4\gamma^{2}+\Delta^{2}]} \bigg[2\gamma \text{e}^{-2\gamma t} - \left(  2\gamma \cos \Delta t + \Delta \sin \Delta t \right) {e}^{-r Q(\frac{\Delta}{\gamma})t} \bigg] \notag
\end{split} \\
\begin{split}
    \rho_{g_{1}g_{2}}^{I}(t) &= \big[-\rho^{R}_{g_{1} g_{2}}(0) \sin\Delta t + \rho^{I}_{g_{1} g_{2}}(0) \cos\Delta t\big] \text{e}^{-r Q(\frac{\Delta}{\gamma})t} \\ & + \frac{pr}{2[4\gamma^{2}+\Delta^{2}]} \bigg[ \Delta \text{e}^{-2\gamma t} + \left( 2\gamma \sin \Delta t - \Delta  \cos \Delta t \right) \text{e}^{-r Q(\frac{\Delta}{\gamma})t}\bigg]. \label{General_Real_Coh_Final}
\end{split}
\end{align}
%In Eqs.~(\ref{General_Real_Coh_Final}) and (\ref{General_Imag_Coh_Final})
Here, $\rho_{ij}(0)$ specify the initial conditions for the density matrix of the $\Lambda$-system at $t=0$, which  we  assume to be in an equal coherent superposition [$\rho_{g_1g_1}(0)=\rho_{g_2g_2}(0)=1/2$].
% and arbitrary compatible values of $\rho_{g_1g_2}^R(0)$ and $\rho_{g_1g_2}^I(0)$. 
%(\textcolor{red}{SUYESH, PLEASE CHECK IF THIS IS CORRECT. WE STILL ASSUME THAT  $\rho_{g_1g_1}(0)=\rho_{g_2g_2}(0)=1/2$, RIGHT? Prof. Timur, this is correct.})

In the important particular case of a fully mixed, coherence-free initial state, we have $\rho_{g_1g_2}(0)=0$ and $\rho_{g_ig_i}(0)=1/2$,  and Eqs.~(\ref{General_Real_Coh_Final}) reduce to
\begin{align} \notag
    \rho_{g_{1}g_{2}}^{R}(t) &= \frac{pr}{2[4\gamma^{2}+\Delta^{2}]} \big[2\gamma {e}^{-2\gamma t} -  ( 2\gamma \cos\Delta t + \Delta \sin\Delta t ){e}^{-r Q(\frac{\Delta}{\gamma})  t} \big],  \\
    \rho_{g_{1}g_{2}}^{I}(t)  &=  \frac{pr}{2[4\gamma^{2}+\Delta^{2}]} [\Delta {e}^{-2\gamma t} + (2\gamma \sin\Delta t - \Delta  \cos\Delta t ) {e}^{-r Q(\frac{\Delta}{\gamma)}} ]. \label{Real_Coh_UD_WPL} 
%    \rho_{g_{1}g_{2}}^{I}(t) &= P \bigg[\Delta \text{e}^{-2\gamma t} + \big\lbrace2\gamma \sin\Delta t - \Delta  \cos\Delta t \big\rbrace \text{e}^{-r q(\Delta/\gamma) }\bigg], \label{Imag_Coh_UD_WPL}
\end{align}
The concomitant population dynamics are given by
\begin{equation} \label{BR_Pop_WPL}
  \rho_{g_{1}g_{1}}(t) = \frac{1}{(3r+2\gamma)} [r + \gamma + \frac{r}{2}{e}^{-2\gamma t} ].
\end{equation}
% imaginary part of Fano coherence $\rho_{g_{1}g_{2}}^{I}(t) = P [\Delta {e}^{-2\gamma t} + (2\gamma \sin\Delta t - \Delta  \cos\Delta t ) {e}^{-r Q(\frac{\Delta}{\gamma)}} ]$. 
\color{black}

%\textcolor{red}{SUYESH, your new equations with general initial conditions (4) do not reduce to Eqs. (5) and (6). Please fix! Prof. Timur, it is fixed now.}

\subsection{Noise-induced coherent dynamics of the $\Lambda$-system}

Figure~\ref{fig:Lambda_sys}(c) shows the ground-state population and coherence dynamics of the incoherently driven $\Lambda$-system initially in the incoherent mixture of the ground states ($\rho_{g_i g_i}(0)=1/2$) for $p=1$ obtained by numerical solution of the BR equations. 
We observe that  incoherent driving produces quantum beats due to Fano coherences {\it in the absence of initial coherence in the system}. The coherent oscillations decay on the timescale $\big[rQ(\frac{\Delta}{\gamma})\big]^{-1}$ in agreement with the analytical result (\ref{Real_Coh_UD_WPL}).  
 % The coherence lifetime $\tau=1/r$ is the same as that of an initial coherent superposition $\rho=\frac{1}{\sqrt{2}}(|g_1\rangle + |g_2\rangle)$ in the absence of incoherent driving. This suggests that decoherence occurs through transitions to the  excited state $|g_i\rangle \to |e\rangle$ followed by spontaneous decay of the excited-state population induced by the vacuum modes  of the electromagnetic field.
As the energy gap between the ground levels narrows down, the function $Q(\frac{\Delta}{\gamma})$ decreases from $1$ to $1/2$  \cite{SM} and the coherence lifetime increases by a factor of two. This is because the incoherent excitations $\ket{g}\leftrightarrow \ket{e_i}$ interfere more effectively at small $\Delta/\gamma$ \cite{Koyu:21}, as signaled by the non-negligible population-to-coherence coupling term $-p\sqrt{r_1r_2}\rho^R_{g_1g_2}$  in Eq.~(\ref{BR_Eqns_Lambda}) in the limit $\Delta/\gamma \ll 1$. 
We note  that if incoherent driving is turned on gradually (rather than suddenly, as assumed here)  the quantum beats will become suppressed, until they eventually  disappear  in the limit of adiabatically turned on incoherent driving \cite{Dodin:16b,Dodin:21b}. However, steady-state Fano coherences generated by polarized incoherent excitation (see Sec.~III) do survive the adiabatic turn-on \cite{Koyu:21}.

If the $\Lambda$-system is initialized in a coherent superposition of its ground states ($\rho_{g_1 g_2}(0)=1/2$) the dynamics under incoherent driving
\textcolor{blue}{is given by Eq.~\eqref{General_Real_Coh_Final}  and contains additional terms proportional to $\rho_{g_1g_2}(0)$,} which arise from the coherent initial condition \cite{SM}.
% the initial due to the  $\rho_{g_1g_2}^U(t)$ terms due to the initial  
From Fig.~\ref{fig:Lambda_sys}(d) we observe that, in the absence of Fano interference ($p=0$), the initially excited coherent superposition decays on the timescale $1/r$ as expected due to incoherent transitions to the excited state $|g_i\rangle \to |e\rangle$ followed by spontaneous decay to the vacuum modes  of the electromagnetic field (indeed, the analytical solution of the BR equations (\ref{BR_Eqns_Lambda}) for $p=0$ is $\rho_{g_1g_2}(t)=\rho_{g_1g_2}(0)e^{-(i\Delta + r)t}$).
Thus, Fano interference generated by incoherent driving ``extends'' the lifetime  of the initial coherent superposition due to the noise-induced contribution (\ref{Real_Coh_UD_WPL}).

\begin{figure}[t!]
	\centering
	\includegraphics[width=0.7\columnwidth, trim = 0 20 0 0]{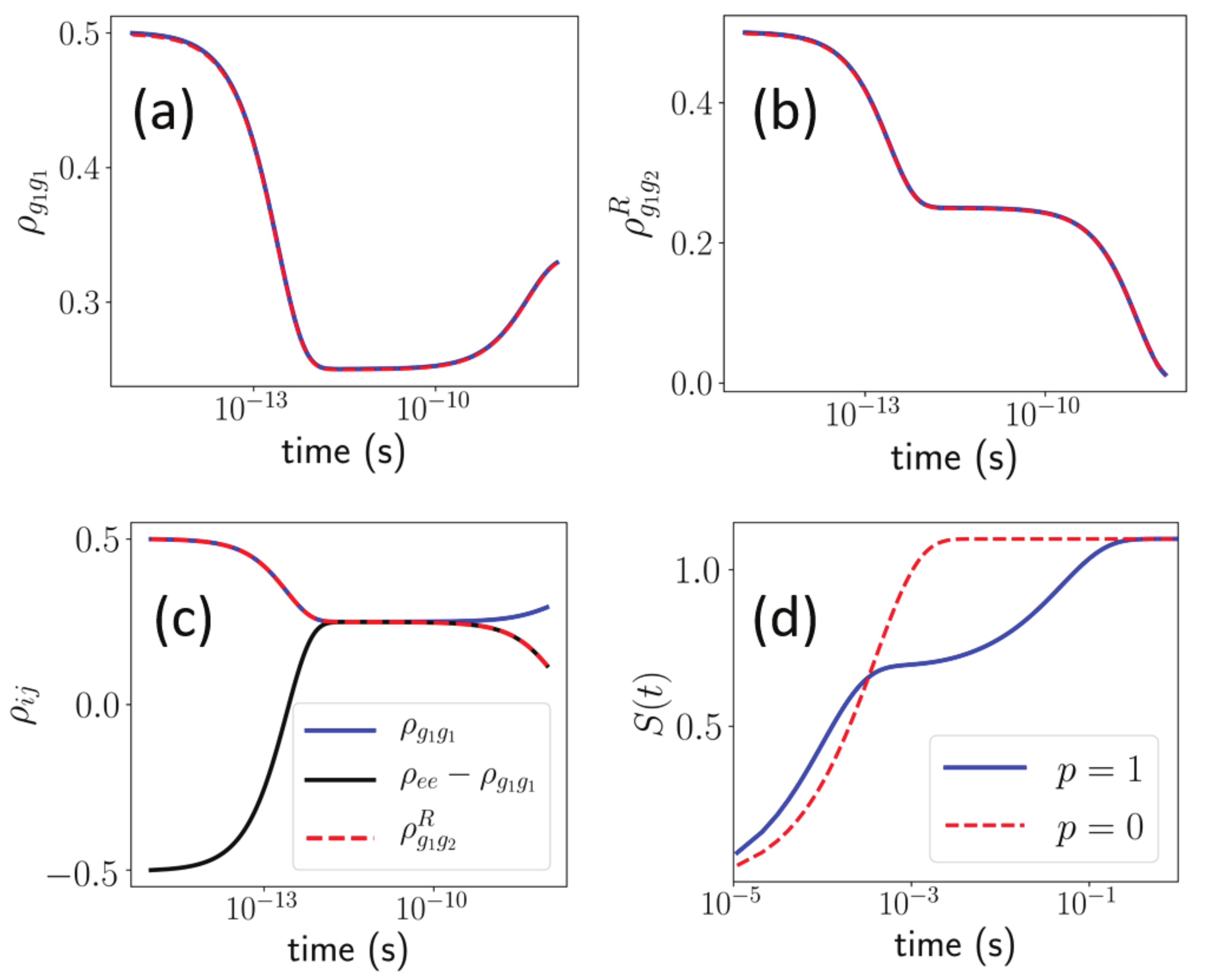}
	\caption{{Ground-state population (a), coherence (b) and population difference (c) dynamics of the $\Lambda$-system strongly driven by incoherent light ($\bar{n}=10^3$, $\gamma = 10^{9} \enspace \text{s}^{-1}$, $\Delta/\gamma=10$ and $p = 1$). \textcolor{blue}{The $\Lambda$ system is initially in the fully mixed state $\rho(0)=\frac{1}{2}(|g_1\rangle\langle g_1| + |g_2\rangle\langle g_2|)$.} (d) Time evolution of the von Neumann entropy with  full Fano coherence [blue solid line] and without Fano coherence [red dashed line] for $\bar{n} = 10^{3}$ and $\Delta/\gamma=10^{2}$. The ground states are initially in the coherent superposition \textcolor{blue}{$\rho(0)=\frac{1}{2}(|g_1\rangle\langle g_1| + |g_2\rangle\langle g_2|+|g_1\rangle\langle g_2| + |g_2\rangle\langle g_1|)$.}}}
%	 ($\rho^{R}_{g_{1}g_{2}} = 1/2$).
	\label{fig:Lambda_dynamics_overdamp}
\end{figure}

We now turn to the overdamped dynamics of the strongly driven $\Lambda$-system defined by the condition $\Delta/r < f(n)$. Expanding the eigenvalues of $\mathbf{A}$ in $1/\bar{n}\ll 1$ we find that  $\lambda_i$ do not depend on $\Delta/\gamma$ for $p<p_c$, where $p_c\simeq 1$ \cite{SM, Koyu:18}. Remarkably, when the transition dipoles are nearly perfectly aligned ($p>p_c$), the scaling  changes dramatically to $\lambda_{2}= -0.75 \frac{\gamma}{\bar{n}} \left({\Delta}/{\gamma}\right)^{2}$, giving rise to a coherent quasi-steady state with the lifetime 
%\begin{equation}\label{tauCoherence}
$\tau_{c} = 1.34 {r}/{\Delta^{2}}$ that increases without limit as $\Delta\to 0$.
 %\qquad (p>p_c).
%\end{equation}
%Thus, similarly to the V-system \cite{Koyu:18}, the strongly driven $\Lambda$-system with $p=1$ exhibits progressively long coherence times as the energy gap between the ground states closes.
This point is illustrated in Figs.~\ref{fig:Lambda_dynamics_overdamp}(a) and (b), where we plot the population and coherence dynamics of the $\Lambda$-system strongly driven by incoherent light. We observe that the coherences rise quickly from zero to an intermediate ``plateau'' value, where they remain for $t=\tau_c$ before eventually decaying back to zero. Our analytical results for coherence dynamics are in excellent agreement with numerical calculations as shown in Figs.~\ref{fig:Lambda_dynamics_overdamp}(a) and (b) (see the Supplementary Material \cite{SM}). They reduce to prior $\Delta=0$ results \cite{Ou:08} in the limit $\Delta\to 0$, where the quasisteady states shown in Figs.~\ref{fig:Lambda_dynamics_overdamp}(a) and (b) become true steady states.
%, and our results reduce to prior $\Delta=0$ values  \cite{Ou:08}).
% with   $\rho_{g_1g_2}(t\to \infty)=X$.  \textcolor{magenta}{[Suyesh, please compare]}.

To understand the physical origin of long-lived Fano coherences in the $\Lambda$-system, we use the effective decoherence rate model \cite{Koyu:18}. As shown in Fig.~\ref{fig:Lambda_dynamics_overdamp}(c)  the decay of the ground-state population $\rho_{g_i g_i}$ is accompanied by a steady growth of the population inversion $\rho_{ee}-\rho_{g_1g_1}$, which drives coherence generation.
% the difference between the excited and ground-state populations.} 
 %In the quasi-steady state the population inversion term $\rho_{ee}-\rho_{g_{1}g_{1}}$ drives coherence generation. 
 %From Fig.~2(c) 
 We observe that in the quasisteady state the time evolution of the population difference is identical to that of   $\rho_{g_{1}g_{2}}^R$ and that $\rho^I_{g_1g_2}$ is time-independent.   
 % incoherent excitation occurs much faster than spontaneous decay ($r\gg \gamma$), which allows us to neglect the 
 % that in formed on the timescale $1/r < t < \tau_c$  (see above)
 % in Fig. 2(a) and (b)
  %Additional insight into the quasisteady state formation can be obtained by 
 Neglecting the terms proportional to $\gamma$ in Eq. (\ref{BR_Eqns_Lambda}), which is a good approximation in the strong pumping limit, and setting the left-land side of the resulting expression to zero, we obtain $\rho_{g_{1}g_{2}}^I=-(\Delta/r)\rho^R_{g_{1}g_{2}}$ \cite{SM}.
 % (we verified this result numerically in the overdamped regime).  
This leads to a simplified  equation of motion for ${\rho}_{g_{1}g_{2}}^R$ valid at $t>1/r$, $\dot{\rho}_{g_{1}g_{2}}^R =  -r\left(1-p + {\Delta^2}/{r^2}\right) \rho_{g_{1}g_{2}}^R$, which implies an exponential decay of the ground-state Fano coherence. The coherence lifetime
% (generated within the first $t \simeq \simeq 1/r$ of coherent evolution). 
%Eq.~(\ref{BRpicture_gamma_eff}) predicts an exponential decay of  the initially formed Fano coherences 
%with an effective decoherence rate  $\gamma_d^\text{eff}(p)=r(1-p+\frac{\Delta^2}{r^2})$, in agreement with the numerical results in Fig. 2(c). Thus, the  {\it effective lifetime }  of  noise-induced Fano coherences 
\begin{equation}\label{BRpicture_tau_eff}
1/\tau_d^\text{eff} = r(1-p + {\Delta^2}/{r^2}),
\end{equation}
is consistent with  the analytical result derived  above.
% up to the factor of 1.34.
There are two distinct contributions to the overall decoherence rate in Eq.~(\ref{BRpicture_tau_eff}) which are similar to those identified in our previous work on the V-system \cite{Koyu:18}: 
(i) the interplay between coherence-generating Fano interference and incoherent stimulated emission  [the term $r(1-p)$] and (ii) the coupling between the real and imaginary parts of the coherence due to the unitary evolution [the term $\Delta^2/r$]. The first mechanism does not contribute in the limit $p\to 1$, explaining the formation of the long-lived coherent quasi-steady state shown in Fig.~\ref{fig:Lambda_dynamics_overdamp}, which decays via mechanism (ii)
%Since $\rho_{g_1g_2}^I$ decays at rate $r$, the decoherence rate due to mechanism (ii) is 
at a rate $\propto r/\Delta^2$.
% the unitary interconversion between 
 %between the real and imaginary parts of the coherence at the rate $\Delta$.  
%As in the case of the V-system \cite{Koyu:18} mechanisms (i) and (ii) contribute equally for $p=p_c^\text{eff}$.
%\color{Blue}
%\color{ForestGreen}

Figure~\ref{fig:Lambda_dynamics_overdamp}(d) shows the time evolution of the von Neumann entropy, 
   \textcolor{blue}{$ \mathcal{S}(t) = - \text{Tr}(\rho \text{ln}\rho)$},
 of the incoherently driven $\Lambda$-system calculated  with ($p=1$) and without $(p=0)$ Fano coherence.
\color{blue}
  The analytic expression for the entropy in terms of the populations and coherences is \cite{SM}
  \begin{equation} \label{von_Neuman_Entropy_eigenbasis2}
    \mathcal{S}(t) = -\Big[\big(\rho_{g_{1}g_{1}} + |\rho_{g_{1}g_{2}}|\big) \text{ln}\big(\rho_{g_{1}g_{1}} + |\rho_{g_{1}g_{2}}|\big) + \big(\rho_{g_{1}g_{1}} - |\rho_{g_{1}g_{2}}|\big) \text{ln}\big(\rho_{g_{1}g_{1}} - |\rho_{g_{1}g_{2}}|\big) + \rho_{ee} \text{ln}\rho_{ee}  \Big]
\end{equation}
with $\rho_{ee}=1-\rho_{g_1g_1}-\rho_{g_2g_2}$.
\color{black}
     We observe that the  entropy of the long-lived coherent quasi-steady state  is two times smaller than that of the corresponding $p=0$ thermal state  due to the presence of substantial coherences in the energy basis. The low-entropy state persists for $\tau_c\propto (r/\Delta^{2})$ before decaying to the high-entropy thermal state. 
%when the entropies of both states equalize.
% equal.
% and without Fano coherence ($p=0$ vs. $p=1$) initially increase   

  \begin{figure}[t!]
	\centering
	\includegraphics[width=0.75\columnwidth, trim = 60 525 0 80]{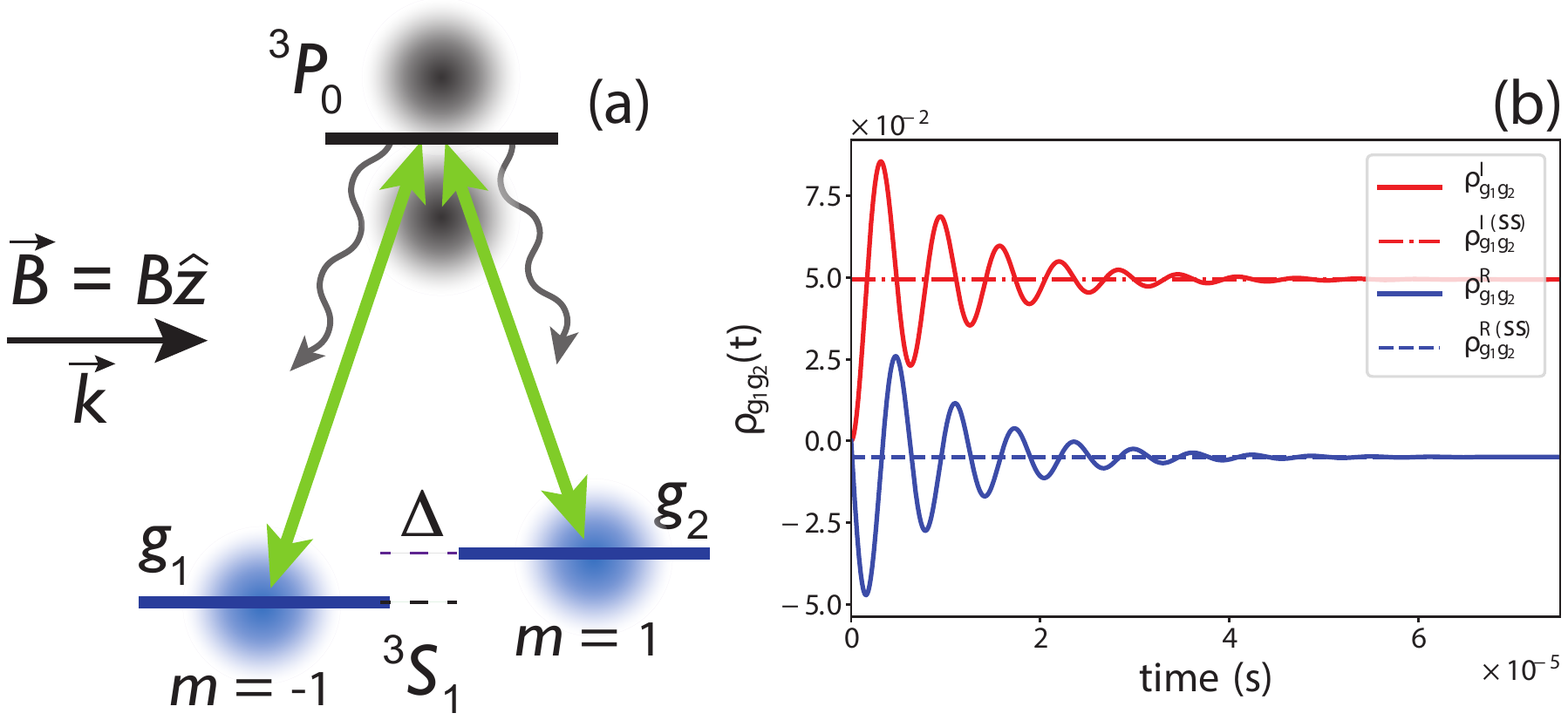}
	\caption{(a) Schematic diagram of the $\Lambda$-system  configuration to detect Fano coherences in He* atoms. The  $^3S_{m=\pm 1}\leftrightarrow {^3}P_{0}$ transitions  are indicated by double-headed arrows. The magnetic field is parallel to the propagation vector of the $x$-polarized incoherent light, which defines the quantization axis. {(b) Ground-state coherence dynamics of He* excited by $x$-polarized incoherent light with the parameters $\gamma=10^8$~s$^{-1}$, $\bar{n} = 10^{-3}$, and $\Delta/\gamma = 10^{-2}$.} }
	\label{fig:exp}
\end{figure}

\section{Proposal for experimental observation of Fano coherences in He$^*$ atoms}

We finally turn to the question of experimental observation of Fano coherences. We suggest metastable He($2^3S_1$) atoms \cite{Vassen:12,Vassen:16} as a readily realizable $\Lambda$-system, in which to observe noise-induced  coherent dynamics. 
% in the incoherently driven $\Lambda$-system.
%could be a promising candidate,  
 The $\Lambda$-system is formed by the $m=\pm 1$ Zeeman sublevels of the metastable $^3S$ state and the nondegenerate excited $^3P_0$ state as shown in Fig.~\ref{fig:exp}(a). The energy gap $\Delta$ between the Zeeman sublevels   is continuously tunable with an external magnetic field, providing access to both the overdamped and underdamped regimes of coherence dynamics. The $^3S_{m=\pm 1}\leftrightarrow {^3}P_{0}$ transitions are driven by a spectrally broadened laser field  polarized in the $x$-direction.
  %as proposed in our previous work on the $V$-system \cite{Dodin:18}. 
  This excitation scheme allows us to (i) neglect radiative transitions involving the $m=0$ ground-state Zeeman sublevel, thereby realizing an ideal three-level $\Lambda$-system (since He$^*$ has no hyperfine structure), and (ii) bypass the $p=1$ condition needed to generate Fano interference via isotropic incoherent excitation \cite{Dodin:18,Koyu:21}.  Because both of the $\Lambda$-system transitions couple to the same polarization mode of the incoherent radiation field, the BR equations (\ref{BR_Eqns_Lambda}) can be simplified by replacing $r+\gamma \to \gamma$ \cite{Dodin:18,Koyu:21,SM}.

Figure~\ref{fig:exp}(b) shows ground-state Fano coherence dynamics of a He$^*$ atom driven by $x$-polarized incoherent  light starting from a coherence-free initial state $\rho_{g_ig_i}(0)=1/2$. As in the case of isotropic incoherent excitation considered above, the coherences exhibit quantum beats with frequency $\Delta$ and lifetime $1/(Qr)$. The coherent evolution  could be probed by applying a $\pi/2$ radiofrequency pulse (as part of the standard Ramsey sequence \cite{Degen:17,Vassen:16})  to convert the coherences to the populations of the $m=\pm 1$ atomic states, which could be measured  by, e.g., state-selective photoionization  \cite{Vassen:12,Vassen:16}.

Significantly, as shown in Fig.~\ref{fig:exp}(b), the Fano coherences generated by polarized incoherent excitation do not vanish in the steady state $(t\to \infty)$ unlike those in isotropic excitation.  This is a result of the imbalance between polarized incoherent excitation and spontaneous emission (the former is directional whereas the latter is isotropic), leading to a breakdown of detailed balance, and the emergence of non-equilibrium steady-states \cite{Agarwal:01,Dodin:18,Koyu:21}. The steady-state populations and coherences \cite{SM}
\begin{align}\notag
\rho_{g_{i}g_{i}}^{(\text{SS})} &= \frac{(r+\gamma) \Delta^{2}+r^{2}\gamma}{(3r+2\gamma)\Delta^{2}+2r^{2}\gamma},
\\
\begin{split}
\rho_{g_{1}g_{2}}^{R(\text{SS})} &= -\frac{r^{2}\gamma}{(3r+2\gamma)\Delta^{2}+2r^{2}\gamma}
  \label{rho_ss}
\end{split} 
\end{align}
deviate from the values expected in thermal equilibrium $[\rho_{g_{i}g_{i}}^{(\text{SS,th})}=(r+\gamma)/(3r+2\gamma)]$. As in the case of the V-system \cite{Koyu:21}, this could be used to detect Fano coherences by measuring the deviation of steady-state populations from their expected values in thermodynamic equilibrium. 
%$(\rho_{g_ig_i}-\rho_{g_ig_i}^\text{th})/\rho_{g_ig_i}^\text{th}$ 
%\cite{Vassen:12,Vassen:16}.

\section{Summary}

We have explored the quantum dynamics of noise-induced Fano coherences in a prototypical three-level $\Lambda$-system driven by BBR. In contrast to its V-system counterpart  \cite{Tscherbul:14,Dodin:16,Dodin:18,Koyu:18} the weakly driven $\Lambda$-system almost always remains in the underdamped regime characterized by oscillatory coherence dynamics that decays on the timescale $1/r$ (for $\Delta/\gamma \gg 1)$ or $2/r$  (for $\Delta/\gamma \ll 1)$, which can be extremely long \textcolor{blue}{for small pumping rates $r/ \gamma  \simeq 10^{-9}$ relevant for photosynthetic solar  light harvesting \cite{Hoki:11,Tscherbul:14}.}
%The coherence time of the $\Lambda$-system initially prepared in a coherent superposition of its ground states is enhanced by a factor of two  in the presence of  Fano coherence induced by  incoherent driving.
This also suggests that Fano coherences may find applications in, e.g., quantum sensing  \cite{Degen:17}, where a qubit's coherence time is a key figure of merit.
Similarly, the long-lived coherent quasi-steady states that arise in the strongly driven $\Lambda$-system [see Fig.~(\ref{fig:Lambda_dynamics_overdamp})] have lower entropy than the corresponding thermal states, motivating further research into the origin and potential utility of these states [as well as the true coherent steady states shown in Fig.~\ref{fig:exp}(b)] in multilevel quantum systems.

\color{blue}
While both the V- and $\Lambda$ systems are three-level systems, the crucial difference between them is that the nearly degenerate levels of the V-system  are subject to spontaneous emission, whereas those of the $\Lambda$ system are not. As a consequence, these systems exhibit very different noise-driven dynamics in the regime, where spontaneous emission rate is much higher than the incoherent driving rate. This occurs in the weakly driven regime $(r/\gamma \ll 1)$ of importance for photosynthetic light-harvesting under ambient sunlight conditions.
On the other hand, if the rate of spontaneous emission is smaller than that of incoherent pumping (the strongly driven regime, $r/\gamma \gg 1$), then one might expect the V and $\Lambda $ systems to display similar dynamical features, as is indeed the case. 
One notable example is provided by the long-lived coherent quasi-steady states, which have exactly the same lifetimes for both the V and $\Lambda$ systems, and can be described by the same effective decoherence rate model \cite{Koyu:18}.
%For example, the lifetimes of noise-induced  coherences in the strongly driven $\Lambda$ and V-systems are the same  and both can be explained by the effective decoherence rate model  \cite{Koyu:18}.

In addition to similar dynamical behavior, the mathematical features of both models in the strongly driven regime are fairly  similar. For instance, the   zero-discriminant equation of the strongly driven $\Lambda$-system given by Eq. (21) of the Supplementary Material \cite{SM} is the same as that of the V-system. Furthermore, the expressions for the population and coherence dynamics of the strongly driven $\Lambda$- and V-systems have the same form when expressed in terms of the coefficients $A_i,B_i$, and $C_i$ (see \cite{SM}, Sec.  I.E2).
Nevertheless, these coefficients have a completely different dependence on the transition dipole alignment factor $p$ for the V and $\Lambda-$systems.  

\color{black}

Finally, we propose an experimental scenario for detecting Fano coherences by driving circularly polarized $^3S_{m=\pm 1}\leftrightarrow {^3}P_{0}$ transitions in metastable He atoms excited by $x$-polarized incoherent radiation and subject to an external magnetic field.
Our results suggest that thermal driving can lead to novel long-lived quantum beats and non-thermal coherent steady states in multilevel atomic and molecular systems, motivating further studies of Fano coherences in more complex multilevel molecular systems and of their role in photosynthetic energy transfer processes and in quantum information science.

\section*{Supplementary Material}
The Supplementary Material contains a detailed derivation of the analytical solutions of the BR equations   presented in the main text.

\section*{Acknowledgements}
We thank Prof. Amar Vutha for suggesting He$^*$ as an experimental realization of the $\Lambda$-system, and  Prof. Paul Brumer for a valuable discussion.
This work was partially supported by NSF PHY-1912668.

 %\section*{Data availability statement}
%The data that support the findings of this study are available from the corresponding author upon reasonable request.

%

\newpage
\widetext
\begin{center}
\textbf{\large Supplemental Material}
\end{center}
%%%%%%%%%% Merge with supplemental materials %%%%%%%%%%
%%%%%%%%%% Prefix a "S" to all equations, figures, tables and reset the counter %%%%%%%%%%
\setcounter{equation}{0}
\setcounter{figure}{0}
\setcounter{table}{0}
\setcounter{page}{1}
\setcounter{section}{0}

This Supplementary Material presents a derivation of analytical solutions of the Bloch-Redfield (BR) quantum master equations for the  $\Lambda$-system driven by isotropic incoherent light (Sec. I). In Sec.~II we derive and solve the BR equations for the $\Lambda$-system driven by $x$-polarized  incoherent light.

Within Sec.~I, section IA provides an overview of the BR  equations and of the theoretical  procedures used to obtain their analytical solutions. The dynamical regimes of the $\Lambda$-system are classified in Sec.~IB. In Secs.~IC and ID we derive the expressions for the eigenvalues and eigenvectors of the coefficient matrix $\mathbf{A}$ (see main text). {Section~IE is devoted to the derivation of analytical solutions to the BR-equations in both the overdamped and the underdamped regimes of the strong pumping limit. The effective decoherence rate model used to explain the physical origin of long-lived Fano coherences is derived in Sec.~IF. 

Section~IG is devoted to the weak pumping limit. We derive the eigenvalues of matrix  $\mathbf{A}$, discuss their qualitative features, and provide analytical solutions for the population and coherence dynamics. In Sec.~IH, we describe the procedure of computing von Neumann entropy of the incoherently driven $\Lambda$-system. }

Section IIA presents the derivation of the BR equations for the $\Lambda$-system driven by polarized incoherent radiation. The steady-state solutions of these equations are discussed in Sec.~IIB. The coefficients relevant to the analytical solutions of the BR equations are listed in Sec.~III.

\section{$\Lambda$-system driven by incoherent light: The Bloch-Redfield equations and their analytical solution}

\subsection{Bloch-Redfield equations and their general solutions}
We consider a symmetric $\Lambda$-system (see  main text and Fig.~1) that interacts with a suddenly turned on incoherent radiation field. Within the framework of the Born-Markov approximation, the time dynamics of a $\Lambda$-system is given by the Bloch-Redfield (BR) quantum master equations \cite{Tscherbul:15b}
\begin{align}
\dot\rho_{g_{i}g_{i}} &= -r_{i} \rho_{g_{i}g_{i}} + (r_{i}+\gamma_{i}) \rho_{ee} - p \sqrt{r_{1} r_{2}} \rho^R_{g_{1}g_{2}} \label{BR_Eqns01_Lambda} \\
\begin{split}
\dot\rho_{g_{1}g_{2}} &= -\frac{1}{2}(r_{1}+r_{2}) \rho_{g_{1}g_{2}} - i\rho_{g_{1}g_{2}} \Delta + p (\sqrt{r_{1} r_{2}}  + \sqrt{\gamma_{1} \gamma_{2}}) \rho_{ee}-\frac{p}{2}\sqrt{r_{1} r_{2}}(\rho_{g_{1}g_{1}}+\rho_{g_{2}g_{2}})  \label{BR_Eqns02_Lambda}
\end{split} 
\end{align}
where
\begin{equation}
\rho_{g_{1}g_{2}} = \rho^{R}_{g_{1}g_{2}} + i \rho^{I}_{g_{1}g_{2}} \label{Coh_Eqn}
\end{equation}
is the coherence between the ground states $\ket{g_{1}}$ and $\ket{g_{2}}$ with the real and imaginary parts $\rho^{R}_{g_{1}g_{2}}$ and $\rho^{I}_{g_{1}g_{2}}$, $\gamma_{i}$ is the radiative decay rate of level $\ket{g_{i}}$ ($i = 1, 2$), $r_{i} = \bar{n} \gamma_{i}$ is the incoherent pumping rate, $p= (\bm{\mu}_{g_{1}e} \cdot \bm{\mu}_{g_{2}e})/ \mu_{g_{1}e}\mu_{g_{2}e}$ is the transition dipole alignment factor, and $\bm{\mu}_{g_{1}e}$, $\bm{\mu}_{g_{2}e}$ are the matrix elements of the transition dipole moment vector between the excited and ground states $\ket{e}$ and $\ket{g_{i}}$. \par 

\begin{figure}[b]
\centering
\includegraphics[width=0.9\textwidth, trim = 0 300 0 370]{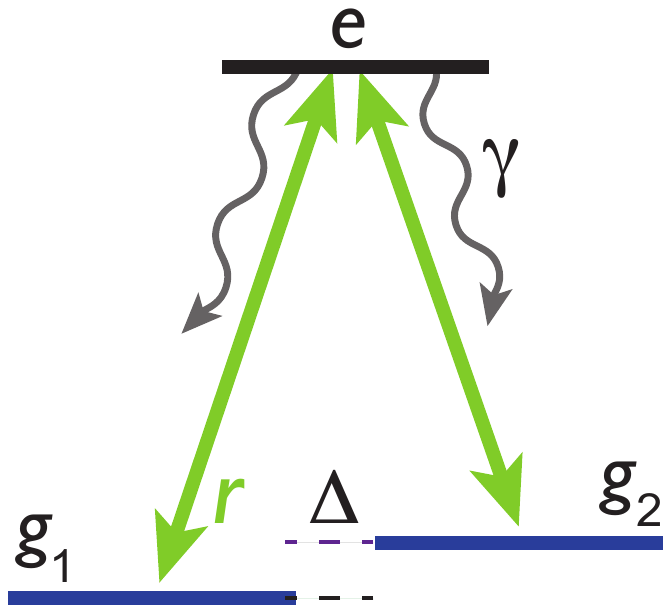}
\caption{Schematic diagram of the $\Lambda$-system with two quasi-degenerate ground states $|g_1\rangle$ and $|g_2\rangle$ and a single excited state $|e\rangle$. The incoherent pumping rates are $r$, the spontaneous decay rate of the excited state is $\gamma$, and the energy gap between the ground states is $\Delta$.}
\label{fig: Lambda-system}
\end{figure}

For a symmetric $\Lambda$-system, where $r_{1} = r_{2} = r$, and $\gamma_{1} = \gamma_{2} = \gamma$, the BR equations ($\ref{BR_Eqns01_Lambda}$) - (\ref{BR_Eqns02_Lambda}) combined with Eq.~(\ref{Coh_Eqn}) reduce to \begin{align}
\dot\rho_{g_{1}g_{1}} &= -r \rho_{g_{1}g_{1}} + (r+\gamma) \rho_{ee} - pr\rho^R_{g_{1}g_{2}} \label{BR_Pop} \\
\dot\rho^{R}_{g_{1}g_{2}} &= -pr\rho_{g_{1}g_{1}} + p(r+\gamma)\rho_{ee} - r\rho^{R}_{g_{1}g_{2}} + \Delta \rho^{I}_{g_{1}g_{2}} \label{BR_Real_Coh} \\
\dot\rho^{I}_{g_{1}g_{2}} &= -\Delta \rho^{R}_{g_{1}g_{2}} - r \rho^{I}_{g_{1}g_{2}} \label{BR_Imag_Coh}
\end{align}

These  equations can be expressed in matrix form 
\begin{equation}
\begin{bmatrix}
           \dot\rho_{g_{1}g_{1}} \\
           \dot\rho^{R}_{g_{1}g_{2}} \\
           \dot\rho^{I}_{g_{1}g_{2}} \\
\end{bmatrix}
=
\begin{bmatrix}
    -(3r+2\gamma) & -pr & 0 \\
    -p(3r+2\gamma) & -r & \Delta  \\
       0 & -\Delta & -r \\
\end{bmatrix}
\begin{bmatrix}
           \rho_{g_{1}g_{1}} \\
           \rho^{R}_{g_{1}g_{2}} \\
           \rho^{I}_{g_{1}g_{2}} \\
\end{bmatrix}
+
\begin{bmatrix}
       (r+\gamma)  \\
       p(r+\gamma) \\
       0  \\
\end{bmatrix} \label{BR_Matrix_Eqn}
\end{equation}
or
\begin{equation}
\dot{\textbf{x}}(t) = \textbf{A} \textbf{x}(t)+ \textbf{d} \label{BREqn_StateVec}
\end{equation}
where $\textbf{x}(t)=\begin{bmatrix}
         \rho_{g_{1}g_{1}} &   \rho_{g_{1}g_{2}}^{R} & \rho_{g_{1}g_{2}}^{I}  
        \end{bmatrix}^{T}$ 
is the state vector in  the Liouville representation,
\begin{equation}
\textbf{A} = 
\begin{bmatrix}
    -(3r+2\gamma) & -pr & 0 \\
    -p(3r+2\gamma) & -r & \Delta  \\
       0 & -\Delta & -r \\
\end{bmatrix} \label{Amatrix_Lambda}
\end{equation}
\textcolor{blue}{
is the matrix of coefficients on the right-hand side of Eqs.~(\ref{BR_Pop})-(\ref{BR_Imag_Coh}), and $\textbf{d}=\begin{bmatrix}
         (r+\gamma) &  p(r+\gamma)  & 0
        \end{bmatrix}^{T}$
is the driving vector.
} 

The inhomogeneous differential equations (\ref{BREqn_StateVec}) can be explicitly solved as %\cite{Boys}
\begin{equation}
    {\textbf{x}}(t) = e^{\textbf{A}t} \textbf{x}_{0} + \int_{0}^{t} ds e^{\textbf{A}(t-s)} \textbf{d}, \label{BR_Gen_Sol}
\end{equation}
where $\textbf{x}_{0}$ defines the initial conditions for the density matrix. Since we are interested in the generation of noise-induced coherences by incoherent driving,  we choose an equal  incoherent mixture of the ground states as our initial state, i.e., $\rho_{g_{1}g_{1}}(t) = \rho_{g_{2}g_{2}}(t) = \frac{1}{2}$ or $\textbf{x}_{0 } = [\frac{1}{2}, 0, 0]^{T}$.

\subsection{Dynamical regimes}
The physical behavior of the $\Lambda$-system as described by the BR equations (\ref{BREqn_StateVec}) is governed by the eigenvalues of the coefficient matrix $\textbf{A}$. Without explicitly finding the eigenvalues, their general features can be understood by  analyzing the discriminant $D$ of the characteristic equation for $\textbf{A}$
\begin{equation}
D= B^{3}+\Big[C-\frac{3}{2}A(B+A^{2})\Big]^{2} \label{Discriminant_Lambda}
\end{equation}
where,
\begin{align}
A &=\frac{1}{3}(5r+2\gamma) \label{A_term} \\
B &=\frac{1}{3}\big[\Delta^{2}+r^{2}+(2-p^{2})r(3r+2\gamma)\big] - A^{2} \label{B_term} \\
C &=\frac{1}{2}(3r+2\gamma)\big[\Delta^{2}+(1-p^{2})r^{2}\big] + A^{3} \label{C_term}
\end{align} 

The dynamical regimes of a $\Lambda$-system can be classified into three types depending on the sign of $D$:
\begin{enumerate}
\item 
\textit{Underdamped regime} ($D>0$) 
The coefficient matrix $\textbf{A}$ given by Eq.~(\ref{Amatrix_Lambda}) has three roots, with one of them being real and the other two being complex conjugates of each other. The normal modes consist of an exponentially decaying eigenmode and two oscillating eigenmodes. 
%This regime is called underdamped regime, using the analogy to a damped harmonic oscillator.
\item 
\textit{Overdamped regime} ($D<0$)
The roots of the coefficient matrix $\textbf{A}$ are all real and negative. The normal modes  all  decay exponentially as a function of time. 
\item 
\textit{Critical regime} ($D=0$)
The coefficient matrix $\textbf{A}$ has three real roots with at least two of them being equal. This regime is at the boundary between the underdamped and overdamped regimes.
\end{enumerate}

It is convenient to express the terms $A$, $B$, $C$ as a function of the occupation number $\bar{n} = r/\gamma$
\begin{align}
A &= \frac{\gamma}{3}\big(5\bar{n}+2\big), \label{A_term_Poly} \\
B &= \frac{\gamma^{2}}{9}\bigg[\Big(3\frac{\Delta^{2}}{\gamma^{2}}-4\Big)-\big(8+6p^{2}\big)\bar{n}-\big(4+9p^{2}\big)\bar{n}^{2}\bigg], \label{B_term_Poly} 
\end{align}
\begin{align}
\begin{split}
C &= \frac{\gamma^{3}}{54}\bigg[\Big(54\frac{\Delta^{2}}{\gamma^{2}}+128\Big)+\Big(81\frac{\Delta^{2}}{\gamma^{2}}+120\Big)\bar{n}+\Big(354-54p^{2}\Big)\bar{n}^{2}+\Big(331-81p^{2}\Big)\bar{n}^{3}\bigg], \label{C_term_Poly}
\end{split}
\end{align}
\begin{align}
C-\frac{3}{2}A(B+A^{2}) &= \frac{\gamma^{3}}{54}\bigg[\Big(36\frac{\Delta^{2}}{\gamma^{2}}+128\Big)+\Big(36\frac{\Delta^{2}}{\gamma^{2}}+48+36p^{2}\Big)\bar{n}+\Big(48+90p^{2}\Big)\bar{n}^{2}+\Big(16+54p^{2}\Big)\bar{n}^{3}\bigg]. \label{E_term_Poly}
\end{align} 

Substituting Eqs.~(\ref{B_term_Poly}) and (\ref{E_term_Poly}) into Eq.~(\ref{Discriminant_Lambda}) we obtain a general expression for $D$ as a polynomial function of the occupation number ($\bar{n}=\frac{r}{\gamma}$)
\begin{equation} \label{Disc_Poly}
D=\frac{\gamma^{6}}{2916}  \sum\limits_{k=0}^{6} d_{k} \bar{n}^{k},
\end{equation} 
where the expansion coefficients $d_{k}$ are listed in Table I.

\subsubsection{Solving the equation $D = 0$ in the strong pumping limit}
For large $\frac{\Delta}{\gamma}$ and $\bar{n}$, the significant terms in Eq.~(\ref{Disc_Poly}) are $d_{0}$, $d_{4}\bar{n}^{4}$, and $d_{6} \bar{n}^{6}$, and the discriminant takes the form
\begin{equation}
D=\frac{\gamma^{6}}{2916}\big(d_{0} + d_{4} \bar{n}^{4} + d_{6} \bar{n}^{6}\big)
\end{equation}

To solve  the equation $D = 0$, we take $\frac{\Delta}{\gamma} = y$, $\bar{n} = x$ and simplify as
\begin{align}
\begin{split}
0 &= \bigg\lbrace 4\big(3y^{2}-4\big)^{3} + \big(36 y^{2}+16\big)^{2}\bigg\rbrace + \bigg\lbrace\big(12\big(4+9p^{2}\big)^{2}(3y^{2}-4\big)-12\big(4+9p^{2}\big)\big(8+6p^{2}\big)^{2}+\big(48+90p^{2}\big)^{2} \\ & \qquad +2\big(16+54p^{2}\big)\big(36y^{2}+48 +36p^{2}\big)\bigg\rbrace x^{4} + \bigg\lbrace\big(16+54p^{2}\big)^{2}-4\big(4+9p^{2}\big)^{3}\bigg\rbrace x^{6}
\end{split} \nonumber 
\end{align}
Dividing on both sides by $x^{6}$, we get
\begin{equation}
\begin{split}
0 &= 108\bigg(\frac{y}{x}\bigg)^{6} + \bigg\lbrace 36\big(4+9p^{2}\big)^{2} + 72\big(16+54p^{2}\big)^{2}\bigg\rbrace \bigg(\frac{y}{x}\bigg)^{2} - \bigg\lbrace 12\big(4+9p^{2}\big)\big(8+6p^{2}\big)^{2}-\big(48+90p^{2}\big)\bigg\rbrace \frac{1}{x^{2}} \\ & \qquad + \bigg\lbrace \big(16+54p^{2}\big)^{2}-4\big(4+9p^{2}\big)^{3}\bigg\rbrace \end{split} \nonumber 
\end{equation}

Neglecting the term proportional to $\frac{1}{x^{2}}$ for large $\bar{n} = x$ and defining $(\frac{y}{x})^{2} = z$, the above equation reduces to the form of a depressed cubic 
\begin{align} \label{Depressed_Cubic}
z^{3} + P z + Q &= 0,
\end{align}
where
\begin{align}
P &= \frac{1}{108}\bigg\lbrace 36\big(4+9p^{2}\big)^{2} + 72\big(16+54p^{2}\big)^{2}\bigg\rbrace = 16+60p^{2}+27p^{4},\\ 
Q &= \frac{1}{108}\bigg\lbrace \big(16+54p^{2}\big)^{2}-\big(4+9p^{2}\big)^{3}\bigg\rbrace = -9p^{4}(1+3p^{2}).
\end{align}
We note that the coefficients $P$ and $Q$ are the same as those derived earlier for a strongly driven V-system \cite{Koyu:18}, which means that the depressed cubic equation (\ref{Depressed_Cubic}) is identical to the equation that defines the zero-discriminant line for the V-system \cite{Koyu:18}. This implies close similarity between the discriminants and, hence, noise-induced  coherent dynamics, of the strongly driven $\Lambda$- and V-systems.

%for large $\Delta/\gamma$ and $\bar{n}$ 
%the large-$\Delta/\gamma$ and $\bar{n}$ 
% implies that the $\Lambda$- and V-system 

To solve the depressed cubic equation, we substitute $z = u + v$ into Eq. ~(\ref{Depressed_Cubic})
which becomes 
\begin{equation} \label{Depressed_Cubic2}
u^{3} + v^{3} + \big(u+v\big) \big(3uv+P\big) + Q = 0 
\end{equation}

The arbitrary variables $u$, $v$ are chosen in such a way that 
\begin{align}
\big(3uv+P\big) &= 0 \nonumber \\
uv = - \frac{P}{3} \label{uv_Product}
\end{align}

Cubing Eq.~(\ref{uv_Product}) on both sides and expressing $v^{3}$ in terms of $u^{3}$, we get
\begin{equation} \label{v_cube}
v^{3} = -\frac{P^{3}}{27} \frac{1}{u^{3}}
\end{equation} 

Using Eqs.~(\ref{uv_Product}) and (\ref{v_cube}) into Eq.~(\ref{Depressed_Cubic2}), we rearrange terms to get a quadratic equation in $u^{3} = t$
\begin{equation}
t^{2} + Q t - \frac{P^{3}}{27} = 0
\end{equation}

The two roots of the above equation are 
\begin{equation} \label{t1_PhaseD}
t_{1} = -\frac{Q}{2} + \sqrt{\frac{Q^{2}}{4}+\frac{P^{3}}{27}}
\end{equation}

\begin{equation} \label{t2_PhaseD}
t_{2} = -\frac{Q}{2} - \sqrt{\frac{Q^{2}}{4}+\frac{P^{3}}{27}}
\end{equation} 

We set $u^{3} = t_{1}$, $v^{3} = t_{2}$ which satisfy the required conditions $u^{3}+v^{3} = -Q$, $u^{3} v^{3} = -\frac{P^{3}}{27}$. This shows that $u = \sqrt[3]{t_{1}}$ and $v = \sqrt[3]{t_{2}}$ are solutions to Eq.~((\ref{Depressed_Cubic2}). As $z = (\frac{y}{x})^{2}$ cannot be complex valued and $\sqrt[3]{t_{1}}$, $\sqrt[3]{t_{1}}$ are real and positive for all $p$, the real solution for $z$ is  
\begin{align}
z &= u + v, \nonumber \\
\bigg(\frac{y}{x}\bigg)^{2} &= \sqrt[3]{t_{1}} + \sqrt[3]{t_{2}}, \nonumber \\ 
\frac{\Delta}{\gamma} &= \sqrt{\sqrt[3]{t_{1}} + \sqrt[3]{t_{2}}} \enspace \bar{n}, \nonumber \\
\frac{\Delta}{\gamma} &= f(p) \enspace \bar{n},
\end{align}  
where 
\begin{equation}
f(p) = \sqrt{\sqrt[3]{t_{1}} + \sqrt[3]{t_{2}}}
\end{equation} 
with $t_{1}$ and $t_{2}$ given by Eqs.~(\ref{t1_PhaseD}) and (\ref{t2_PhaseD}).

This shows that in the strong pumping limit and for large values of $\frac{\Delta}{\gamma}$, and $\bar{n}$ the critical line behaves as a straight line with the slope given by $m = f(p)$, a function of $p$ only.

%\subsection{Dynamical regimes}
%The dynamical regimes in the strong pumping regime is governed by the general discriminant (\ref{General_Disc_Lambda})
%\begin{equation} \label{General_Disc_SPL}
%D = \frac{\gamma^{6}}{2916} \sum_{k=0}^{6} d_{6} \bar{n}^{k}
%\end{equation}
%The dynamical regimes of the strongly driven $\Lambda$-system comprises all the three regimes as discussed in Sec.~2.2 in contrast to the dynamics in the weakly pumped regime (see Sec.~4.1). This is clearly demonstrated in the phase diagram of the general discriminant (\ref{General_Disc_Lambda}) in Fig.~(\ref{fig: Contour_Plot_Large_X_Y}).

\begin{figure}[ht]
\centering
\includegraphics[width=1.0\linewidth]{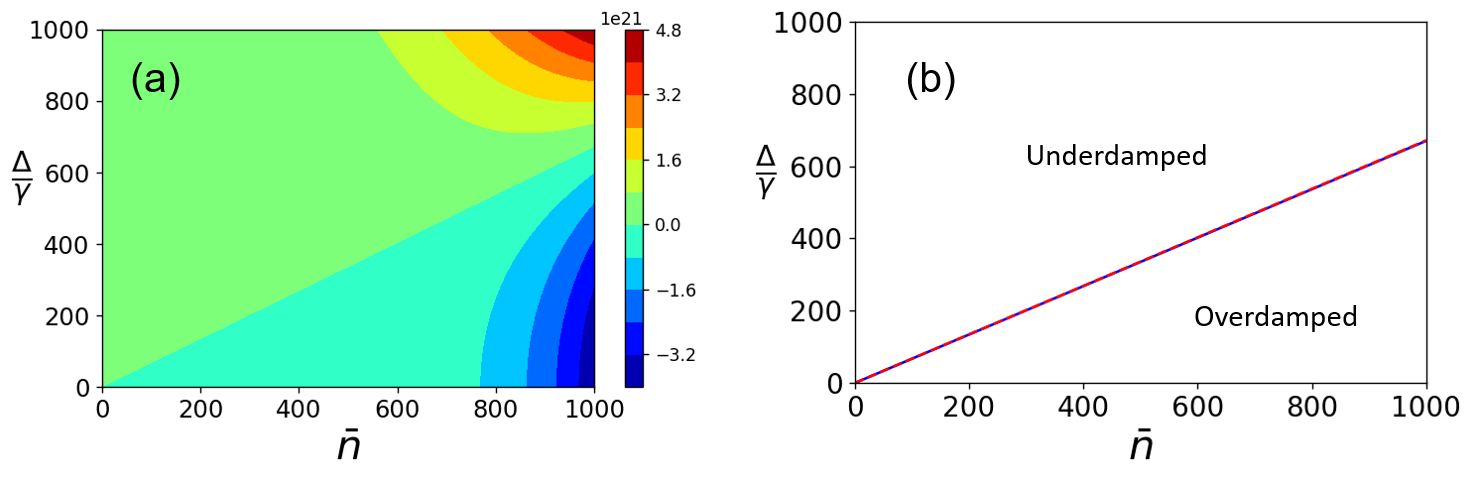}
\caption{(a) A contour plot of the discriminant $D$ for large $\Delta/\gamma$ and $\bar{n}$ and $p = 1$ (b) Lines of zero discriminant that separates underdamped ($D > 0$) and overdamped regions as a function of $\Delta/\gamma$ and $\bar{n}$.}
\label{fig: Contour_Plot_Large_X_Y}
\end{figure}

Figure~\ref{fig: Contour_Plot_Large_X_Y} displays the lines of zero discriminant (\textit{D} = 0) as a function of $\bar{n} \gg 1$, and ground-state energy splitting $\Delta/\gamma$ for $p = 1$. The zero discriminant line separates the underdamped (\textit{D} > 0) and overdamped (\textit{D} < 0) regimes of the $\Lambda$-system.

%It is convenient to express the discriminant (\ref{Discriminant_Lambda}) as a function of physical parameters $p$, $\gamma$, $\Delta$, and $\bar{n}$. It is thus expressed as a polynomial function of the average photon number ($\bar{n}=\frac{r}{\gamma}$) of the radiation field
%\begin{equation} \label{General_Disc_Lambda}
%D = \frac{\gamma^{6}}{2916} \sum_{k=0}^{6} d_{6} \bar{n}^{k}
%\end{equation}
%where the expansion coefficients $d_{k}$'s (listed in Table 1) are functions of $p$, and $\Delta/\gamma$.

\subsection{Eigenvalues and coherence lifetimes}
The general solution of the BR equations (\ref{BREqn_StateVec}) requires the evaluation of the exponential of the coefficient matrix $\textbf{A}$. To this end, we diagonalize $\textbf{A}$ using the symbolic Python package SymPy to find its eigenvalues $\lambda_{j}$, the inverse of which represent the lifetimes (or decay rates) of the corresponding normal modes $\textbf{V}_{j}$.  
\begin{equation}
\lambda_{j} = -A + \alpha_{j} \frac{B}{\mathcal{T}} - \beta_{j} \mathcal{T} \label{Eigval_Gen}
\end{equation}
where 
\begin{align}
    \mathcal{T} &= \sqrt[3]{E + \sqrt{D}}, \label{T_term} \\
         E &= \Big[C - \frac{3}{2}A(B+A^{2}) \Big], \label{E_term} \\
        \omega &= \frac{-1+i\sqrt{3}}{2}, \\
        \omega^{2} &=\frac{-1-i\sqrt{3}}{2}.
\end{align}
Here, the terms $A$, $B$, $C$ are defined in Eqs.~(\ref{A_term}) - (\ref{C_term}) and $\alpha_{j}$, $\beta_{j}$ are the cube roots of unity with the values $(\alpha_{1}, \beta_{1}) = (1, 1)$, $(\alpha_{2}, \beta_{2}) = (\omega^{2}, \omega)$, $(\alpha_{3}, \beta_{3}) = (\omega, \omega^{2})$. 

The expressions for the discriminant (\ref{Discriminant_Lambda}) and the eigenvalues (\ref{Eigval_Gen}) of $\mathbf{A}$ (\ref{Amatrix_Lambda}) are completely general.  Below we will consider  two extreme regimes of weak pumping ($\bar{n} \ll 1$) and strong pumping ($\bar{n} \gg 1$).
% which simplify the mathematical details retaining the essential physics. In the following two sections, we explore the behavior of a $\Lambda$-system irradiated by incoherent light in the strong and the weak pumping limits. 

\subsubsection{Strong pumping limit [$(\bar{n} \gg 1$]}
{\it \bf Overdamped regime [$\Delta / (\bar{n} \gamma) < f(p)$]}.

Defining a new variable $x = 1 / \bar{n} \ll 1$, expressing the terms $\mathcal{D}$ and $E$ in the polynomial form of $x= 1 / \bar{n}= \gamma / r$, and rearranging Eq.~(\ref{E_term_Poly}), we obtain
\begin{equation} \label{E_Poly}
E = \frac{r^{3}}{6}\sum\limits_{k=0}^{3} c_{k} x^{3-k}
\end{equation}
where the $p$-dependent expansion coefficients $c_{k}$ ($k$ = 0, 1, 2, 3) are listed in Table I.

In order to simplify the term $\mathcal{T}$ in the eigenvalue expression, we express $\sqrt{D}$ where $D$ in the following form
\begin{equation} \label{Dsqrt}
\sqrt{D}=r^{3}\frac{\sqrt{d_{6}}}{54}\big(1+\alpha(x)\big)^{\frac{1}{2}},
\end{equation}
where
\begin{equation}
\alpha(x) = \frac{1}{d_{6}} \sum\limits_{k=1}^{6} d_{6-k} x^{k} .
\end{equation}

In the strong pumping limit ($x=\frac{1}{\bar{n}} \ll 1$), all the terms $\frac{d_{6-k}}{d_{6}}x^{k}$ are small compared to 1, and we have $|\alpha(x)| << 1$. Expanding $\sqrt{1+\alpha(x)}$,
\begin{equation} \label{alpha_binomial}
    (1 +\alpha)^{1/2} = 1 + \frac{1}{2} \alpha -\frac{1}{8} \alpha^{2} + \frac{3}{48} \alpha^{3} - \frac{15}{384} \alpha^{4} + \frac{105}{3840} \alpha^{5} - ...,
\end{equation}
and evaluating the terms $\alpha_{k}$ with $k \leq 4$ by using multinomial expansion
\begin{align}
\alpha^{2}&=\bigg(\frac{d_{5}}{d_{6}}\bigg)^{2}x^{2} + \bigg(2\frac{d_{4}d_{5}}{d_{6}^{2}}\bigg)x^{3}+\bigg[{2\frac{d_{3}d_{5}}{d_{6}^{2}}+\bigg(\frac{d_{4}}{d_{6}}\bigg)^{2}}\bigg]x^{4}+\bigg(2\frac{d_{2}d_{5}}{d_{6}^{2}}+2\frac{d_{3}d_{4}}{d_{6}^{2}}\bigg)x^{5}+\bigg[2\frac{d_{1}d_{5}}{d_{6}^{2}}+2\frac{d_{2}d_{4}}{d_{6}^{2}}+(\frac{d_{4}}{d_{6}})^{3}\bigg]x^{6}, \nonumber \\
\alpha^{3}&=\bigg(\frac{d_{5}}{d_{6}}\bigg)^{3}x^{3}+\bigg(3\frac{d_{4}d_{5}^{2}}{d_{6}^{3}}\bigg)x^{4}+\bigg(3\frac{d_{3}d_{5}^{2}}{d_{6}^{3}}+3\frac{d_{4}^{2}d_{5}}{d_{6}^{3}}\bigg)x^{5}+\bigg[3\frac{d_{2}d_{5}^{2}}{d_{6}^{3}}+6\frac{d_{3}d_{4}d_{5}}{d_{6}^{3}}+\bigg(\frac{d_{4}}{d_{6}}\bigg)^{3}\bigg]x^{6}, \nonumber \\
\alpha^{4}&=\bigg(\frac{d_{5}}{d_{6}}\bigg)^{4}x^{4}+\bigg(4\frac{d_{4}d_{5}^{3}}{d_{6}^{4}}\bigg)x^{5}+\bigg(4\frac{d_{3}d_{5}^{3}}{d_{6}^{4}}+6\frac{d_{4}^{2}d_{5}^{2}}{d_{6}^{4}}\bigg)x^{6}, \notag
\end{align}
we find the expression of $\sqrt{D}$ by substituting Eq.~(\ref{alpha_binomial}) into (\ref{Dsqrt})
\begin{equation} \label{Dsqrt2}
\sqrt{D}=r^{3}\bigg(\frac{\sqrt{d_{6}}}{54}\bigg) \bigg[1+\sum_{k=1}^{6}u_{k} x^{k} \bigg]
\end{equation}
where the expansion coefficients $u_{k}$ are listed in Table II. We can now evaluate the term $\mathcal{T}$ by substituting Eqs.~(\ref{E_term_Poly}) and (\ref{Dsqrt2}) into Eq.~(\ref{T_term})
\begin{align}
    \mathcal{T} &= \bigg[\frac{r^{3}}{54} \sum_{k = 0}^{3} c_{k} x^{3-k}+\frac{r^{3}}{54} \sqrt{d_{6}} \bigg(1+\sum_{k=1}^{6}u_{k} x^{k}\bigg)\bigg]^{1/3} \nonumber \\
    &= \frac{r}{\sqrt[3]{54}} \sqrt[3]{c_{3}+\sqrt{d_{6}}}\bigg[1+ \frac{\sum_{k = 1}^{3} \big(c_{3-k} + \sqrt{d_{6}} u_{k}\big) x^{k}}{ (c_{3}+\sqrt{d_{6}})}+ \frac{\sum_{k=4}^{6} \sqrt{d_{6}} u_{k} x^{k}}{ (c_{3}+\sqrt{d_{6}})}\bigg]^{1/3} \nonumber \\
    &= r K \bigg[1+\sum_{k=1}^{6} b_{k} x^{k} \bigg]^{1/3} \label{T_term2}
\end{align}
where the term $K$ and the expansion coefficients $b_{k}$ are listed in Table I. To further simplify the cube root, we express the above equation in the following form
\begin{equation}
\mathcal{T} = rK[1+\beta(x)]^{\frac{1}{3}}
\end{equation}\\
where
\begin{equation}
    \beta(x) = \sum_{k=1}^{6} b_{k} x^{k}
\end{equation}
In the strong pumping limit $x=\frac{1}{\bar{n}} \ll 1$, the terms $b_{k} x^{k}, \enspace k\geq 1$ are all negligible compared to 1, and we have $ |\beta(x)| \ll 1$. This allows us to use the binomial expansion of $(1+\beta)^{1/3}$,
\begin{equation}
(1+\beta)^{\frac{1}{3}}=1+\frac{1}{3}\beta-\frac{1}{9}\beta^{2}+\frac{5}{81}\beta^{3}-\frac{10}{243}\beta^{4}+\frac{22}{729}\beta^{5}-\frac{154}{6561}\beta^{6}+\frac{374}{6561}\beta^{7} - ... \label{beta_binomial}
\end{equation}
and evaluate the higher order terms $\beta_{k}$ with $k \leq 3$ using the multinomial expansion
\begin{align}
  \beta^{2} &= b_{1}^{2}x^{2}+2b_{1}b_{2}x^{3}+(2b_{1}b_{3}+b_{2}^{2})x^{4}+(2b_{1}b_{4}+2b_{2}b_{3})x^{5}+(2b_{1}b_{5}+2b_{2}b_{4}+b_{3}^{2})x^{6} + ... \nonumber \\
\beta^{3} &= b_{3}x^{3}+(3b_{1}^{2}b_{2})x^{4}+(3b_{1}^{2}b_{3}+3b_{1}b_{2}^{2})x^{5}+(3b_{1}^{2}b_{4}+6b_{1}b_{2}b_{3}+b_{2}^{3})x^{6}+ ...  \nonumber 
\end{align}
Using Eq.~(\ref{beta_binomial}) up to the third order in Eq.~(\ref{T_term2}), we obtain
\begin{equation}
\mathcal{T} = rK\bigg[1+\sum_{k=1}^{6} v_{k} x^{k}\bigg]
\end{equation}
where the coefficients $v_{k}$ are listed in Table II. \par 

Next we simplify the term $\frac{1}{\mathcal{T}}$
\begin{equation} \label{T_inv}
\frac{1}{\mathcal{T}} = \frac{1}{rK} \bigg[1+\sum_{k=1}^{6}v_{k} x^{k} \bigg]^{-1} = \frac{1}{rK} \bigg[1+ \Lambda(x) \bigg]^{-1},
\end{equation}
where
\begin{equation}
    \Lambda(x) = \sum_{k=1}^{6}v_{k} x^{k}.
\end{equation}

In the strong pumping limit $x=\frac{1}{\bar{n}} \ll 1$, the terms $v_{k} x^{k}, \enspace k \geq 1$ are all negligible compared to 1, and we find $ |\Lambda(x)| \ll 1$. Taking the binomial expansion of $(1+\Lambda)^{-1}$
\begin{equation} \label{Lambda_binomial}
(1+\Lambda)^{-1}=1-\Lambda+\Lambda^{2}-\Lambda^{3}+\Lambda^{4}-\Lambda^{5}+\Lambda^{6}- ...,
\end{equation}
we compute the terms $\Lambda^{k}$ with $k \leq 4$ by using the multinomial expansion
\begin{align}
\Lambda^{2}&=v_{1}^{2}x^{2}+2v_{1}v_{2}x^{3}+(2v_{1}v_{3}+v_{2}^{2})x^{4}+(2v_{1}v_{4}+2v_{2}v_{3})x^{5}+(2v_{1}v_{5}+2v_{2}v_{4}+v_{3}^{2})x^{6}, \nonumber \\
\Lambda^{3}&=v_{1}^{3}x^{3}+3v_{1}^{2}v_{2}x^{4}+(3v_{1}^{2}v_{3}+3v_{1}v_{2}^{2})x^{5}+(3v_{1}^{2}v_{4}+6v_{1}v_{2}v_{3}+v_{3}^{2})x^{6}, \nonumber \\
\Lambda^{4}&=v_{1}^{4}x^{4}+4v_{1}^{3}v_{2}x^{5}+(4v_{1}^{3}v_{3}+6v_{1}^{2}v_{2}^{2})x^{6}. \nonumber 
\end{align}

Substituting Eq.~(\ref{Lambda_binomial}) up to the fourth order into Eq.~(\ref{T_inv}), we find
\begin{equation}
\frac{1}{\mathcal{T} }= \frac{1}{rK}\bigg[1+ \sum_{k=1}^{6} \mathcal{W}_{6} x^{6} \bigg],
\end{equation}
where the coefficients $\mathcal{W}_{k}$ are listed in Table III. Finally we compute the second term $\frac{B}{\mathcal{T}}$ of the eigenvalue expression (\ref{Eigval_Gen}). 
\begin{equation} \label{B_term2}
B=\frac{r^{2}}{9}\bigg[\bigg(3\frac{\Delta^{2}}{\gamma^{2}}-4\bigg)x^{2}-(8+6p^{2})x-(4+9p^{2})\bigg]
\end{equation}
Multiplying Eqs.~(\ref{T_inv}) and (\ref{B_term2}), we obtain
\begin{align}
\frac{B}{\mathcal{T}} &= \frac{r^{2}}{9}\bigg[\bigg(3\frac{\Delta^{2}}{\gamma^{2}}-4\bigg)x^{2}-(8+6p^{2})x-(4+9p^{2})\bigg]\frac{1}{rK}\bigg[1+\sum_{k=1}^{6} \mathcal{W}_{6} x^{6}\bigg] \nonumber \\
\begin{split}
&=\frac{r}{9K}\bigg\lbrace-(4+9p^{2})-\bigg[(8+6p^{2})+(4+9p^{2})\mathcal{W}_{1}\bigg]x+\bigg[\bigg(3\frac{\Delta^{2}}{\gamma^{2}}-4\bigg)-(8+6p^{2})\mathcal{W}_{1}-(4+9p^{2})\mathcal{W}_{2}\bigg]x^{2}\\ \qquad & +\bigg[\bigg(3\frac{\Delta^{2}}{\gamma^{2}}-4\bigg)\mathcal{W}_{1}-(8+6p^{2})\mathcal{W}_{2}-(4+9p^{2})\mathcal{W}_{3}\bigg]x^{3}+\bigg[\bigg(3\frac{\Delta^{2}}{\gamma^{2}}-4\bigg)\mathcal{W}_{2}-(8+6p^{2})\mathcal{W}_{3}-(4+9p^{2})\mathcal{W}_{4}\bigg]x^{4}\\ \qquad &+\bigg[\bigg(3\frac{\Delta^{2}}{\gamma^{2}}-4\bigg)\mathcal{W}_{3}-(8+6p^{2})\mathcal{W}_{4}-(4+9p^{2})\mathcal{W}_{5}\bigg]x^{5}+\bigg[\bigg(3\frac{\Delta^{2}}{\gamma^{2}}-4\bigg)\mathcal{W}_{4}-(8+6p^{2})\mathcal{W}_{5}-(4+9p^{2})\mathcal{W}_{6}\bigg]x^{6}\\ \qquad &+\bigg[\bigg(3\frac{\Delta^{2}}{\gamma^{2}}-4\bigg)\mathcal{W}_{5}-(8+6p^{2})\mathcal{W}_{6}\bigg]x^{7}+\bigg(3\frac{\Delta^{2}}{\gamma^{2}}-4\bigg)\mathcal{W}_{6}x^{8} \bigg\rbrace \label{B_over_T}
\end{split}
\end{align}

Using Eqs.~(\ref{A_term}), (\ref{T_term2}) and (\ref{B_over_T}) into Eq.~(\ref{Eigval_Gen}), we obtain the eigenvalue expression as a polynomial function of $x = 1/\bar{n} \ll 1$
\begin{equation} \label{lambda_SPL}   
\lambda_{j}=r\sum_{k=0}^{8}z_{jk} x^{k} \enspace (j = 1 - 3)
\end{equation}
where the expansion coefficients $z_{jk}$ are listed in Table III.  The above expression is valid for $x \leq 0.01$ and $p > 0.1$ (for $\Delta/\gamma < 1$) and $0.89 < p < 1$ (for $\Delta/\gamma > 1$).

In the strong pumping limit ($x = 1/\bar{n} \ll 1$), we can neglect the cubic and higher-order terms. The eigenvalue can thus be represented by the following quadratic equation
\begin{equation}
    \lambda_{j} = r [z_{j0}+ z_{j1} x + z_{j2} x^{2}] \label{expansion3}
\end{equation}
where the expansion coefficients $z_{jk} \enspace k = 0 - 2$ are
\begin{align} 
    K &= \sqrt[3]{\frac{c_{3}}{54}+\frac{\sqrt{d_{6}}}{54}}, \label{K_term}\\
    z_{j0} &= \frac{5}{3} - \frac{\alpha_{j}}{9K} (4+9p^{2}) - \beta_{j} K, \label{zj0} \\
    z_{j1} &= \frac{2}{3} - \frac{\alpha_{j}}{9K} [(8+6p^{2})+(4+9p^{2})\mathcal{W}_{1}] - \beta_{j} K v_{1}, \label{zj1} \\
    z_{j2} &= f_{j1} \Big(\frac{\Delta}{\gamma}\Big)^{2} + f_{j2}. \label{zj2}
\end{align}
where the term $K$ is defined in Table~I, $v_{1}$ in Table~II. $f_{j1}$ and $f_{j2}$ are listed in Table~IV. These parameters depend on $p$ only, and the coefficients $z_{j0}$ and $z_{j1}$ are independent of the ratio of the excited-state splitting to the radiative decay rate ${\Delta}/{\gamma}$. Conversely, the coefficient of $x^2$ in Eq. (\ref{expansion3}) carries an explicit quadratic dependence on  $\Delta/\gamma$.

\begin{figure}[b]
%\captionsetup{singlelinecheck = false, format= hang, justification=raggedright, font=footnotesize, labelsep=space}
\begin{center}
\includegraphics[width=1.0\textwidth]{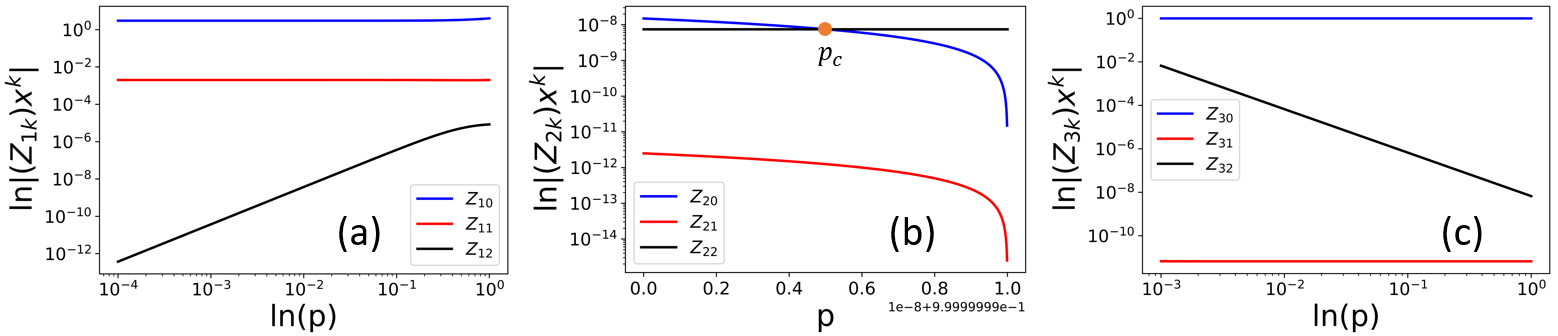}
\end{center}
\caption{Relative contributions of the different terms to the eigenvalues $\lambda_1$ (a), $\lambda_2$ (b) and $\lambda_3$ (c)  plotted as a function of  $p$ for $\Delta/\gamma=10^{-1}$, $\bar{n}=10^{3}$. The absolute value of $z_{jk} x^{k}$ are plotted because the $z_{jk}$ can take negative values. Note the unusual notation in the $x$-axis caption in panel (b) used  to represent values of $p$ very close to unity, {\it i.e.}, 0.5 $\times$ 1e-8 + 9.9999999e-1 = 0.999999995.}
\label{fig:zjk_terms}
\end{figure}

\begin{figure}[t!]
%\captionsetup{singlelinecheck = false, format= hang, justification=raggedright, font=footnotesize, labelsep=space}
     \includegraphics[width=0.8\linewidth]{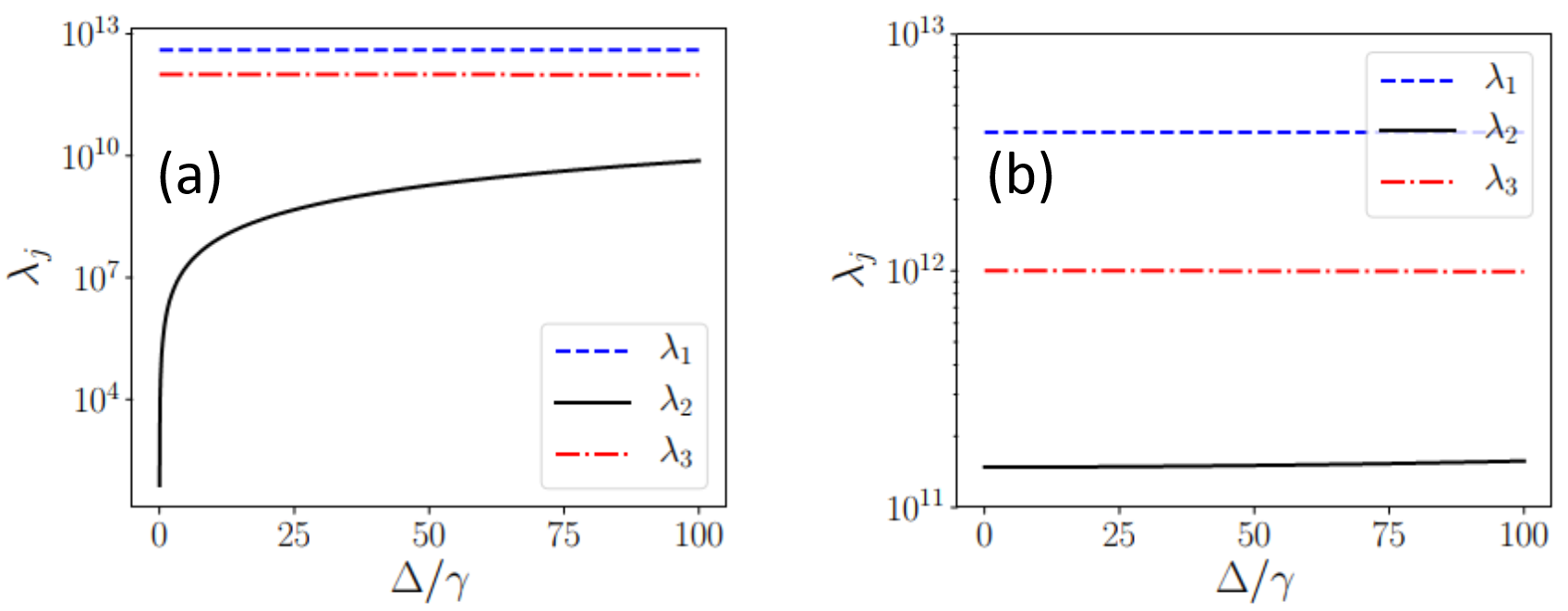} 
     \caption{(a) The eigenvalues $\lambda_{j} (j = 1 -3)$ of matrix $\textbf{A}$ as a function of $\Delta/\gamma$ for $\bar{n} = 10^{3}$ and $p = 1.0$. (b) Same as in (a) but for $\bar{n}=10^{3}$ and $p = 0.9$.}
\label{fig:dyn_Delta_over_gamma}     
\end{figure}

In Fig.~\ref{fig:zjk_terms}, we compare the different terms in Eq.~(\ref{expansion3}), as a function of $p$ for $k =0$ - 2. Figures~\ref{fig:zjk_terms}(a) and \ref{fig:zjk_terms}(c) displays that $z_{10}$, $z_{30}$ are the dominant contributions to $\lambda_1$ and $\lambda_3$  for all $p$. This approximation allows us to write
\begin{equation}\label{zj0dominant}
\lambda_{j}=r z_{j0}=(\gamma z_{j0}) \bar{n}, \enspace  (j=1,3)
\end{equation}
where $z_{j0}(p)$ are given by Eq. (\ref{zj0}). The scaling behavior given by Eq. (\ref{zj0dominant}) is illustrated in Fig.~\ref{fig:dyn_Delta_over_gamma}(b), which shows that the eigenvalues $\lambda_1$ and $\lambda_3$ are independent of ${\Delta}/{\gamma}$ regardless of the value of $p$.  \par 

 On the other hand, the behavior of the eigenvalue $\lambda_2$ depend on the interplay of the terms $z_{20}(p)$ and $z_{22}(p) x^{2}$. Figure~\ref{fig:zjk_terms}(b) shows that there is a critical value of $p=p_c$ at which the curves $z_{20}(p)$ and $z_{22}(p) x^{2}$ intersect, and relative contributions of the different terms to $\lambda_2$ change dramatically. For $p < p_{c}$, $z_{20}(p)$  is the leading term, and $\lambda_{2}$ scales in the same way as the other eigenvalues (\ref{zj0dominant}). When $p > p_{c}$, the dominant term is $z_{22} x^{2}$ and the scaling of $\lambda_{2}$ with $\Delta/\gamma$ is quadratic, similar to that of $z_{22}$ [see Eq.~(\ref{zj2})]. \par 

The critical value of $p$ is a function of $\Delta/\gamma$ and $\bar{n}$ and ranges from 0.995 to 1.0 for $\bar{n}=10^{3}$ and $\Delta / \gamma = 10^{2} - 10^{-2}$ [See Fig. ~\ref{fig:zjk_terms}(b)]. The different scaling relation of $\lambda_{2}$ for $p<p_c$ and $p>p_c$ leads to different population and coherence dynamics in these regimes (see below). \par 

Figure~\ref{fig:zjk_terms}(b) illustrates that the quadratic contribution to the second eigenvalue is much larger than the linear and constant terms, and hence we can approximate $\lambda_{2} \simeq rz_{j2}({\gamma}/{r})^2$. Combining Eqs.~(\ref{expansion3}) and (\ref{zj2}) and noting that $f_{22}(p)\rightarrow 0$ for $p > p_{c}$, we find
%, so that $f_{21}(p) (\frac{\Delta}{\gamma})^{2}\gg f_{22}(p)$
%\begin{equation}\label{lambda2expr}
%\lambda_{2}=\gamma \frac{1}{\bar{n}} \left[ f_{21}(p)\left(\frac{\Delta}{\gamma}\right)^{2}+f_{22}(p)\right] 
%\end{equation}
\begin{equation}\label{lambda2expr}
\lambda_{2}= \frac{\gamma}{\bar{n}} f_{21}(p)\left(\frac{\Delta}{\gamma}\right)^{2}.
\end{equation}

The distinct quadratic scaling of $\lambda_{2}$ with $\Delta/\gamma$  is illustrated in Fig.~\ref{fig:dyn_Delta_over_gamma}(a). The function $f_{21}(p)$ (see the Appendix) increases monotonously approaching the value $-0.749$ in the limit $p\to 1$. As the second eigenvalue gives  the decay rate of the real part of the coherence (see Sec.~3.2.1) the coherence lifetime $\tau_c = 1/|\lambda_2|$ is given by
\begin{equation}\label{tauCoherence}
\tau_{c} = 1.34\frac{\bar{n}}{\gamma} \left(\frac{\Delta}{\gamma}\right)^{-2} \qquad (p>p_c).
\end{equation}

%Mathematically, the linear scaling of the coherence lifetime is a direct consequence of the second eigenvalue being dominated by a single $z_{22}x^2$ contribution as shown in Fig.~\ref{fig:zjk_terms}(b). This characteristic scaling occurs only for $p>p_c$ with $p_c = 1-\epsilon$ very close to unity [typical values of  $\epsilon$ range from  $5\times 10^{-11}$ to $5\times 10^{-3}$  for $\bar{n} = 10^{3}$. Thus, the $\Lambda$-system with nearly parallel transition dipole moments can exhibit very long coherence times in the strong pumping limit.

{\it \bf Underdamped regime [$\Delta / (\bar{n} \gamma) < f(p)$]}.
In the underdamped regime, we can neglect the population to coherence coupling in Eq.~(\ref{BR_Pop}). This allows us to simplify the coefficient matrix $\textbf{A}$, and find the eigenvalues without using the general eigenvalue expressions given by Eq.~(\ref{Eigval_Gen}). The coefficient matrix in Eq.~(\ref{Amatrix_Lambda}) can be further simplified to give
\begin{equation}
\textbf{A} = 
\begin{bmatrix}
    -(3r+2\gamma) & 0 & 0 \\
    -p(3r+2\gamma) & -r & \Delta  \\
       0 & -\Delta & -r \\
\end{bmatrix} \label{Amatrix_Lambda_UD}
\end{equation}

The eigenvalues of matrix $\textbf{A}$ (\ref{Amatrix_Lambda_UD}) are
\begin{align}
    \lambda_{1} &= -(3r+2\gamma) \label{EigVal1_SPL} \\
    \lambda_{2} &= -r + i \Delta \label{EigVal2_SPL} \\
    \lambda_{3} &= -r - i \Delta \label{EigVal3_SPL}
\end{align}

These expressions show that there is one real and two complex conjugate eigenvalues, as expected in the underdamped regime.

\subsubsection{Weak pumping regime [$(\bar{n} \ll 1$]}
The dynamics of a weakly driven $\Lambda$-system is mostly in the underdamped regime as the discriminant (\ref{Discriminant_Lambda}) is always positive for $\Delta \neq 0$ and {$\Delta/r > p/2$}. In the weak pumping limit, where $\bar{n} \ll 1$, we find the expression of $E$ by rearranging Eq.~(\ref{E_term_Poly}) as
\begin{equation} \label{E_Poly_WPL}
E = \frac{\gamma^{3}}{54}\sum\limits_{k=0}^{2} c_{k} \bar{n}^{k}
\end{equation}
where we have neglected the terms higher than the second order in $\bar{n}$, and the expansion coefficients $c_{k}$ ($k$ = 0, 1, 2, 3) are listed in Table I.

In order to simplify the term $\mathcal{T}$ in the eigenvalue expression, we find $\sqrt{D}$ where $D$ is the discriminant from Eq.~(\ref{Disc_Poly}), by expressing it in the following form
\begin{equation} \label{Dsqrt_WPL}
\sqrt{D} = \frac{\gamma^{3}}{54} \sqrt{d_{0}} \bigg(1+\frac{d_{1}}{d_{0}}\bar{n}\bigg)^{\frac{1}{2}} = \frac{\gamma^{3}}{54} \sqrt{d_{0}} \bigg(1+\frac{d_{1}}{2d_{0}}\bar{n}\bigg)
\end{equation}
where we have retained the terms up to the first order in $\bar{n}$ and used the binomial theorem to simplify the last equation. 

We can now evaluate the term $\mathcal{T}$ by substituting Eqs.~(\ref{E_term_Poly}) and (\ref{Dsqrt_WPL}) into Eq.~(\ref{T_term})
\begin{align}
    \mathcal{T} &= \bigg[\frac{\gamma^{3}}{54} \sum_{k = 0}^{3} c_{k} \bar{n}^{k}+\frac{\gamma^{3}}{54} \sqrt{d_{0}} \bigg(1+\frac{d_{1}}{2d_{0}}\bar{n}\bigg)\bigg]^{1/3} \nonumber \\
    &= \frac{\gamma\Tilde{K}}{3} \bigg[1+\sum_{k=1}^{3} t_{k} \bar{n}^{k} \bigg]^{1/3} = \frac{\gamma\Tilde{K}}{3} \bigg[1+\frac{1}{3}\sum_{k=1}^{3} t_{k} \bar{n}^{k} \bigg] \label{T_term2_WPL}
\end{align}
where the term $\tilde{K}$ and expansion coefficients $t_{k}$ are listed in Table XIII. 

Next we simplify the term $\frac{1}{\mathcal{T}}$ in the second term of the eigenvalue expression (\ref{Eigval_Gen})
\begin{equation} \label{T_inv_WPL}
\frac{1}{\mathcal{T}} = \frac{3}{\gamma\tilde{K}} \bigg[1+\frac{1}{3}\sum_{k=1}^{3}t_{k} \bar{n}^{k} \bigg]^{-1} = \frac{3}{\gamma\tilde{K}} \bigg[1 -\frac{1}{3}\sum_{k=1}^{3}t_{k} \bar{n}^{k} \bigg]
\end{equation}

The second term $\frac{B}{\mathcal{T}}$ of the eigenvalue expression (\ref{Eigval_Gen}) takes the form
\begin{equation} \label{B_term2_WPL}
B=\frac{\gamma^{2}}{9}\bigg[\bigg(3\frac{\Delta^{2}}{\gamma^{2}}-4\bigg)-(8+6p^{2})\bar{n}-(4+9p^{2}) \bar{n}^{2}\bigg]
\end{equation}
Multiplying Eqs.~(\ref{T_inv_WPL}) and (\ref{B_term2_WPL}), we obtain
\begin{align}
\frac{B}{\mathcal{T}} &= \frac{\gamma^{2}}{9}\bigg[\bigg(3\frac{\Delta^{2}}{\gamma^{2}}-4\bigg)-(8+6p^{2})\bar{n}-(4+9p^{2}) \bar{n}^{2}\bigg]\frac{3}{\gamma\tilde{K}}\bigg[1-\frac{1}{3}\sum_{k=1}^{3} {t}_{k} \bar{n}^{k}\bigg] \nonumber \\
\begin{split}
&=\frac{\gamma}{3K}\bigg[\bigg(3\frac{\Delta^{2}}{\gamma^{2}}-4\bigg) -\bigg\lbrace \big(8+6p^{2}\big)+\frac{1}{3}\bigg(3\frac{\Delta^{2}}{\gamma^{2}}-4\bigg)t_{1}\bigg\rbrace\bar{n}-\bigg\lbrace\big(4+9p^{2}\big)-\frac{1}{3}\big( 8+9p^{2}\big)t_{1}+\frac{1}{3}\bigg(3\frac{\Delta^{2}}{\gamma^{2}}-4\bigg)t_{2}\bigg\rbrace \bar{n}^{2} \\ & \qquad +\frac{1}{3}\bigg\lbrace\big(4+9p^{2}\big)t_{1}+\big( 8+9p^{2}\big)t_{2}-\frac{1}{3} \bigg(3\frac{\Delta^{2}}{\gamma^{2}}-4\bigg)t_{3}\bigg\rbrace \bar{n}^{3}+\frac{1}{3}\bigg\lbrace\big(4+9p^{2}\big)t_{2}+\big( 8+9p^{2}\big)t_{3}\bigg\rbrace\bar{n}^{4}\\ & \qquad +\frac{1}{3}\bigg\lbrace\big(4+9p^{2}\big)t_{3}\bar{n}^{5}\bigg] \label{B_over_T_WPL}
\end{split}
\end{align}

Using Eqs.~(\ref{A_term}), (\ref{T_term2_WPL}) and (\ref{B_over_T_WPL}) into Eq.~(\ref{Eigval_Gen}), we finally find the eigenvalue expression as a polynomial function of $\bar{n} \ll 1$
\begin{equation} \label{lambda_WPL}
\lambda_{j}=\gamma \sum_{k=0}^{5}\tilde{z}_{jk} \bar{n}^{k} \enspace (j = 1 - 3)
\end{equation}
where the expansion coefficients $\tilde{z}_{jk}$ are listed in Table XIII.

\subsection{Eigenvectors in the strong pumping limit}
The general expression for the eigenvectors of the coefficient matrix $\mathbf{A}$ is obtained by solving the system of linear equations $\big(\mathbf{A} - \lambda_{j}\big) V_{j} = 0$.
\begin{equation} \label{Evect_SPL}
V_{j} =
\begin{bmatrix}
       \frac{\Delta p(r+\gamma)}{\mathcal{D}_{j}}  \\
       -\frac{\Delta (3r+\gamma+\lambda_{j})}{\mathcal{D}_{j}} \\
       1  \\
\end{bmatrix} \enspace ; \enspace j =\enspace 1,\enspace 2,\enspace 3
\end{equation}\\
where
\begin{equation} \label{Denom_Evect}
\mathcal{D}_{j}=-\lambda_{j}^{2}-2(\gamma+2r)\lambda_{j}-(1-p^{2})\gamma^{2}-4\gamma r(1-p^{2})+3r^{2}(1-p^{2})
\end{equation}

\subsubsection{Eigenvectors in the overdamped regime [$\Delta / (\bar{n} \gamma) < f(p)$]}
In the strongly pumped regime ($x\ll 1$), the terms $x^{n}$ with $n \geq 3$ can be neglected in Eq.~(\ref{lambda_SPL}) and the eigenvalues are given by
\begin{equation} \label{lambda_SPL_Quad}
\lambda_{j}=r\Big[z_{j0}+z_{j1}x+z_{j2}x^{2}\Big]
\end{equation}

To evaluate the term $\mathcal{D}_{j}$, we evaluate the square of $\lambda_{j}$ as 
\begin{equation} \label{lambda_SPL_Quad_Square}
\lambda_{j}^{2} = r^{2}\Big[z_{j0}^{2}+2z_{j0}z_{j1}x+\Big(z_{j1}^{2}+2z_{j0}z_{j2}\Big)x^{2}+2z_{j1}z_{j2}x^{3}+z_{j2}^{2}x^{4}\Big]
\end{equation}

Substituting Eqs.~(\ref{lambda_SPL_Quad}), (\ref{lambda_SPL_Quad_Square}) in Eq.~(\ref{Denom_Evect}), we obtain
\begin{equation}
\mathcal{D}_{j} = -r^{2} \sum_{k=0}^{4}L_{jk} x^{k},
\end{equation}
where the expansion coefficients $L_{jk}$ are listed in Table VI.

For the first eigenvector $\mathbf{V}_{1}$, we find
\begin{equation} \label{D1_Evect}
\mathcal{D}_{1} = -r^{2} \sum_{k=0}^{4}L_{1k} x^{k}
\end{equation}

In particular, the terms $L_{10}$, $L_{11} x$ are negligible compared to other terms in Eq.~(\ref{D1_Evect}) and are dropped.

To find $\frac{1}{\mathcal{D}_{1}}$ required to evaluate Eq.~(\ref{Evect_SPL}), we proceed as follows
\begin{align} \label{D1_inv_Evect}
\frac{1}{\mathcal{D}_{1}} &=-\frac{1}{r^{2}L_{12}x^{2}\Big[1+\frac{L_{13}}{L_{12}}x+\frac{L_{14}}{L_{12}}x^{2}\Big]} \nonumber \\
&=-\frac{1}{r^{2}L_{12}x^{2}} \Big[1+\alpha(x)\Big]^{-1}
\end{align} 
where 
\begin{equation}  \label{alpha_Evect}
\alpha(x) = \frac{L_{13}}{L_{12}}x+\frac{L_{14}}{L_{12}}x^{2}
\end{equation}

For $x \ll 1$,\enspace $\alpha(x)=\Big(\frac{L_{13}}{L_{12}}x+\frac{L_{14}}{L_{12}}x^{2}\Big) \ll 1$ and we can use the binomial expansion to get
\begin{equation} \label{alpha_Evect_Binomial}
(1+\alpha)^{-1}=1-\alpha+\alpha^{2}-\alpha^{3}+ \ldots
\end{equation}

Evaluating the terms up to the third order in $\alpha$ 
\begin{align}  
\alpha^{2} &=\Big(\frac{L_{13}}{L_{12}}\Big)^{2}x^{2}+\frac{2L_{13}L_{14}}{L_{12}^{2}}x^{3}+\Big(\frac{L_{14}}{L_{12}}\Big)^{2}x^{4} \label{alpha_sq_Evect} \\
\alpha^{3} &=\Big(\frac{L_{13}}{L_{12}}\Big)^{3}x^{3}+\frac{3L_{13}^{2}L_{14}}{L_{12}^{3}}x^{4}+\frac{3L_{13}L_{14}^{2}}{L_{12}^{3}}x^{5}+\Big(\frac{L_{14}}{L_{12}}\Big)^{3}x^{6} \label{alpha_cube_Evect}
\end{align}
and using Eqs.~(\ref{alpha_Evect}) to (\ref{alpha_cube_Evect}) into Eq.~(\ref{D1_inv_Evect}), we get 
\begin{equation}
\frac{1}{\mathcal{D}_{1}}=-\frac{1}{r^{2}L_{12}x^{2}} \Big[1+\sum_{m=1}^{6} k_{1m} x^{m}\Big]
\end{equation}
where the expansion coefficients $k_{1m}$ are listed in Table VI.

We can now evaluate the first component of the eigenvector $\mathbf{V_{1}}$ as follows
\begin{align}
V_{11} &=\frac{\Delta p (\gamma+r)}{\mathcal{D}_{1}}
=-\frac{p}{L_{12}}\frac{\Delta r(1+\frac{\gamma}{r})}{r^{2}(\frac{\gamma}{r})^{2}}\Big[1+\sum_{m=1}^{6} k_{1m} x^{m}\Big] =-\frac{p}{L_{12}}\Big(\frac{\Delta}{\gamma}\Big)\bar{n} \sum_{m=0}^{7}a_{1m}x^{m} \label{V_11}
\end{align}
where the expansion coefficients $a_{1m}$ are listed in Table VI.

Proceeding in a similar way, we find the second component of the eigenvector $\mathbf{V}_{1}$ as
\begin{equation} \label{V_12}
V_{12}=\frac{1}{L_{12}}\Big(\frac{\Delta}{\gamma}\Big)\bar{n}\sum_{m=0}^{8}b_{1m}x^{m}
\end{equation}
where the expansion coefficients $b_{1m}$ are listed in Table VI. The third component of the eigenvector $\mathbf{V}_{1}$ is $V_{13}=1$.

Combining the expressions for $V_{11}$, $V_{12}$ and $V_{13}$, we obtain the first eigenvector as
\begin{equation} \label{Evect1}
\mathbf{V}_{1} = 
\begin{bmatrix}
     -\frac{p}{L_{12}}\Big(\frac{\Delta}{\gamma}\Big)\bar{n} \sum_{m=0}^{7}a_{1m}x^{m}\\
    \frac{1}{L_{12}}\Big(\frac{\Delta}{\gamma}\Big)\bar{n}\sum_{m=0}^{8}b_{1m}x^{m}  \\
       1 \\
\end{bmatrix}
\end{equation}

Proceeding in a similar way as for the first eigenvector $\mathbf{V}_{1}$, we evaluate $\mathcal{D}_{2}$ and the components of the second eigenvector $\mathbf{V}_{2}$
\begin{equation} \label{Evect2}
\mathbf{V}_{2} = 
\begin{bmatrix}
     -\frac{p}{L_{22}}\Big(\frac{\Delta}{\gamma}\Big)\bar{n} \sum_{m=0}^{7}a_{2m}x^{m}\\
    \frac{1}{L_{22}}\Big(\frac{\Delta}{\gamma}\Big)\bar{n}\sum_{m=0}^{8}b_{2m}x^{m}  \\
       1 \\
\end{bmatrix}
\end{equation} \\
where the coefficients $L_{2k}$, $k_{2m}$, $a_{2m}$, $b_{2m}$ are listed in Table VI, and $z_{2k}$,\enspace $k=\enspace 0,\enspace 1,\enspace 2$ are evaluated with $\alpha_{2}=\enspace \omega^{2},\enspace \beta_{2}=\enspace \omega$.

For the third eigenvector $\mathbf{V}_{3}$, the term $\mathcal{D}_{3}$ is given by
\begin{equation} \label{D3_Evect}
\mathcal{D}_{3} = -r^{2} \sum_{k=0}^{4} L_{3k} x^{k}
\end{equation}
where the coefficients $L_{3k}$ are listed in Table VI, and $z_{3k}$,\enspace $ k=\enspace 0,\enspace 1,\enspace 2$ are evaluated with $\alpha_{3}=\enspace \omega,\enspace \beta_{3}=\enspace \omega^{2}$.

In contrast to the first and second eigenvectors, the term $L_{30}$ and $L_{31}x$ are not negligible compared to other $L_{3k} x^{k}$ terms. To find $\frac{1}{\mathcal{D}_{3}}$, we proceed as follows
\begin{align} \label{D3_Inv_Evect}
\frac{1}{\mathcal{D}_{3}} =-\frac{1}{r^{2}L_{30}\Big[1+\frac{1}{L_{30}} \sum_{k=1}^{4}L_{3k} x^{k}\Big]} 
=-\frac{1}{r^{2}L_{30}} \Big[1+\alpha(x)\Big]^{-1},
\end{align}
where
\begin{equation} \label{alpha_D3}
\alpha(x)= \frac{1}{L_{30}} \sum_{k=1}^{4}L_{3k} x^{k}.
\end{equation}

For $x \ll 1$, $\alpha(x)= \frac{1}{L_{30}} \sum_{k=1}^{4}L_{3k} x^{k} \ll 1$. Taking the binomial expansion of Eq.~(\ref{D3_Inv_Evect}) and evaluating the terms up to the fifth order in $\alpha$ and $x$ we find
\begin{align}
\alpha^{2} &=(\frac{L_{31}^{2}}{L_{30}^{2}})x^{2}+\frac{2L_{31}L_{32}}{L_{30}^{2}}x^{3}+(\frac{2L_{31}L_{33}}{L_{30}^{2}}+\frac{L_{32}^{2}}{L_{30}^{}{2}})x^{4}+(\frac{2L_{31}L_{34}}{L_{30}^{2}}+\frac{2L_{32}L_{33}}{L_{30}^{2}})x^{5}+ ..., \\
\alpha^{3} &=\frac{L_{31}^{3}}{L_{30}^{3}}x^{3}+(\frac{3L_{31}^{2}L_{32}}{L_{30}^{3}}+\frac{3L_{31}L_{32}^{2}}{L_{30}^{3}})x^{4}+\frac{3L_{31}^{2}L_{33}}{L_{30}^{3}}x^{5}+ ..., \\
\alpha^{4} &=\frac{L_{31}^{4}}{L_{30}^{4}}x^{4}+\frac{4L_{31}^{3}L_{32}}{L_{30}^{4}}x^{5}+ ..., \\
\alpha^{5} &= \frac{L_{31}^{5}}{L_{30}^{5}} x^{5}+ ... . \label{alpha_fifth_D3}
\end{align}
Substituting Eqs.~(\ref{alpha_D3}) to (\ref{alpha_fifth_D3}) into Eq. (\ref{D3_Inv_Evect}), we get
\begin{equation}
\frac{1}{\mathcal{D}_{3}}=-\frac{1}{r^{2}L_{30}} \Big[1+\sum_{m=1}^{5}k_{3m}x^{m}\Big]
\end{equation}
where the expansion coefficients $k_{3m}$ are listed in Table VII.

The first component of the eigenvector $\mathbf{V}_{3}$ is computed as,
\begin{align}
V_{31} &=\frac{\Delta p (\gamma+r)}{\mathcal{D}_{3}} =-\frac{p}{L_{30}}\frac{\Delta r(1+x)}{r^{2}}\Big[1+\sum_{m=1}^{5}k_{3m}x^{m}\Big] =-\frac{p}{L_{30}}\Big(\frac{\Delta}{\gamma}\Big)\frac{1}{\bar{n}} \sum_{m=0}^{5}a_{3m}x^{m} \label{V_31}
\end{align} 
where the coefficients $a_{3m}$ are listed in Table VII. Proceeding in the same way, the second component of the eigenvector $\mathbf{V}_{3}$ is evaluated as
\begin{equation} \label{V_32}
V_{32}=\frac{1}{L_{30}}\Big(\frac{\Delta}{\gamma}\Big) \frac{1}{\bar{n}} \sum_{m=0}^{8}b_{3m}x^{m}
\end{equation}
where the expansion coefficients $b_{3m}$ are listed in Table VII. The third component of the eigenvector $\mathbf{V}_{3}$ is $V_{33}=1$.

The third eigenvector is thus
\begin{equation} \label{Evect3}
\mathbf{V}_{3} = 
\begin{bmatrix}
     -\frac{p}{L_{30}}\Big(\frac{\Delta}{\gamma}\Big) \frac{1}{\bar{n}} \sum_{m=0}^{5}a_{3m}x^{m}\\
    \frac{1}{L_{30}}\Big(\frac{\Delta}{\gamma}\Big) \frac{1}{\bar{n}} \sum_{m=0}^{5}b_{3m}x^{m}  \\
       1 \\
\end{bmatrix}
\end{equation}

Combining the expressions for the eigenvectors $\mathbf{V_{1}}$, $\mathbf{V_{2}}$ and $\mathbf{V_{3}}$ we obtain the matrix of eigenvectors of $\mathbf{A}$ as \\
\begin{equation} \label{M_matrix}
\bf{M} = 
\begin{bmatrix}
    -\frac{p}{L_{12}}\Big(\frac{\Delta}{\gamma}\Big)\bar{n} \sum_{m=0}^{7}a_{1m}x^{m} & -\frac{p}{L_{22}}\Big(\frac{\Delta}{\gamma}\Big)\bar{n} \sum_{m=0}^{7}a_{2m}x^{m} & -\frac{p}{L_{30}}\Big(\frac{\Delta}{\gamma}\Big) \frac{1}{\bar{n}} \sum_{m=0}^{5}a_{3m}x^{m} \\
    \frac{1}{L_{12}} \Big(\frac{\Delta}{\gamma}\Big) \bar{n}\sum_{m=0}^{8}b_{1m}x^{m}
 & \frac{1}{L_{22}} \Big(\frac{\Delta}{\gamma}\Big) \bar{n}\sum_{m=0}^{8}b_{2m}x^{m}  & \frac{1}{L_{30}}\Big(\frac{\Delta}{\gamma}\Big) \frac{1}{\bar{n}} \sum_{m=0}^{5}b_{3m}x^{m}  \\
       1 & 1 & 1 \\
\end{bmatrix}
\end{equation} 

\subsubsection{The determinant and the inverse of the eigenvector matrix M}
Expanding Eq.~(\ref{M_matrix}) through second order in $x << 1$ and neglecting the insignificant terms, we get
\begin{equation}
\bf{M} = 
\begin{bmatrix}
    -\frac{p}{L_{12}}(\frac{\Delta}{\gamma})\bar{n} (a_{10}+a_{12}x^{2}) & -\frac{p}{L_{22}}(\frac{\Delta}{\gamma})\bar{n} a_{20} & -\frac{p}{L_{30}}(\frac{\Delta}{\gamma})(\frac{1}{\bar{n}}) a_{30} \\
    \frac{1}{L_{12}}(\frac{\Delta}{\gamma})\bar{n} (b_{10}+b_{12}x^{2})
 & \frac{1}{L_{22}}(\frac{\Delta}{\gamma})\bar{n} b_{20}  & \frac{1}{L_{30}}(\frac{\Delta}{\gamma})(\frac{1}{\bar{n}}) b_{30}  \\
       1 & 1 & 1 \\
\end{bmatrix}
\end{equation}
We take,
\begin{equation}
\bf{M} = 
\begin{bmatrix}
    V_{11} & V_{21} & V_{31}  \\
    V_{12} & V_{22} & V_{32}  \\
    V_{13} & V_{23} & V_{33}  \\ 
\end{bmatrix}  
\end{equation}
where the components $V_{ij}$; $i, j = 1, 2, 3$ are listed in Table IX. 

To simplify the expression for $V_{11}$ in Eq.~(\ref{M_matrix}), we begin with the $L_{12}$ term in the denominator
\begin{align}
L_{12} &= z_{11}^{2}+2z_{10}z_{12}+2z_{11}+4z_{12}+(1-p^{2}) \nonumber \\
&= F_{11}(p)\Big(\frac{\Delta}{\gamma}\Big)^{2}+F_{12}(p) 
\end{align}
where the terms $F_{11}(p)$, $F_{12}(p)$ are listed in Table VIII. \\
We find that $F_{11}(p)>>F_{12}(p)$ for all $p$ and hence
\begin{equation} \label{L12}
L_{12} \approx F_{11}\bigg(\frac{\Delta}{\gamma}\bigg)^{2}
\end{equation} 

To further simplify Eq.~(\ref{M_matrix}) we substitute the expressions for $a_{10}$ and $a_{12}$ from Table VI to the expression for $V_{11}$ in Eq.~(\ref{M_matrix}) and use Eq.~(\ref{L12}) to get
\begin{align}
a_{10} &= 1 \\ 
a_{12} &=k_{11}+k_{12} \nonumber \\
&=-\frac{L_{13}}{L_{12}}-\frac{L_{14}}{L_{12}}+\bigg(\frac{L_{13}}{L_{12}}\bigg)^{2} 
\approx -\frac{L_{14}}{L_{12}} 
= -\frac{f_{11}}{(4+2z_{10})}\bigg(\frac{\Delta}{\gamma}\bigg)^{2}
\end{align}

Using
\begin{equation}
\frac{a_{10}+a_{12}x^{2}}{L_{2}} = \frac{1-\frac{f_{11}}{(4+2z_{10})}\Big(\frac{\Delta}{\gamma}\Big)^{2} \frac{1}{\bar{n}^{2}}}{F_{11}(p) \Big(\frac{\Delta}{\gamma}\Big)^{2}} \\
\end{equation}
we finally obtain a simplified expression for $V_{11}$ 
\begin{align}
V_{11} &= -p \bar{n} \bigg(\frac{\Delta}{\gamma}\bigg) \Bigg[\frac{1-\frac{f_{11}}{(4+2z_{10})}(\frac{\Delta}{\gamma})^{2}(\frac{1}{\bar{n}})^{2}}{F_{11}(p)(\frac{\Delta}{\gamma})^{2}}\Bigg] \nonumber \\
&= p\bigg[-\frac{1}{F_{11}}+\frac{f_{11}}{F_{11}}\frac{1}{(2z_{10}+4)}\frac{1}{\bar{n}^{2}}\bigg(\frac{\Delta}{\gamma}\bigg)^{2}\bigg]\bar{n}\bigg(\frac{\Delta}{\gamma}\bigg)^{-1} 
\end{align}

Proceeding in a similar way we obtain analytic expressions for the other matrix elements $V_{ij}$ listed in Table IX.

Now that we found the analytic expression for the elements of matrix $\bf{M}$, we need to find its inverse. 
\begin{equation}
\mathbf{M}^{-1} = \frac{1}{\text{det}(\mathbf{M})} \text{adj}(\mathbf{M})
\end{equation}
where $\text{det}(\bf{M})$ is the determinant of $\bf{M}$ and $\text{adj}(\bf{M})$ is the adjoint of $\bf{M}$.\\
Using the expressions of $V_{ij}$ in Table IX, we evaluate the minors of $\bf{M}$ i.e. $T_{ij}$ in order to find its adjoint.
\begin{align}
T_{11} &= V_{22}V_{33}-V_{23}V_{32} \nonumber \\
&=\frac{(3+z_{20})}{F21}\bar{n}\Big(\frac{\Delta}{\gamma}\Big)^{-1}-\frac{(3+z_{30})}{L_{30}}\frac{1}{\bar{n}}\Big(\frac{\Delta}{\gamma}\Big) \nonumber \\
&=\bigg[\frac{(3+z_{20})}{F21}-\frac{(3+z_{30})}{F31}\frac{1}{\bar{n}^{2}}\bigg(\frac{\Delta}{\gamma}\bigg)^{2}\bigg]\bar{n}\bigg(\frac{\Delta}{\gamma}\bigg)^{-1} \nonumber \\
&=\bigg[m_{1}(p)+m_{2}(p)\frac{1}{\bar{n}^{2}}\bigg(\frac{\Delta}{\gamma}\bigg)^{2}\bigg]\bar{n}\bigg(\frac{\Delta}{\gamma}\bigg)^{-1} 
\end{align}
where the coefficients $m_{1}(p)$, $m_{2}(p)$ are listed in Table XI.

Proceeding in the same way, we evaluate the remaining minors of $\bf{M}$ which are listed in Table X. The coefficients $m_{i}(p)$ that define $T_{ij}$ are listed in Table XI. All of the coefficients $m_{i}(p)$, ($i$ = 1 to 16) are functions of $p$ only. \\
The adjoint of $\bf{M}$ is given by 
\begin{align}
\text{adj}(\mathbf{M}) &= \mathbf{M}^{\rm T}
= 
\begin{bmatrix}
    T_{11} & T_{12} & T_{13} \\
    T_{21} & T_{22} & T_{23} \\
    T_{31} & T_{32} & T_{33} \\
\end{bmatrix},
\end{align} 
and the determinant of $\bf{M}$ is
\begin{equation} \label{detM_general}
\text{det}(\mathbf{M}) = V_{11}T_{11}+V_{21}T_{21}+V_{31}T_{31}.
\end{equation}

%So, we do not present the eigenvectors for the underdamped regime. 
%The derivation of the analytical solutions in both regimes is provided in the next section.

\subsection{Density Matrix Evolution} % in the strong pumping limit}
Having evaluated the eigenvector matrix $\mathbf{M}$, as well as its determinant and inverse in the previous section,  are now in a position to derive analytic expressions for the time evolution of the density matrix of the strongly driven $\Lambda$-system. This is done in  {\it Subsection~1} below for the overdamped regime. In  {\it Subsection~2} we will derive approximate solutions of the BR equations in the underdamped regime  without evaluating the eigenvectors. 
%\textcolor{magenta}{Suyesh, please check this paragraph.}

\subsubsection{Analytical expression in the overdamped regime [$\Delta / (\bar{n} \gamma) < f(p)$]}

%  by a thermal environment. 
%The eigenvectors, along with the eigenvalues, determine the exact analytical solutions in the overdamped regime. 
%In this section, we derive analytic expressions for the time evolution of the density matrix.

 The general solution of the Bloch-Redfield equations can be obtained from Duhamel's formula 
%[See Ref.~X of the main text].
\begin{equation} 
\textbf{x}(t)(t)=e^{\mathbf{A} t} \textbf{x}_{0}+\int_{0}^{t} e^{\mathbf{A} (t-s)} \textbf{d} \enspace ds \label{BR_SolnGreenFunction}
\end{equation}
where,
\begin{equation}
\textbf{x}(t) = \begin{bmatrix}
         \rho_{g_{1}g_{1}} &   \rho_{g_{1}g_{2}}^{R} & \rho_{g_{1}g_{2}}^{I}  
        \end{bmatrix}^{T}
\end{equation}
\begin{equation}
\textbf{x}_{0}=\begin{bmatrix}
         \rho_{g_{1}g_{1}}(0) &  \rho_{g_{1}g_{2}}^{R}(0) & \rho_{g_{1}g_{2}}^{I}(0)  
        \end{bmatrix}^{T} 
\end{equation} is the initial vector, and 
\begin{equation}
\textbf{d}=\begin{bmatrix}
%[(r+gamma)  p(r+gamma)  0]^T
        (r+\gamma)  &  p(r+\gamma)  & 0
        \end{bmatrix}^{T}
\end{equation}
is the driving vector.

To solve Eq.~(\ref{BR_SolnGreenFunction}) we evaluate the exponential of the coefficient matrix $\bf{A}$
\begin{align}
e^{t\bf{A}} &= \mathbf{M} e^{t \mathbf{\Lambda}} \mathbf{M}^{-1} \nonumber \\
&=\frac{1}{\text{det}(\bf{M})}
\begin{bmatrix}
    V_{11} & V_{21} & V_{31} \\
    V_{12} & V_{22} & V_{32} \\
    V_{13} & V_{23} & V_{33} \\
\end{bmatrix}
\begin{bmatrix}
    e^{\lambda_{1}t} & 0 & 0 \\
    0 & e^{\lambda_{2}t} & 0 \\
    0 & 0 & e^{\lambda_{3}t} \\
\end{bmatrix}
\begin{bmatrix}
    T_{11} & T_{12} & T_{13} \\
    T_{21} & T_{22} & T_{23} \\
    T_{31} & T_{32} & T_{33} \\
\end{bmatrix} \nonumber \\ 
&=\frac{1}{\text{det}(\bf{M})}
\begin{bmatrix}
    \phi_{11} & \phi_{21} & \phi_{31} \\
    \phi_{12} & \phi_{22} & \phi_{32} \\
    \phi_{13} & \phi_{23} & \phi_{33} \\
\end{bmatrix} \label{expA_Sym}
\end{align}
where $\bf{M}$ is the eigenvector matrix found in the previous section, $\bf{\Lambda} = \begin{bmatrix}
    \lambda_{1} & 0 & 0 \\
    0 & \lambda_{2} & 0 \\
    0 & 0 & \lambda_{3} \\
\end{bmatrix}$ is the eigenvalue matrix and
\begin{equation}
\phi_{ij}=\sum_{k=1}^{3}e^{\lambda_{k}t}V_{kj}T_{ki} \enspace ; \enspace i,\enspace j =\enspace 1,\enspace 2,\enspace 3.
\end{equation}
where $\lambda_{j}$, $j = 1, 2, 3$ are the eigenvalues of the coefficient matrix $\mathbf{A}$. 

We first find the contribution from the initial vector $\textbf{x}_{0}$ in the Eq.~(\ref{BR_SolnGreenFunction})
\begin{align}
    e^{\mathbf{A} t} \textbf{x}_{0} &= \frac{1}{\text{det}(\bf{M})}
\begin{bmatrix}
    \phi_{11} & \phi_{21} & \phi_{31} \\
    \phi_{12} & \phi_{22} & \phi_{32} \\
    \phi_{13} & \phi_{23} & \phi_{33} \\
\end{bmatrix}\begin{bmatrix}
         \rho_{g_{1}g_{1}}(0) \\  
         \rho_{g_{1}g_{2}}^{R}(0) \\
         \rho_{g_{1}g_{2}}^{I}(0)  
        \end{bmatrix} \nonumber \\
        &= \frac{1}{\text{det}(\bf{M})}
\begin{bmatrix}
    \phi_{11}\rho_{g_{1}g_{1}}(0) + \phi_{21}\rho_{g_{1}g_{2}}^{R}(0)  + \phi_{31}\rho_{g_{1}g_{2}}^{I}(0) \\
    \phi_{12}\rho_{g_{1}g_{1}}(0) + \phi_{22}\rho_{g_{1}g_{2}}^{R}(0)  + \phi_{32}\rho_{g_{1}g_{2}}^{I}(0) \\
    \phi_{13}\rho_{g_{1}g_{1}}(0) + \phi_{23}\rho_{g_{1}g_{2}}^{R}(0)  + \phi_{33}\rho_{g_{1}g_{2}}^{I}(0) \\
\end{bmatrix} \label{StateVec_Init}
\end{align}

Using Eqs.~(\ref{expA_Sym}) and (\ref{StateVec_Init}) in Eq.~(\ref{BR_SolnGreenFunction}), we get
\begin{align}
\textbf{x}(t) &= \frac{1}{\text{det}(\bf{M})} \int_{0}^{t} 
\begin{bmatrix}
\phi_{11} & \phi_{21} & \phi_{31} \\
\phi_{12} & \phi_{22} & \phi_{32} \\
\phi_{13} & \phi_{23} & \phi_{33} \\
\end{bmatrix}
\begin{bmatrix}
    r \\
    pr \\
    0 \\
\end{bmatrix}  
ds  \nonumber \\  
&=\frac{r}{\text{det}(\bf{M})} \int_{0}^{t}
\begin{bmatrix}
\phi_{11} + p \phi_{21} \\
\phi_{12} + p \phi_{22} \\
\phi_{13} + p \phi_{23} \\
\end{bmatrix} 
ds  \label{BR_SolnGreenFunction2}
\end{align}

%\color{red}
%(SKEdits: Added the initial state contribution in the general solutions.) 
%SUEYSH, 	PLEASE DOUBLE-CHECK YOUR NEW SECTION BELOW IN RED. I AM NOT SURE IT IS RELATED TO COHERENT INITIAL CONDITIONS!

\subsubsection{Analytical solutions for the strongly driven $\Lambda$ system in the overdamped regime}

Evaluating the integrals of $\textbf{x}(t)$ from Eq.~(\ref{BR_SolnGreenFunction2}) and considering the coherece-free initial state ($\bm{x}_{0} = [\frac{1}{2}, 0, 0]^{T}$), we find the expressions for $\rho_{g_{1}g_{1}}(t)$, $\rho_{g_{1}g_{2}}^{R}(t)$, and $\rho_{g_{1}g_{2}}^{I}(t)$
\begin{align}
\rho_{g_{1}g_{1}}(t) &= \frac{1}{\text{det}(\bf{M})}\big[\phi_{11}\rho_{g_{1}g_{1}}(0) + \phi_{21}\rho_{g_{1}g_{2}}^{R}(0)  + \phi_{31}\rho_{g_{1}g_{2}}^{I}(0)\big]+\frac{r}{\text{det}(\bf{M})} \sum_{k=1}^{3}\frac{(1-e^{\lambda_{k}t})}{-\lambda_{k}}V_{k1}(T_{k1}+pT_{k2}) \notag \\
&= \frac{1}{\text{det}(\bf{M})}\sum_{k=1}^{3} \frac{1}{2}V_{k1}T_{k1} +\frac{r}{\text{det}(\bf{M})} \sum_{k=1}^{3}\frac{(1-e^{\lambda_{k}t})}{-\lambda_{k}}V_{k1}(T_{k1}+pT_{k2}) \label{rho_gg_Pop}\\
\rho_{g_{1}g_{2}}^{R}(t) &=\frac{1}{\text{det}(\bf{M})}\big[ \phi_{12}\rho_{g_{1}g_{1}}(0) + \phi_{22}\rho_{g_{1}g_{2}}^{R}(0)  + \phi_{32}\rho_{g_{1}g_{2}}^{I}(0)\big]+\frac{r}{\text{det}(\bf{M})} \sum_{k=1}^{3}\frac{(1-e^{\lambda_{k}t})}{-\lambda_{k}}V_{k2}(T_{k1}+pT_{k2}) \notag \\
&=\frac{1}{\text{det}(\bf{M})}\sum_{k=1}^{3} \frac{1}{2}V_{k2}T_{k1}+\frac{r}{\text{det}(\bf{M})} \sum_{k=1}^{3}\frac{(1-e^{\lambda_{k}t})}{-\lambda_{k}}V_{k2}(T_{k1}+pT_{k2}) \label{rho_g1g2_Real} \\
\rho_{g_{1}g_{2}}^{I}(t) &=\frac{1}{\text{det}(\bf{M})}\big[ \phi_{13}\rho_{g_{1}g_{1}}(0) + \phi_{23}\rho_{g_{1}g_{2}}^{R}(0)  + \phi_{33}\rho_{g_{1}g_{2}}^{I}(0)\big]+\frac{r}{\text{det}(\bf{M})} \sum_{k=1}^{3}\frac{(1-e^{\lambda_{k}t})}{-\lambda_{k}}V_{k3}(T_{k1}+pT_{k2}) \notag \\
&=\frac{1}{\text{det}(\bf{M})}\sum_{k=1}^{3} \frac{1}{2}V_{k3}T_{k1}+\frac{r}{\text{det}(\bf{M})} \sum_{k=1}^{3}\frac{(1-e^{\lambda_{k}t})}{-\lambda_{k}}V_{k3}(T_{k1}+pT_{k2}) \label{rho_g1g2_Imag}
\end{align} 

To express these general solutions in terms of the physical parameters, we need to express the determinant of matrix $\bf{M}$ in terms of these parameters. The determinant of matrix $\bf{M}$ is obtained from Eq.~(\ref{detM_general}) using the required coefficients $V_{ij}$, $T_{ij}$ which are listed in Tables IX and X, respectively.
\begin{equation} \label{detM}
\begin{split}
\text{det}(\textbf{M}) = p\Big[\Big(-\frac{m_{1}}{F_{11}}+\Big(-\frac{m_{2}}{F_{11}}+\frac{f_{11}}{F_{11}}\frac{1}{(2z_{10}+4)}\Big)\frac{1}{\bar{n}^{2}}\Big(\frac{\Delta}{\gamma}\Big)^{2}\Big)+p\Big(-\frac{m_{5}}{F_{21}}-\frac{m_{6}}{F_{21}}\frac{1}{\bar{n}^{2}}\Big(\frac{\Delta}{\gamma}\Big)^{2}\Big)\\
+p\Big(-\frac{m_{11}}{L_{30}}\frac{1}{\bar{n}^{2}}\Big(\frac{\Delta}{\gamma}\Big)^{2}-\frac{m_{12}}{L_{30}}\frac{1}{\bar{n}^{4}}\Big(\frac{\Delta}{\gamma}\Big)^{4}\Big)\Big]\bar{n}^{2}\Big(\frac{\Delta}{\gamma}\Big)^{-2} \\
\end{split}
\end{equation}

The term proportional to $\frac{1}{\bar{n}^{4}}(\frac{\Delta}{\gamma})^{4})$ can be neglected for large $\bar{n}$ and we find
\begin{align} \label{detM2}
\text{det}(\bf{M}) &= p\Big[\Big(-\frac{m_{1}}{F_{11}}-\frac{m_{5}}{F_{21}}\Big)+\Big(-\frac{m_{2}}{F_{11}}+\frac{f_{11}}{F_{11}}\frac{1}{(2z_{10}+4)}-\frac{m_{6}}{F_{21}}-\frac{m_{11}}{L_{30}})\frac{1}{\bar{n}^{2}}\Big(\frac{\Delta}{\gamma}\Big)^{2}\Big]\bar{n}^{2}\Big(\frac{\Delta}{\gamma}\Big)^{-2} \nonumber \\
&=\Big[T_{1}(p)+T_{2}(p)\frac{1}{\bar{n}^{2}}\Big(\frac{\Delta}{\gamma}\Big)^{2}\Big]\bar{n}^{2}\Big(\frac{\Delta}{\gamma}\Big)^{-2}
\end{align}
where the coefficients $T_{1}(p)$, $T_{2}(p)$ are listed in Table VIII.

To find the expression for $\rho_{g_{1}g_{1}}(t)$, we calculate the following terms in Eq.~(\ref{rho_gg_Pop})      
\begin{align} 
V_{11}(T_{11}+pT_{12}) &= \Big[A_{1}+A_{2}\frac{1}{\bar{n}^{2}}\Big(\frac{\Delta}{\gamma}\Big)^{2}\Big]\bar{n}^{2}\Big(\frac{\Delta}{\gamma}\Big)^{-2}, \label{V11_T11_T12_Term} \\
V_{21}(T_{21}+pT_{22}) &= \Big[A_{3}+A_{4}\frac{1}{\bar{n}^{2}}\Big(\frac{\Delta}{\gamma}\Big)^{2}\Big]\bar{n}^{2}\Big(\frac{\Delta}{\gamma}\Big)^{-2}, \label{V21_T21_T22_Term} \\
V_{31}(T_{31}+pT_{32}) &= \Big[A_{5}+A_{6}\frac{1}{\bar{n}^{2}}\Big(\frac{\Delta}{\gamma}\Big)^{2}\Big], \label{V31_T31_T32_Term}
\end{align}
where the coefficients $A_{i}(p)$ , $i$ = 1 to 6 are listed in Table XII. All the terms $A_{i}(p)$ are only functions of $p$.

Using Eqs.~(\ref{V11_T11_T12_Term}) to (\ref{V31_T31_T32_Term}) in Eq.~(\ref{rho_gg_Pop}) the general expression of $\rho_{g_{1}g_{1}}(t)$ can be recast in the form
\begin{equation} \label{rho_gg_Pop2}
\begin{split}
\rho_{g_{1}g_{1}}(t)= \frac{1}{2\text{det}(\bf{M})} \Big[A_{1} \bar{n}^{2}\Big(\frac{\Delta}{\gamma}\Big)^{-2}e^{\lambda_{1}t}+A_{3} \bar{n}^{2}\Big(\frac{\Delta}{\gamma}\Big)^{-2}e^{\lambda_{2}t}+ A_{5}e^{\lambda_{3}t}\Big]+\frac{r}{\text{det}(\bf{M})}\Big[\Big(A_{1}+A_{2}\frac{1}{\bar{n}^{2}}\Big(\frac{\Delta}{\gamma}\Big)^{2}\Big)\bar{n}^{2}\Big(\frac{\Delta}{\gamma}\Big)^{-2}\Big(\frac{1-e^{\lambda_{1}t}}{-\lambda_{1}}\Big) \\ +\Big(A_{3}+A_{4}\frac{1}{\bar{n}^{2}}\Big(\frac{\Delta}{\gamma}\Big)^{2}\Big)\bar{n}^{2}\Big(\frac{\Delta}{\gamma}\Big)^{-2}\Big(\frac{1-e^{\lambda_{2}t}}{-\lambda_{2}}\Big)+
\Big(A_{5}+A_{6}\frac{1}{\bar{n}^{2}}\Big(\frac{\Delta}{\gamma}\Big)^{2}\Big)\Big(\frac{1-e^{\lambda_{3}t}}{-\lambda_{3}}\Big)\Big]
\end{split}
\end{equation}

Proceeding in the same way as for $\rho_{g_{1}g_{1}}(t)$, we find the expressions for $\rho^{R}_{g_{1}g_{2}}(t)$ and $\rho^{I}_{g_{1}g_{2}}(t)$ as
\begin{equation} \label{rho_g1g2_Real2}
\begin{split}
\rho^{R}_{g_{1}g_{2}}(t)=\frac{1}{2\text{det}(\bf{M})} \Big[B_{1} \bar{n}^{2}\Big(\frac{\Delta}{\gamma}\Big)^{-2}e^{\lambda_{1}t}+B_{3} \bar{n}^{2}\Big(\frac{\Delta}{\gamma}\Big)^{-2}e^{\lambda_{2}t}+ B_{5}e^{\lambda_{3}t}\Big]+\frac{r}{\text{det}(\bf{M})}\Big[\Big(B_{1}+B_{2}\frac{1}{\bar{n}^{2}}\Big(\frac{\Delta}{\gamma}\Big)^{2}\Big)\bar{n}^{2}\Big(\frac{\Delta}{\gamma}\Big)^{-2}\Big(\frac{1-e^{\lambda_{1}t}}{-\lambda_{1}}\Big) \\+\Big(B_{3}+B_{4}\frac{1}{\bar{n}^{2}}\Big(\frac{\Delta}{\gamma}\Big)^{2}\Big)\bar{n}^{2}\Big(\frac{\Delta}{\gamma}\Big)^{-2}\Big(\frac{1-e^{\lambda_{2}t}}{-\lambda_{2}}\Big)+
\Big(B_{5}+B_{6}\frac{1}{\bar{n}^{2}}\Big(\frac{\Delta}{\gamma}\Big)^{2}\Big)\Big(\frac{1-e^{\lambda_{3}t}}{-\lambda_{3}}\Big)\Big]
\end{split}
\end{equation}
\begin{equation} \label{rho_g1g2_Imag2}
\begin{split}
\rho^{I}_{g_{1}g_{2}}(t) =\frac{1}{2\text{det}(\bf{M})} \Big[C_{1}e^{\lambda_{1}t}+C_{3}e^{\lambda_{2}t}+C_{5}e^{\lambda_{3}t}\Big]\bar{n}\Big(\frac{\Delta}{\gamma}\Big)^{-1}+\frac{r}{\text{det}(\bf{M})}\Big[\Big(C_{1}+C_{2}\frac{1}{\bar{n}^{2}}\Big(\frac{\Delta}{\gamma}\Big)^{2}\Big)\Big(\frac{1-e^{\lambda_{1}t}}{-\lambda_{1}}\Big) \\ +\Big(C_{3}+C_{4}\frac{1}{\bar{n}^{2}}\Big(\frac{\Delta}{\gamma}\Big)^{2}\Big)\Big(\frac{1-e^{\lambda_{2}t}}{-\lambda_{2}}\Big) +\Big
(C_{5}+C_{6}\frac{1}{\bar{n}^{2}}\Big(\frac{\Delta}{\gamma}\Big)^{2}\Big)\Big(\frac{1-e^{\lambda_{3}t}}{-\lambda_{3}}\Big)\Big]\bar{n}\Big(\frac{\Delta}{\gamma}\Big)^{-1}
\end{split}
\end{equation}
where the coefficients $B_{i}(p)$, $C_{i}(p)$, $i$=1 to 6 are listed in Table XII. All the terms $B_{i}(p)$ and $C_{i}(p)$ are only functions of $p$.

The general expressions for $\rho_{g_{1}g_{1}}(t)$, $\rho^{R}_{g_{1}g_{2}}(t)$ and $\rho^{I}_{g_{1}g_{2}}(t)$ can take two forms depending on the value of the alignment factor $p$. If the alignment factor is greater than the critical value $i.e.$ $p>p_{c}$, then
\begin{align} \label{lambda13_scaling}
\lambda_{j} &= \gamma z_{j0} \bar{n}
= -r |z_{j0}| \enspace ;\enspace j =\enspace 1,\enspace 3
\end{align}

However, the second eigenvalue has a different scaling relation as shown in Sec.~IC
\begin{equation} \label{lambda2_scaling}
\lambda_{2} = -\gamma |f_{21}| \frac{1}{\bar{n}}\bigg(\frac{\Delta}{\gamma}\bigg)^{2}.
\end{equation}

Substituting the eigenvalues from Eqs.~(\ref{lambda13_scaling}), (\ref{lambda2_scaling}) into Eqs.~(\ref{rho_gg_Pop2}) to (\ref{rho_g1g2_Imag2}), using Eq.~(\ref{detM2}) for $\text{det}(\mathbf{M}$) and neglecting the small term proportional to $\frac{1}{\bar{n}^{4}}(\frac{\Delta}{\gamma})^{4}$ for large $\bar{n}$, we get
\begin{equation} \label{rho_gg_Pop3}
\begin{split}	
\rho_{g_{1}g_{1}}(t) = \frac{1}{\Big[T_{1}(p)+T_{2}(p)\frac{1}{\bar{n}^{2}}\Big(\frac{\Delta}{\gamma}\Big)^{2}\Big]}\Bigg\lbrace \frac{1}{2}\bigg[A_{1} e^{-\gamma|z_{10}|\bar{n}t}+A_{3}e^{-\gamma|f_{21}|\frac{1}{\bar{n}}\Big(\frac{\Delta}{\gamma}\Big)^{2}t}+A_{5}\frac{1}{\bar{n}^{2}}\Big(\frac{\Delta}{\gamma}\Big)^{2}e^{-\gamma|z_{30}|\bar{n}t}\bigg] + \bigg[\Big(A_{1}+A_{2}\frac{1}{\bar{n}^{2}}\Big(\frac{\Delta}{\gamma}\Big)^{2}\Big) \\ \times \Big(\frac{1-e^{-\gamma|z_{10}|\bar{n}t}}{ |z_{10}|}\Big)
+\Big(A_{3}+A_{4}\frac{1}{\bar{n}^{2}}\Big(\frac{\Delta}{\gamma}\Big)^{2}\Big)\bar{n}^{2}\Big(\frac{\Delta}{\gamma}\Big)^{-2} \Big(\frac{1-e^{-\gamma|f_{21}|\frac{1}{\bar{n}}\Big(\frac{\Delta}{\gamma}\Big)^{2}t}}{|f_{21}|}\Big) +
A_{5}\frac{1}{\bar{n}^{2}}\Big(\frac{\Delta}{\gamma}\Big)^{2}\Big(\frac{1-e^{-\gamma|z_{30}|\bar{n}t}}{ |z_{30}|}\Big)\bigg]\Bigg\rbrace
\end{split}
\end{equation}
\begin{equation} \label{rho_g1g2_Real3}
\begin{split}
\rho^{R}_{g_{1}g_{2}}(t)= \frac{1}{\Big[T_{1}(p)+T_{2}(p)\frac{1}{\bar{n}^{2}}\Big(\frac{\Delta}{\gamma}\Big)^{2}\Big]}\Bigg\lbrace \frac{1}{2}\bigg[B_{1} e^{-\gamma|z_{10}|\bar{n}t}+B_{3}e^{-\gamma|f_{21}|\frac{1}{\bar{n}}\Big(\frac{\Delta}{\gamma}\Big)^{2}t}+B_{5}\frac{1}{\bar{n}^{2}}\Big(\frac{\Delta}{\gamma}\Big)^{2}e^{-\gamma|z_{30}|\bar{n}t}\bigg]+ \bigg[\Big(B_{1}+B_{2}\frac{1}{\bar{n}^{2}}\Big(\frac{\Delta}{\gamma}\Big)^{2}\Big) \\ \times \Big(\frac{1-e^{-\gamma|z_{10}|\bar{n}t}}{ |z_{10}|}\Big)
+\Big(B_{3}+B_{4}\frac{1}{\bar{n}^{2}}\Big(\frac{\Delta}{\gamma}\Big)^{2}\Big)\bar{n}^{2}\Big(\frac{\Delta}{\gamma}\Big)^{-2} \Big(\frac{1-e^{-\gamma|f_{21}|\frac{1}{\bar{n}}\Big(\frac{\Delta}{\gamma}\Big)^{2}t}}{|f_{21}|}\Big)+
B_{5}\frac{1}{\bar{n}^{2}}\Big(\frac{\Delta}{\gamma}\Big)^{2}\Big(\frac{1-e^{-\gamma|z_{30}|\bar{n}t}}{ |z_{30}|}\Big)\bigg]\Bigg\rbrace
\end{split}
\end{equation}
\begin{equation} \label{rho_g1g2_Imag3}
\begin{split}
\rho^{I}_{g_{1}g_{2}}(t)=\frac{1}{\Big[T_{1}(p)+T_{2}(p)\frac{1}{\bar{n}^{2}}\Big(\frac{\Delta}{\gamma}\Big)^{2}\Big]} \frac{1}{\bar{n}} \Big(\frac{\Delta}{\gamma}\Big)\Bigg\lbrace \frac{1}{2}\bigg[C_{1} e^{-\gamma|z_{10}|\bar{n}t}+C_{3}e^{-\gamma|f_{21}|\frac{1}{\bar{n}}\Big(\frac{\Delta}{\gamma}\Big)^{2}t}+C_{5} e^{-\gamma|z_{30}|\bar{n}t}\bigg] + \bigg[\Big(C_{1}+C_{2}\frac{1}{\bar{n}^{2}}\Big(\frac{\Delta}{\gamma}\Big)^{2}\Big) \\ \times \Big(\frac{1-e^{-\gamma|z_{10}|\bar{n}t}}{ |z_{10}|}\Big)
+\Big(C_{3}+C_{4}\frac{1}{\bar{n}^{2}}\Big(\frac{\Delta}{\gamma}\Big)^{2}\Big)\bar{n}^{2}\Big(\frac{\Delta}{\gamma}\Big)^{-2} \Big(\frac{1-e^{-\gamma|f_{21}|\frac{1}{\bar{n}}\Big(\frac{\Delta}{\gamma}\Big)^{2}t}}{|f_{21}|}\Big)+
\Big(C_{5}+C_{6}\frac{1}{\bar{n}^{2}}\Big(\frac{\Delta}{\gamma}\Big)^{2}\Big)\Big(\frac{1-e^{-\gamma|z_{30}|\bar{n}t}}{ |z_{30}|}\Big)\bigg]\Bigg\rbrace
\end{split}
\end{equation}

The solutions (\ref{rho_gg_Pop3}), (\ref{rho_g1g2_Real3}) and (\ref{rho_g1g2_Imag3}) are plotted in Figs.~2(a) and (b) of the main text.

If the alignment factor is less than the critical value ($p<p_{c}$), then
\begin{align}
\lambda_{j} &= \gamma z_{j0} \bar{n}
= -r |z_{j0}| \enspace ;\enspace j =\enspace 1,\enspace 2,\enspace 3 \label{lambda_j_scaling}.
\end{align}

Substituting the eigenvalues into Eqs.~(\ref{rho_gg_Pop2}) to (\ref{rho_g1g2_Imag2}), using Eq.~(\ref{detM2}) for $\text{det}(\textbf{M})$ and neglecting the small term proportional to $\frac{1}{\bar{n}^{4}}(\frac{\Delta}{\gamma})^{4}$ for large $\bar{n}$, we get
\begin{equation} \label{rho_gg_Pop4}
\begin{split}
\rho_{g_{1}g_{1}}(t) = \frac{1}{\Big[T_{1}(p)+T_{2}(p)\frac{1}{\bar{n}^{2}}\Big(\frac{\Delta}{\gamma}\Big)^{2}\Big]}\Bigg\lbrace \frac{1}{2}\bigg[A_{1} e^{-\gamma|z_{10}|\bar{n}t}+A_{3}e^{-\gamma|z_{20}|\bar{n}t}+A_{5}\frac{1}{\bar{n}^{2}}\Big(\frac{\Delta}{\gamma}\Big)^{2}e^{-\gamma|z_{30}|\bar{n}t}\bigg] + \bigg[\Big(A_{1}+A_{2}\frac{1}{\bar{n}^{2}}\Big(\frac{\Delta}{\gamma}\Big)^{2}\Big) \\ \times \Big(\frac{1-e^{-\gamma|z_{10}|\bar{n}t}}{ |z_{10}|}\Big)
+\Big(A_{3}+A_{4}\frac{1}{\bar{n}^{2}}\Big(\frac{\Delta}{\gamma}\Big)^{2}\Big)\Big(\frac{1-e^{-\gamma|z_{20}|\bar{n}t}}{ |z_{20}|}\Big)+ 
A_{5}\frac{1}{\bar{n}^{2}}\Big(\frac{\Delta}{\gamma}\Big)^{2}\Big(\frac{1-e^{-\gamma|z_{30}|\bar{n}t}}{ |z_{30}|}\Big)\bigg] \Bigg\rbrace
\end{split}
\end{equation}
\begin{equation} \label{rho_g1g2_Real4}
\begin{split}
\rho^{R}_{g_{1}g_{2}}(t)=\frac{1}{\Big[T_{1}(p)+T_{2}(p)\frac{1}{\bar{n}^{2}}\Big(\frac{\Delta}{\gamma}\Big)^{2}\Big]}\Bigg\lbrace \frac{1}{2}\bigg[B_{1} e^{-\gamma|z_{10}|\bar{n}t}+B_{3}e^{-\gamma|z_{20}|\bar{n}t}+B_{5}\frac{1}{\bar{n}^{2}}\Big(\frac{\Delta}{\gamma}\Big)^{2}e^{-\gamma|z_{30}|\bar{n}t}\bigg]+ \bigg[\Big(B_{1}+B_{2}\frac{1}{\bar{n}^{2}}\Big(\frac{\Delta}{\gamma}\Big)^{2}\Big) \\ \times \Big(\frac{1-e^{-\gamma|z_{10}|\bar{n}t}}{ |z_{10}|}\Big)
+\Big(B_{3}+B_{4}\frac{1}{\bar{n}^{2}}\Big(\frac{\Delta}{\gamma}\Big)^{2}\Big)\Big(\frac{1-e^{-\gamma|z_{20}|\bar{n}t}}{ |z_{20}|}\Big)+ 
B_{5}\frac{1}{\bar{n}^{2}}\Big(\frac{\Delta}{\gamma}\Big)^{2}\Big(\frac{1-e^{-\gamma|z_{30}|\bar{n}t}}{ |z_{30}|}\Big)\bigg]\Bigg\rbrace
\end{split}
\end{equation} \\
\begin{equation} \label{rho_g1g2_Imag4}
\begin{split}
\rho^{I}_{g_{1}g_{2}}(t)= \frac{1}{\Big[T_{1}(p)+T_{2}(p)\frac{1}{\bar{n}^{2}}\Big(\frac{\Delta}{\gamma}\Big)^{2}\Big]} \frac{1}{\bar{n}} \Big(\frac{\Delta}{\gamma}\Big)\Bigg\lbrace \frac{1}{2}\bigg[C_{1} e^{-\gamma|z_{10}|\bar{n}t}+C_{3}e^{-\gamma|z_{20}|\bar{n}t}+C_{5} e^{-\gamma|z_{30}|\bar{n}t}\bigg]+\bigg[(C_{1}+C_{2}\frac{1}{\bar{n}^{2}}\Big(\frac{\Delta}{\gamma}\Big)^{2}\Big) \\ \times \Big(\frac{1-e^{-\gamma|z_{10}|\bar{n}t}}{ |z_{10}|}\Big)+\Big(C_{3}+C_{4}\frac{1}{\bar{n}^{2}}\Big(\frac{\Delta}{\gamma}\Big)^{2}\Big)\Big(\frac{1-e^{-\gamma|z_{20}|\bar{n}t}}{ |z_{20}|}\Big)+\Big(C_{5}+C_{6}\frac{1}{\bar{n}^{2}}\Big(\frac{\Delta}{\gamma}\Big)^{2}\Big)\Big(\frac{1-e^{-\gamma|z_{30}|\bar{n}t}}{ |z_{30}|}\Big)\bigg]\Bigg\rbrace 
\end{split}
\end{equation}
 
In the limit of small energy level spacing ($\frac{\Delta}{\gamma} \ll 1$) and large $\bar{n}$, we can neglect terms proportional to $\frac{1}{\bar{n}^{2}}(\frac{\Delta}{\gamma})^{-2}$ and the general solution further simplifies.

For $p>p_{c}$, we obtain
\begin{align}
\rho_{g_{1}g_{1}}(t)&=\frac{1}{T_{1}(p)}\Bigg\lbrace\frac{1}{2}\bigg[A_{1} e^{-\gamma|z_{10}|\bar{n}t}+A_{3}e^{-\gamma|f_{21}|\frac{1}{\bar{n}}\Big(\frac{\Delta}{\gamma}\Big)^{2}t}\bigg]+\bigg[A_{1}\Big(\frac{1-e^{-\gamma|z_{10}|\bar{n}t}}{ |z_{10}|}\Big)
+A_{4}\Big(\frac{1-e^{-\gamma|f_{21}|\frac{1}{\bar{n}}\big(\frac{\Delta}{\gamma}\big)^{2}t}}{|f_{21}|}\Big)\bigg] \Bigg\rbrace \label{rho_gg_Pop5} \\
\rho^{R}_{g_{1}g_{2}}(t) &= \frac{1}{T_{1}(p)}\Bigg\lbrace\frac{1}{2}\bigg[B_{1} e^{-\gamma|z_{10}|\bar{n}t}+B_{3}e^{-\gamma|f_{21}|\frac{1}{\bar{n}}\Big(\frac{\Delta}{\gamma}\Big)^{2}t}\bigg]+\bigg[B_{1}\Big(\frac{1-e^{-\gamma|z_{10}|\bar{n}t}}{ |z_{10}|}\Big)
+B_{4}\Big(\frac{1-e^{-\gamma|f_{21}|\frac{1}{\bar{n}}\big(\frac{\Delta}{\gamma}\big)^{2}t}}{|f_{21}|}\Big)\bigg]\Bigg\rbrace \label{rho_g1g2_Real5}
\end{align}
\begin{equation}
\begin{split}
\rho^{I}_{g_{1}g_{2}}(t) = \frac{1}{T_{1}(p)}  \Big(\frac{\Delta}{\bar{n}\gamma}\Big)\Bigg\lbrace \frac{1}{2}\bigg[C_{1} e^{-\gamma|z_{10}|\bar{n}t}+C_{3}e^{-\gamma|f_{21}|\frac{1}{\bar{n}}\Big(\frac{\Delta}{\gamma}\Big)^{2}t}+C_{5} e^{-\gamma|z_{30}|\bar{n}t}\bigg]+\bigg[C_{1}\Big(\frac{1-e^{-\gamma|z_{10}|\bar{n}t}}{ |z_{10}|}\Big) \\
+C_{4}\Big(\frac{1-e^{-\gamma|f_{21}|\frac{1}{\bar{n}}\big(\frac{\Delta}{\gamma}\big)^{2}t}}{|f_{21}|}\Big) +C_{5}\Big(\frac{1-e^{-\gamma|z_{30}|\bar{n}t}}{ |z_{30}|}\Big)\bigg]\Bigg\rbrace \label{rho_g1g2_Imag5}
\end{split}
\end{equation}

For $p<p_{c}$, we find
\begin{align}
\rho_{g_{1}g_{1}}(t) &= \frac{1}{T_{1}(p)}\Bigg\lbrace\frac{1}{2}\bigg[A_{1} e^{-\gamma|z_{10}|\bar{n}t}+A_{3}e^{-\gamma|z_{20}|\bar{n}t}\bigg]+\bigg[A_{1}\Big(\frac{1-e^{-\gamma|z_{10}|\bar{n}t}}{ |z_{10}|}\Big)+A_{3} \Big(\frac{1-e^{-\gamma|z_{20}|\bar{n}t}}{ |z_{20}|}\Big)\bigg]\Bigg\rbrace \label{rho_gg_Pop6} \\
\rho^{R}_{g_{1}g_{2}}(t) &= \frac{1}{T_{1}(p)}\Bigg\lbrace\frac{1}{2}\bigg[B_{1} e^{-\gamma|z_{10}|\bar{n}t}+B_{3}e^{-\gamma|z_{20}|\bar{n}t}\bigg]+\bigg[B_{1}\Big(\frac{1-e^{-\gamma|z_{10}|\bar{n}t}}{ |z_{10}|}\Big)
+B_{3}\Big(\frac{1-e^{-\gamma|z_{20}|\bar{n}t}}{ |z_{20}|}\Big)\bigg]\Bigg\rbrace \label{rho_g1g2_Real6}
\end{align}
\begin{equation}
\begin{split}
\rho^{I}_{g_{1}g_{2}}(t) = \frac{1}{T_{1}(p)}\Big(\frac{\Delta}{\bar{n}\gamma}\Big)\Bigg\lbrace \frac{1}{2}\bigg[C_{1} e^{-\gamma|z_{10}|\bar{n}t}+C_{3}e^{-\gamma|z_{20}|\bar{n}t}+C_{5} e^{-\gamma|z_{30}|\bar{n}t}\bigg]+ \bigg[C_{1}\Big(\frac{1-e^{-\gamma|z_{10}|\bar{n}t}}{ |z_{10}|}\Big) \\ +C_{3}\Big(\frac{1-e^{-\gamma|z_{20}|\bar{n}t}}{ |z_{20}|}\Big)+C_{5}\Big(\frac{1-e^{-\gamma|z_{30}|\bar{n}t}}{ |z_{30}|}\Big)\bigg]\Bigg\rbrace \label{rho_g1g2_Imag6}
\end{split}
\end{equation}

\color{black}

\subsubsection{Analytical expression in the underdamped regime [$\Delta / (\bar{n} \gamma) > f(p)$]; Approximate solutions}
In the underdamped regime of the $\Lambda$-system, where the ground-state splitting is large compared to the incoherent pumping rate $i.e. \enspace \Delta \gg r$, the real part of coherence ($\rho^{R}_{g_{1}g_{2}}(t)$) is small in comparison to the ground state populations [$\rho_{g_{1}g_{1}}(t)$ in Eq.~(1) of the main text]. We can neglect the real part of coherence to write Eq.~(1) in the following form
\begin{align} \label{BR_UD_pop}
    \dot\rho_{g_{1}g_{1}} &= -\textit{r} \rho_{g_{1}g_{1}} + (\textit{r}+\gamma) (1-2 \rho_{g_{1}g_{1}}) \nonumber \\
    &= -(3r+2\gamma) \rho_{g_{1}g_{1}} + (r+\gamma)
\end{align} 
where we have expressed the excited state population in terms of ground state populations.

We consider the situation where the system evolves from zero coherences and all the populations are in the ground state at time $t = 0$ ($i.e. \enspace \rho_{g_{1}g_{1}}= \rho_{g_{2}g_{2}} = 1/2$ and $\rho_{g_{1}g_{2}} = 0$ ). The first order linear differential equation (\ref{BR_UD_pop}) can then be solved to yield
\begin{equation} \label{GeneralSolnUD1}
    \rho_{g_{1}g_{1}}(t) = \frac{r+\gamma}{(3r+2\gamma)} + \frac{r}{2(3r+2\gamma)} e^{-(3r+2\gamma)t}
\end{equation}

The explicit expression for $\rho_{g_{1}g_{1}}$ obtained above is used to simplify the coherence equation (2) of the main text which for the real part of coherence ($\rho_{g_{1}g_{2}}^{R}$) takes the form
\begin{equation} \label{CohEqnUD}
    \dot\rho^{R}_{g_{1}g_{2}} =  -\textit{r} \rho^{R}_{g_{1}g_{2}} + \Delta \rho^{I}_{g_{1}g_{2}} - \frac{pr}{2} e^{-(3r+2\gamma)t}
\end{equation}
where the incoherent pumping rate $r = \bar{n} \gamma$. \par

We observe that the assumption that the real part of coherence is small compared to the the excited state population [see Eq.~(1) of the main text] decouples the coherence from the excited-state population. This is a valid assumption in the underdamped regime. As a result Eq.~(2) of the main text reduces to a system of coupled equations with only two variables ($\rho_{ab}^{R}$ and $\rho_{ab}^{I}$) along with Eq.~(\ref{BR_UD_pop})
\begin{align}
    \dot\rho^{R}_{g_{1}g_{2}} &=  -\textit{r} \rho^{R}_{g_{1}g_{2}} + \Delta \rho^{I}_{g_{1}g_{2}} - \frac{pr}{2} e^{-(3r+2\gamma)t} \nonumber \\
    \dot\rho^{I}_{g_{1}g_{2}} &= -\Delta \rho^{R}_{g_{1}g_{2}} - \textit{r}\rho^{I}_{g_{1}g_{2}} \label{CohEqn_2by2}
\end{align} \par 

Eq.~(\ref{CohEqn_2by2}), in the Liouville representation takes the form
\begin{equation} \label{DEOM_StateVec_2by2}
    \dot{\textbf{x}}(t) = \mathbf{A} \textbf{x} + \textbf{d}
\end{equation}
where $\textbf{x}(t) = [\rho^{R}_{g_{1}g_{2}}, \enspace \rho_{g_{1}g_{2}}^{I}]^{T}$ is the state vector, $\textbf{d} = [- \frac{pr}{2} e^{-(3r+2\gamma)t}, \enspace 0]^{T}$ is the driving vector and $\mathbf{A}$ is the coefficient matrix
\begin{equation}
    \mathbf{A} = 
\begin{bmatrix} 
-r & \Delta \\
-\Delta & -r 
\end{bmatrix}
\end{equation} \par 

The solution of Eq.~(\ref{DEOM_StateVec_2by2}) can be written as %\cite{BoyceDiPrima}
\begin{equation} \label{GreensFunc_SV}
    \dot{\textbf{x}}(t) = e^{\mathbf{A}t} \textbf{x}_{0} + \int_{0}^{t} e^{\mathbf{A}(t-s)}\enspace \textbf{d} \enspace ds
\end{equation}
where $\textbf{x}_{0} = [0, \enspace 0]^{T}$ is the initial vector, so the first term on the right hand side is zero. To evaluate the second term, we need to find the integrand which is an exponential function of the coefficient matrix $\mathbf{A}$. Let $\lambda_{i}$ and $\mathbf{v}_{i}$ be the eigenvalue and the corresponding eigenvector of the matrix $\mathbf{A}$ such that $\mathbf{A} \mathbf{v}_{i} = \lambda_{i}\mathbf{v}_{i}; \enspace i =1, 2$. Then, using the similarity transformation, the coefficient matrix can be recast as
\begin{equation}
    \mathbf{A} = \mathbf{M} \Lambda \mathbf{M}^{-1}
\end{equation}
where $\Lambda = 
\begin{bmatrix} 
\lambda_{1} & 0 \\
0 & \lambda_{2} 
\end{bmatrix}$ is the diagonal eigenvalue matrix, $\mathbf{M} = [\mathbf{v}_{1} \enspace \mathbf{v}_{2}]^{T}$ is the eigenvector matrix with the corresponding eigenvectors as columns and $\mathbf{M}^{-1}$ is the inverse of $\mathbf{M}$. \par

Diagonalizing the coefficient matrix $\mathbf{A}$, we find the eigenvalues $\lambda_{1} = -r + i\Delta$, $\lambda_{2} = -r - i\Delta$ with corresponding eigenvectors $\mathbf{v}_{1} = [1 \enspace i]^{T}$ and $\mathbf{v}_{2} = [1 \enspace -i]^{T}$. So, the eigenvalue matrix and its inverse are given by
\begin{align}
    \mathbf{M} = 
\begin{bmatrix} 
1 & 1 \\
i & -i 
\end{bmatrix}; \enspace
    \mathbf{M}^{-1} = -\frac{1}{2i} 
\begin{bmatrix} 
-i & -1 \\
-i & 1 
\end{bmatrix}
\end{align}  

The exponential function of the coefficient matrix in Eq.~(\ref{GreensFunc_SV}) now simplifies as
\begin{align}
    \text{e}^{\mathbf{A}(t-s)} &= \mathbf{M} \text{e}^{\Lambda (t-s)} \mathbf{M}^{-1} \nonumber \\
    &= -\frac{1}{2i} 
\begin{bmatrix} 
1 & 1 \\
i & -i 
\end{bmatrix}  
\begin{bmatrix} 
\text{e}^{\lambda_{1} (t-s)} & 0 \\
0 & \text{e}^{\lambda_{2} (t-s)} 
\end{bmatrix}   
\begin{bmatrix} 
-i & -1 \\
-i & 1 
\end{bmatrix} \nonumber \\
&= -\frac{1}{2i} 
\begin{bmatrix} 
-i(\text{e}^{\lambda_{1} (t-s)}+\text{e}^{\lambda_{2} (t-s)}) & -(\text{e}^{\lambda_{1} (t-s)}-\text{e}^{\lambda_{2} (t-s)}) \\
(\text{e}^{\lambda_{1} (t-s)}-\text{e}^{\lambda_{2} (t-s)}) & -i (\text{e}^{\lambda_{1} (t-s)}+\text{e}^{\lambda_{2} (t-s)}) 
\end{bmatrix} \nonumber \\
&= -\frac{1}{2i} 
\begin{bmatrix} 
-i\text{e}^{-r(t-s)}(\text{e}^{i \Delta (t-s)}+\text{e}^{-i \Delta (t-s)}) & -\text{e}^{-r(t-s)}(\text{e}^{i \Delta (t-s)}-\text{e}^{-i \Delta (t-s)}) \\
\text{e}^{-r(t-s)}(\text{e}^{i \Delta (t-s)}-\text{e}^{-i \Delta (t-s)}) & -i \text{e}^{-r(t-s)}(\text{e}^{i \Delta (t-s)}+\text{e}^{-i \Delta (t-s)})
\end{bmatrix} \nonumber \\
&= -\frac{1}{2i} \text{e}^{-r(t-s)} 
\begin{bmatrix} 
-2i \cos \Delta (t-s) & -2i \sin \Delta (t-s) \\
2i \sin \Delta (t-s) & -2i \cos \Delta (t-s) 
\end{bmatrix} \nonumber \\
&= \text{e}^{-r(t-s)} \begin{bmatrix} 
\cos \Delta (t-s) & \sin \Delta (t-s) \\
-\sin \Delta (t-s) & \cos \Delta (t-s) 
\end{bmatrix} 
\end{align} \par 

Substituting this result and the driving vector in the line below Eq.~(\ref{DEOM_StateVec_2by2}) in Eq.~(\ref{GreensFunc_SV}), we obtain
\begin{align}
    \textbf{x}(t) &= \int_{0}^{t} \text{e}^{-r(t-s)} 
    \begin{bmatrix} 
\cos \Delta (t-s) & \sin \Delta (t-s) \\
-\sin \Delta (t-s) & \cos \Delta (t-s) 
\end{bmatrix} 
\begin{bmatrix} 
-\frac{pr}{2}\text{e}^{-(3r+2\gamma)s}  \\
0  
\end{bmatrix} \enspace ds \nonumber \\
&= \int_{0}^{t} \text{e}^{-r(t-s)} 
    \begin{bmatrix} 
-\frac{pr}{2}\text{e}^{-(3r+2\gamma)s} \cos \Delta (t-s) \\
\frac{pr}{2}\text{e}^{-(3r+\gamma)s} \sin \Delta (t-s) 
\end{bmatrix}  \enspace ds \nonumber \\
&= -\frac{pr}{2} \text{e}^{-rt} \int_{0}^{t}  \text{e}^{-2(r+\gamma)s}
    \begin{bmatrix} 
\cos \Delta (t-s) \\
-\sin \Delta (t-s) 
\end{bmatrix}  \enspace ds \label{SS_2by2}
\end{align} 

Comparing the corresponding components of $\textbf{x}(t)$ on both sides of Eq.~(\ref{SS_2by2}), we find
\begin{align}
    \rho_{g_{1}g_{2}}^{R}(t) &=  -\frac{pr}{2} \text{e}^{-rt} \int_{0}^{t} \text{e}^{-2 (r+\gamma) s} \cos \Delta (t-s) \enspace ds, \nonumber \\
    \rho_{g_{1}g_{2}}^{I}(t) &= \frac{pr}{2} \text{e}^{-rt} \int_{0}^{t} \text{e}^{-2 (r+\gamma) s} \sin \Delta (t-s) \enspace ds. \label{Coh_For_2by2}
\end{align}

%We use the following standard integrals to evaluate Eq. (16) 
%\begin{align}
%\int \text{e}^{ax} \cos bx \enspace dx &= \frac{1}{a^{2}+b^{2}} \text{e}^{ax} (a \cos bx +b \sin bx) \nonumber \\
%\int \text{e}^{ax} \sin bx \enspace dx &= \frac{1}{a^{2}+b^{2}} \text{e}^{ax} (a \sin bx - b \cos bx) \nonumber 
%\end{align} \par

%To evaluate the integrals in Eq.~(\ref{Coh_For_2by2}), it is convenient to consider a more general integral
%\begin{align}
%    \text{I} &= \int_{0}^{t} \text{e}^{-2(r+\gamma) s} \text{e}^{i \Delta (t-s)} \enspace ds 
 %   = \text{e}^{i \Delta t} \int_{0}^{t} \text{e}^{-(2r+2\gamma+i\Delta)s} \enspace ds 
  %  = \text{e}^{i \Delta t} \left[ -\frac{1}{(2r+2\gamma+i\Delta)} \text{e}^{-(2r+2\gamma+i\Delta)s}\right]_0^t \nonumber \\
  %  &= \text{e}^{i \Delta t}\frac{-1}{(2r+2\gamma+i\Delta)} \big[\text{e}^{-(2r+2\gamma+i\Delta)t} - 1\big] 
  %  = \frac{2r+2\gamma-i\Delta}{4(r+\gamma)^{2}+\Delta^{2}} \big[\text{e}^{i \Delta t} - \text{e}^{-(2r+2\gamma)t} \big] \nonumber \\
%    &= \frac{2r+2\gamma-i\Delta}{4(r+\gamma)^{2}+\Delta^{2}} \big[\cos \Delta t + i \sin \Delta t - \text{e}^{-(2r+2\gamma)t} %\big] \nonumber \\
%    \begin{split}
 %   &= \Big\lbrace \frac{2(r+\gamma)}{4(r+\gamma)^{2}+\Delta^{2}} \big[\cos \Delta t - \text{e}^{-(2r+2\gamma)t}\big] +\frac{\Delta}{4(r+\gamma)^{2}+\Delta^{2}} \sin \Delta t \Big\rbrace \\ &\qquad + i\Big\lbrace \frac{2(r+\gamma)}{4(r+\gamma)^{2}+\Delta^{2}} \sin \Delta t  - \frac{\Delta}{4(r+\gamma)^{2}+\Delta^{2}} \big[\cos \Delta t - \text{e}^{-(2r+2\gamma)t}\big] \Big \rbrace
  %  \end{split} \label{Gen_Int_Iso}
%\end{align}

Evaluating the integrals we obtain 
\begin{align}
    \int_{0}^{t} \text{e}^{-2 (r+\gamma) s} \cos \Delta (t-s) \enspace ds &=  \frac{2(r+\gamma)}{4(r+\gamma)^{2}+\Delta^{2}} \big[\cos \Delta t - \text{e}^{-(2r+2\gamma)t}\big] +\frac{\Delta}{4(r+\gamma)^{2}+\Delta^{2}} \sin \Delta t, \nonumber \\
    \int_{0}^{t} \text{e}^{-2 (r+\gamma) s} \sin \Delta (t-s) \enspace ds &= \frac{2(r+\gamma)}{4(r+\gamma)^{2}+\Delta^{2}} \sin \Delta t  - \frac{\Delta}{4(r+\gamma)^{2}+\Delta^{2}} \big[\cos \Delta t - \text{e}^{-(2r+2\gamma)t}\big]. \label{Gen_Int_Iso2}
\end{align} \par

Finally, substituting Eq.~(\ref{Gen_Int_Iso2}) into Eq.~(\ref{Coh_For_2by2}) we find the explicit expressions for the real part of coherence
\begin{align} \label{GeneralSolnUD2}
    \rho_{g_{1}g_{2}}^{R}(t) &=  -\frac{pr}{2} \text{e}^{-rt} \bigg[ \frac{2(r+\gamma)}{4(r+\gamma)^{2}+\Delta^{2}} \Big\lbrace\cos \Delta t - \text{e}^{-(2r+2\gamma)t}\Big\rbrace +\frac{\Delta}{4(r+\gamma)^{2}+\Delta^{2}} \sin \Delta t \bigg] \nonumber \\
    &= \frac{pr}{2[4(r+\gamma)^{2}+\Delta^{2}]} \bigg[2(r+\gamma) \text{e}^{-(3r+2\gamma)t} - \Big\lbrace 2(r+\gamma) \cos \Delta t + \Delta \sin \Delta t \Big\rbrace \text{e}^{-rt} \bigg]. 
\end{align} \par

Similarly,
\begin{align} \label{GeneralSolnUD3}
    \rho_{g_{1}g_{2}}^{I}(t) &= \frac{pr}{2} \text{e}^{-rt} \bigg[ \frac{2(r+\gamma)}{4(r+\gamma)^{2}+\Delta^{2}} \sin \Delta t  - \frac{\Delta}{4(r+\gamma)^{2}+\Delta^{2}} \Big\lbrace\cos \Delta t - \text{e}^{-(2r+2\gamma)t}\Big\rbrace \bigg] \nonumber \\
    &= \frac{pr}{2[4(r+\gamma)^{2}+\Delta^{2}]} \bigg[ \Delta \text{e}^{-(3r+2\gamma)t} + \Big\lbrace2(r+\gamma) \sin \Delta t - \Delta  \cos \Delta t \Big\rbrace \text{e}^{-rt}\bigg]. 
\end{align} 

These equations provide  closed-form analytic solutions for the population and coherence dynamics of the $\Lambda$-system in the underdamped regime, in both the strong and weak pumping limits.
For the weakly driven $\Lambda$-system ($r \ll \gamma$ and $r \ll \Delta$), we can further simplify the expressions (\ref{GeneralSolnUD2}) and (\ref{GeneralSolnUD3}) as 
\begin{align}
%    \rho_{g_{1}g_{1}}(t) &= \frac{r+\gamma}{(3r+2\gamma)} + \frac{r}{2(3r+2\gamma)}\text{e}^{-2\gamma t} \label{BR_Pop_WPL} \\
    \rho_{g_{1}g_{2}}^{R}(t) &= \frac{pr}{2[4\gamma^{2}+\Delta^{2}]} \bigg[2\gamma \text{e}^{-2\gamma t} - \big\lbrace 2\gamma \cos\Delta t + \Delta \sin\Delta t \big\rbrace \text{e}^{-rt} \bigg] \label{Real_Coh_UD_WPL} \\
    \rho_{g_{1}g_{2}}^{I}(t) &= \frac{pr}{2[4\gamma^{2}+\Delta^{2}]} \bigg[\Delta \text{e}^{-2\gamma t} + \big\lbrace2\gamma \sin\Delta t - \Delta  \cos\Delta t \big\rbrace \text{e}^{-rt}\bigg] \label{Imag_Coh_UD_WPL}
\end{align} 

\color{blue}
\subsubsection{Evolution from a coherent initial state}
We finally consider a more general case, where the $\Lambda$ system evolves from a coherent initial state defined by the vector $\textbf{x}_{0} = [\rho^{R}_{g_{1} g_{2}}(0), \enspace \rho^{I}_{g_{1} g_{2}}(0)]^{T}$. In that case, the general solution contains an additional contribution from the initial vector in Eq.~(\ref{GreensFunc_SV}). 
%to Eqs.~(\ref{GeneralSolnUD2}) and (\ref{GeneralSolnUD3}). 
To evaluate the contribution from the initial vector $\textbf{x}_{0}$, we perform matrix multiplication
\begin{align}
  e^{At}\textbf{x}_{0} &= \text{e}^{-rt} \begin{bmatrix} 
\cos \Delta t & \sin \Delta t \\
-\sin \Delta t & \cos \Delta t \end{bmatrix}   \begin{bmatrix}
\rho^{R}_{g_{1} g_{2}}(0) \\
\rho^{I}_{g_{1} g_{2}}(0)
\end{bmatrix}
 = \text{e}^{-rt} \begin{bmatrix} 
 \rho^{R}_{g_{1} g_{2}}(0) \cos\Delta t + \rho^{I}_{g_{1} g_{2}}(0)  \sin\Delta t \\
-\rho^{R}_{g_{1} g_{2}}(0) \sin\Delta t + \rho^{I}_{g_{1} g_{2}}(0) \cos\Delta t \end{bmatrix}
\end{align}

Adding the corresponding terms from the above equation to (\ref{GeneralSolnUD2}) and (\ref{GeneralSolnUD3}), we obtain the following analytical solutions for the population and coherence dynamics starting from a coherent initial state 
\begin{align}
\begin{split}
    \rho_{g_{1}g_{2}}^{R}(t) &= \big[\rho^{R}_{g_{1} g_{2}}(0) \cos\Delta t + \rho^{I}_{g_{1} g_{2}}(0)  \sin\Delta t)\big] \text{e}^{-rt} \\ & + \frac{pr}{2[4(r+\gamma)^{2}+\Delta^{2}]} \bigg[2(r+\gamma) \text{e}^{-(3r+2\gamma)t} - \Big\lbrace 2(r+\gamma) \cos \Delta t + \Delta \sin \Delta t \Big\rbrace \text{e}^{-rt} \bigg] 
\end{split} \nonumber \\
\begin{split}
    \rho_{g_{1}g_{2}}^{I}(t) &= \big[-\rho^{R}_{g_{1} g_{2}}(0) \sin\Delta t + \rho^{I}_{g_{1} g_{2}}(0) \cos\Delta t\big] \text{e}^{-rt} \\ & + \frac{pr}{2[4(r+\gamma)^{2}+\Delta^{2}]} \bigg[ \Delta \text{e}^{-(3r+2\gamma)t} + \Big\lbrace2(r+\gamma) \sin \Delta t - \Delta  \cos \Delta t \Big\rbrace \text{e}^{-rt}\bigg] \label{GeneralSolnFinal}
\end{split}
\end{align}

When the ground states of the $\Lambda$-system are initially in the equal coherent superposition with zero phase $[\rho^{R}_{g_{1} g_{2}}(0) =1/2,\rho_{g_{i} g_{i}}(0) =1/2]$, the initial vector is $\textbf{x}_{0} = [1/2, \enspace 0]^{T}$, and the general solutions in Eq.~(\ref{GeneralSolnFinal}) become
\begin{align}
    \rho_{g_{1}g_{2}}^{R}(t) &=  \frac{1}{2}\cos\Delta t \text{e}^{-rt} + \frac{pr}{2[4(r+\gamma)^{2}+\Delta^{2}]} \bigg[2(r+\gamma) \text{e}^{-(3r+2\gamma)t} - \Big\lbrace 2(r+\gamma) \cos \Delta t + \Delta \sin \Delta t \Big\rbrace \text{e}^{-rt} \bigg] \nonumber \\
    \rho_{g_{1}g_{2}}^{I}(t) &= -\frac{1}{2} \sin\Delta t \text{e}^{-rt} + \frac{pr}{2[4(r+\gamma)^{2}+\Delta^{2}]} \bigg[ \Delta \text{e}^{-(3r+2\gamma)t} + \Big\lbrace2(r+\gamma) \sin \Delta t - \Delta  \cos \Delta t \Big\rbrace \text{e}^{-rt}\bigg] \label{GeneralSolnFinal2}
\end{align}
\color{black}
\textcolor{blue}{These equations  are valid in the underdamped limit of the strongly driven $\Lambda$-system with the ground states initially in an equal coherent superposition. The general solutions for the coherent dynamics of a weakly driven $\Lambda$-system in the underdamped regime are discussed in Sec.~IG.3.}

\subsection{Physical basis for long-lived Fano coherences: The effective decoherence rate model} 

In this section we derive an approximate equation of motion, which accurately describes the time evolution of Fano coherences at $t>1/r$ in the strong-pumping limit ($r\gg \gamma$). As in our previous work on the strongly driven V-system \cite{Koyu:18}, this equation admits a transparent physical interpretation in terms of the various coherence-generating and destroying processes as described in the main text.

% physical origin of the long coherence lifetimes observed in Figs.~\ref{fig: Lambda_System_Dynamics_OD_SPL}(b) and (c). We study the BR master equations (\ref{BR_Eqns01_Lambda}) and (\ref{BR_Eqns02_Lambda}), and show the emergence of long-lived coherences as a result of the competition between the different generation and decay processes. 

In the limit $r\gg \gamma$ we can neglect the terms proportional to $\gamma$ and further simplify the BR equations for the real and imaginary parts of the Fano coherence $\rho_{g_{1}g_{2}}=\rho_{g_{1}g_{2}}^R+i\rho_{g_{1}g_{2}}^I$ to find
\begin{align}
%\dot{\rho}_{aa} &= -r\rho_{aa} + r\rho_{cc} - pr \rho^R_{ab}, \\ 
\label{BRpicture1}
\dot{\rho}_{g_{1}g_{2}}^R &= -r\rho_{g_{1}g_{2}}^R + pr(\rho_{ee} - \rho_{g_{1}g_{1}}) +\rho_{g_{1}g_{2}}^I\Delta \\  
\dot{\rho}_{g_{1}g_{2}}^I &= -r\rho_{g_{1}g_{2}}^I  - \rho_{g_{1}g_{2}}^R\Delta \label{BRpicture2}
\end{align}

The above equations describe the production and decay of Fano coherences $\rho_{g_{1}g_{2}}$ due to quantum interference between the incoherent excitation pathways $e \to g_{1}$ and $e \to g_{2}$ originating from the excited state. The coherence generation rate $pr\rho_{ee}$ is  proportional to the transition dipole alignment parameter $p$, which quantifies the degree of interference between the two transitions. We note that the {\it real part of coherence} is generated directly via the interference process. The imaginary coherence is decoupled from the population dynamics, but contributes nontrivially as a result of its coupling to the real part of the coherence [the $\rho_{g_{1}g_{2}}^R\Delta$ term in Eq. (\ref{BRpicture2})]. The coherences evolve unitarily in time according to the terms proportional to  $\Delta$ and decay via stimulated emission described by the term $-r\rho_{g_{1}g_{2}}$ in Eq.~(\ref{BRpicture2}). An additional interference term $pr(\rho_{ee} - \rho_{g_{1}g_{1}})$  in Eq. (\ref{BRpicture1})  causes decoherence at a rate proportional to the population difference between the ground and excited levels.

\begin{figure}[ht!]
%\captionsetup{singlelinecheck = false, format= hang, justification=raggedright, font=footnotesize, labelsep=space}
     \includegraphics[width=0.9\linewidth]{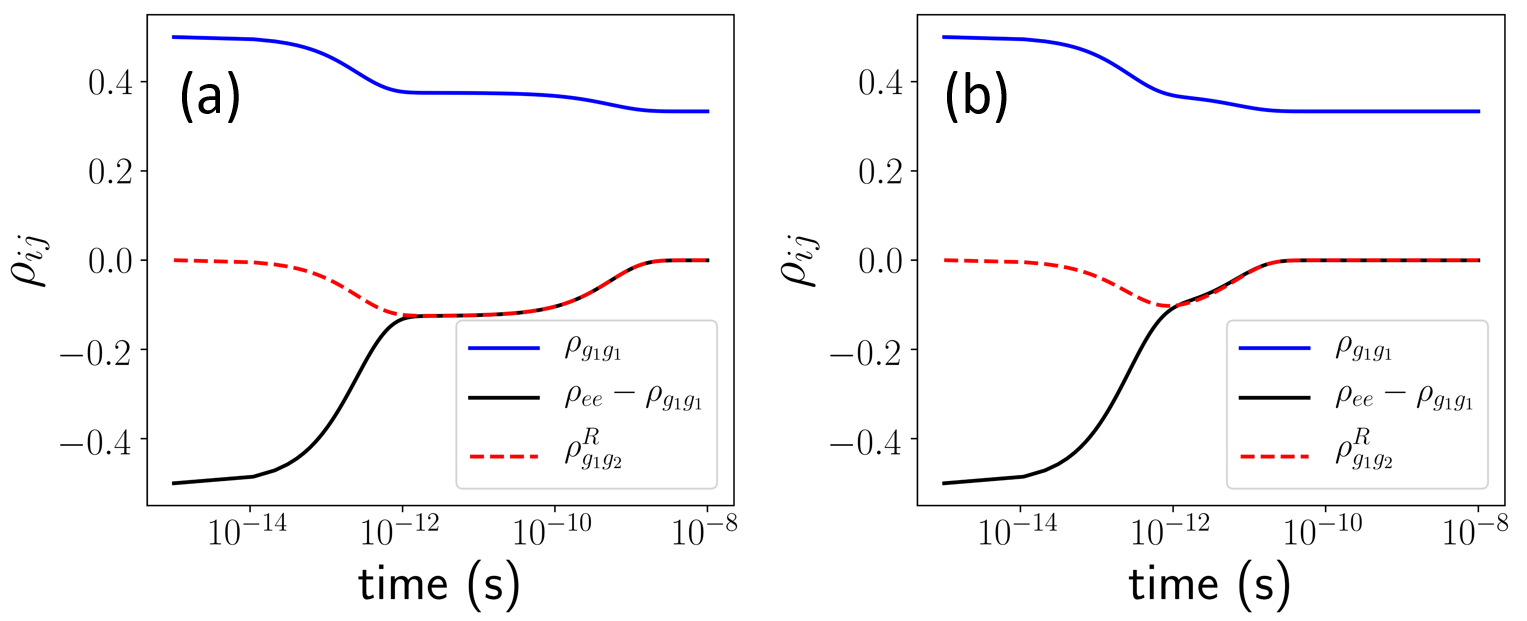} 
     \caption{Population and coherence dynamics of the symmetric $\Lambda$-system with coherence-free initial state ($\rho^{R}_{g_{1}g_{2}} = 0$) in the strong-pumping limit ( $\bar{n}=10^{3}$) for ${\Delta}/{\gamma}=50$. The transition dipole alignment parameter $p$ is set to 1 [panel (a)] and to 0.9 [panel (b)].}
 \label{fig:P1} 
\end{figure}
 
To illustrate the interplay between the different coherence-generating and coherence-destroying mechanisms, we plot in Fig.~\ref{fig:P1} the time evolution of the populations and coherences that enter Eq.~(\ref{BRpicture1}). The decay of the ground-state population $\rho_{ee}$ is accompanied by a steady growth of excited-state populations and coherences.  In the quasi-steady state formed on the timescale $1/r < t < \tau_c$ the population inversion term $\rho_{ee}-\rho_{g_{1}g_{1}}$ drives coherence generation. From Fig.~\ref{fig:P1}, we observe the quasisteady state (1) the time evolution of the population difference $\rho_{ee}-\rho_{g_{1}g_{1}}$ is identical to that of the real part of the coherence  $\rho_{g_{1}g_{2}}^R$ and (2) the imaginary part of the coherence remains constant in time.   Setting the left-land side of Eq. (\ref{BRpicture2}) to zero, we obtain the imaginary part of the quasisteady coherence as $\rho_{g_{1}g_{2}}^I=-(\Delta/r)\rho^R_{g_{1}g_{2}}$ (we verified this result numerically in the overdamped regime).  These considerations allow us to simplify Eq.~(\ref{BRpicture1}) to yield
\begin{equation}\label{BRpicture_gamma_eff}
\dot{\rho}_{g_{1}g_{2}}^R =  -r\left(1-p + \frac{\Delta^2}{r^2}\right) \rho_{g_{1}g_{2}}^R   \qquad (t>1/r),
\end{equation}
which describes coherence decay on the timescale $t > 1/r$ (note that coherence generation occurs on  shorter timescales given by $t \simeq 1/r$). The simple form of Eq. (\ref{BRpicture_gamma_eff}) enables us to introduce an {\it effective decoherence rate}  $\gamma_d^\text{eff}(p)=r(1-p+\frac{\Delta^2}{r^2})$ and the   {\it effective coherence time }
\begin{equation}\label{BRpicture_tau_eff}
\tau_d^\text{eff} =  \frac{1}{r(1-p + \frac{\Delta^2}{r^2})} 
\end{equation}

The effective decoherence rate model thus establishes that the lifetime of noise-induced coherences  is determined by two mechanisms: (1) the interplay between coherence-generating Fano interference and incoherent stimulated emission discussed above [the term $r(1-p)$] and (2) the coupling between the real and imaginary parts of the coherence  [the term $\Delta^2/r$].  The second mechanism  is due to the unitary interconversion between the real and imaginary parts of the coherence, which occurs at a rate $\Delta$.  Because the imaginary coherence decays at a rate $r$ [Eq. (\ref{BRpicture2})], this unitary interconversion makes a small second-order  contribution to the overall decay rate proportional to $r\Delta^2$. The two mechanisms contribute equally for $p=p_c^\text{eff}$ where
\begin{equation}
p_c^\text{eff} = 1-\frac{\Delta^2}{r^2},
\end{equation}
which is close to the exact result derived above $p_c = \sqrt{1-\Delta^2/r^2} \simeq 1-\frac{1}{2}\frac{\Delta^2}{r^2}$ in the overdamped limit $(\Delta/(\bar{n}\gamma) \ll 1)$.

Our expression for the effective decoherence rate  (\ref{BRpicture_tau_eff}) clarifies the physical origin of the different regimes of coherent dynamics.  At $p>p_c^\text{eff}$ the term $r(1-p)$ is small compared with $\Delta^2/r^2$, the rate of coherence generation (via Fano interference) and decay (via incoherent stimulated emission) are almost exactly balanced. As a result,  the  effective decoherence rate is dominated by the second-order mechanism (2) and Eq. (\ref{BRpicture_tau_eff}) yields  $\tau_d^\text{eff} =  {r}/{\Delta^2}=\bar{n}\gamma/\Delta^2$, which is identical to  the exact result to within the factor 1.34. Remarkably, the effective decoherence model correctly predicts the linear scaling of the coherence lifetime with both $\bar{n}$ and $\Delta/\gamma$ in the $p>p_c$ regime.

%At $p < p_c^\text{eff}$, the rate of coherence decay exceeds that of coherence generation,  the term $r(1-p)$ dominates, and the effective coherence time is determined by the interplay between Fano interference and incoherent stimulated emission. From Eq. (\ref{BRpicture_tau_eff}) we obtain  $\tau_d^\text{eff} = \frac{1}{r(1-p)}$, again consistent with the exact result (XX) to within a numerical factor. As in the limit $p\to 1$,  the effective decoherence model correctly predicts the inverse scaling of the coherence lifetime with $\bar{n}$ in the $p<p_c$ regime.

We note that the transition dipole alignment parameter $p$ can be thought of as controlling the relative contribution of mechanisms (1) and (2) leading to the overall effective decoherence rate  (\ref{BRpicture_tau_eff}). In the strong-pumping regime, mechanism (1) is much more efficient in destroying the coherence than mechanism (2). Significantly, mechanism (1) is $p$-dependent and can be suppressed by taking $p\to 1$, leading to the long-lived coherent regime governed by mechanism (2).

\subsection{Weak pumping regime}

In this section we will specifically consider the weak pumping limit, where  incoherent driving is much slower than  spontaneous decay ($r \ll \gamma$). In this regime, the average photon number of the radiation field is small ($\bar{n} \ll 1$).

\subsubsection{Solving the equation $D = 0$ in the weak pumping limit}
For small $\frac{\Delta}{\gamma}$ and $\bar{n}$, the discriminant in Eq.~(\ref{Disc_Poly}) is well represented by the zeroth and second order terms $d_{0}$, $d_{2}\bar{n}^{2}$ only, and it can be written as
\begin{equation} \label{D_wpl}
D=\frac{\gamma^{6}}{2916}\big(d_{0} + d_{2} \bar{n}^{2} \big)
\end{equation}
where $d_{0}$, and $d_{2}$ are the expansion coefficients listed in Table~I, which are further simplified as \begin{align}
    d_{0} &= 108 \Big(\frac{\Delta}{\gamma}\Big)^{2} \Big[\Big(\frac{\Delta}{\gamma}\Big)^{2} + 4 \Big]^{2} \nonumber \\
    d_{2} &= -9(47+108p^{2})\Big(\frac{\Delta}{\gamma}\Big)^{4} + 72(64+87p^{2}+18p^{4}) \Big(\frac{\Delta}{\gamma}\Big)^{2} - 432p^{4} \label{d_coeffs_WPL}
\end{align}

Our goal is to find a relationship between $\frac{\Delta}{\gamma}$ and $\bar{n}$ by solving the equation $D = 0$. Using the expansion coefficients in Eq.~(\ref{d_coeffs_WPL}), Eq.~(\ref{D_wpl}) can be recast
\begin{align}
    \bar{n}^{2} &= -\frac{d_{0}}{d_{2}}
                = \frac{108 \Big(\frac{\Delta}{\gamma}\Big)^{2} \Big[\Big(\frac{\Delta}{\gamma}\Big)^{2} + 4 \Big]^{2}}{\Big[9(47+108p^{2})\Big(\frac{\Delta}{\gamma}\Big)^{4} - 72(64+87p^{2}+18p^{4}) \Big(\frac{\Delta}{\gamma}\Big)^{2} + 432p^{4}\Big]}
\end{align}
We rearrange the above equation to obtain the ratio of $(\Delta/\gamma)$ to $\bar{n}$ in the following form 
\begin{align}
    \frac{1}{\bar{n}^{2}}\Big(\frac{\Delta}{\gamma}\Big)^{2} &= \frac{\Big[9(47+108p^{2})\Big(\frac{\Delta}{\gamma}\Big)^{4} - 72(64+87p^{2}+18p^{4}) \Big(\frac{\Delta}{\gamma}\Big)^{2} + 432p^{4}\Big]}{108 \Big[\Big(\frac{\Delta}{\gamma}\Big)^{2} + 4 \Big]^{2}}
\end{align}

For small values of $\Delta/\gamma$, we can neglect the $(\Delta/\gamma)^{2}$ and $(\Delta/\gamma)^{4}$ terms on the right-hand side  to obtain
\begin{equation} \label{CR_Eq}
    \Big(\frac{\Delta}{r}\Big)^{2} = \frac{p^{4}}{4}
\end{equation}

Finally, we take the square root on both sides of Eq.~(\ref{CR_Eq}) to find the equation for the critical line 
\begin{equation}
    \frac{\Delta}{r} = \frac{p^{2}}{2}
\end{equation}

\begin{figure}[b]
\centering
\includegraphics[width=0.95\linewidth]{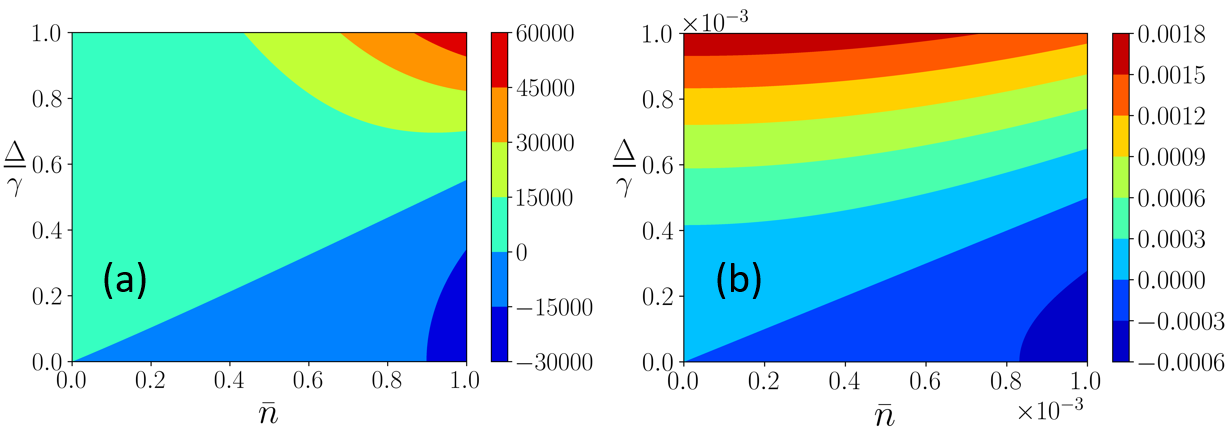}
\caption{(a) A contour plot of the discriminant $D$ for small $\Delta/\gamma$ and $\bar{n}$ and $p = 1.0$. (b) An expanded view of the region close to the origin ($\Delta/\gamma \ll 1$, $\bar{n} \ll 1$ and $p = 1.0$).}
\label{fig: Contour_Plot_WPL}
\end{figure}

%Using the general results in Sec.~IA and neglecting the terms $\bar{n}^k$ with $k\ge 1$ in the general discriminant (\ref{Disc_Poly}) in the weak pumping limit becomes
%\begin{align}
%D = \frac{\gamma^{6}}{2916} \Big[4\Big(3\frac{\Delta^{2}}{\gamma^{2}}-4\Big)^{3}+ \Big(36\frac{\Delta^{2}}{\gamma^{2}}+16\Big)^{2}\Big] &= \frac{\gamma^{6}}{27}\Big(\frac{\Delta}{\gamma}\Big)^{2}\Big[\Big(\frac{\Delta}{\gamma}\Big)^{4}+8\Big(\frac{\Delta}{\gamma}\Big)^{2}+16\Big] \nonumber \\
%&= \frac{\gamma^{6}}{27}\Big(\frac{\Delta}{\gamma}\Big)^{2}\Big[\Big(\frac{\Delta}{\gamma}\Big)^{2}+4\Big]^{2} \label{Discriminant_WPL}
%\end{align}

%The discriminant (\ref{Discriminant_WPL}) is always positive, which implies that the weakly driven $\Lambda$-system is always in the underdamped regime characterized by oscillatory coherence dynamics. This is in contrast to the weakly driven V-system, where all three dynamical regimes are present \cite{Tscherbul:14,Dodin:16}. This feature of the $\Lambda$-system is further seen in the contour plot of Fig.~\ref{fig: Contour_Plot_WPL}(a), which shows that the line of zero discriminant passes through the origin, and only the region of positive discriminant exists in the limit $\bar{n} \rightarrow 0$.
%This implies that the dynamics of a weakly driven $\Lambda$-system is strictly constrained in the underdamped regime. %Additionally, the discriminant (\ref{Discriminant_WPL}) is independent of the dipole alignment factor $p$. 

{Therefore, for small values of $\frac{\Delta}{\gamma}$ and $\bar{n}$, the critical line is a straight line with the slope given by $m = p^{2}/2$. The dynamical regimes  for small $\bar{n}$ are illustrated in the contour plot of Fig.~\ref{fig: Contour_Plot_WPL}(a), (b), which displays that the line of zero discriminant passes through the origin. The underdamped and overdamped regimes are defined by the conditions $ \frac{\Delta}{r} > \frac{p^{2}}{2}$ and $ \frac{\Delta}{r} < \frac{p^{2}}{2}$, respectively. Because one expects $\Delta/r >1$ to be  the case for very small $r$, overdamped dynamics are strongly suppressed in the weak pumping limit. }
%The dynamics of a weakly driven $\Lambda$-system is principally in the underdamped regime characterized by oscillatory coherence dynamics.}

For completeness, we also consider the case of the weakly driven degenerate $\Lambda$-system ($\Delta = 0$), where
%analyze the discriminant (\ref{Disc_Poly}). 
 the zeroth and first order coefficients $d_{0}(p)$ and $d_{1}(p)$ in Eq.~(\ref{Disc_Poly}) vanish and the discriminant becomes
 \begin{equation} \label{Discriminant_WPL_Degenerate}
    D = -432p^{2}\bar{n}^{2}
\end{equation}
The discriminant  is always negative for $p \ne 0$. This implies that the degenerate $\Lambda$-system always exhibits dynamics either in the overdamped regime ($p \neq 0$) or the critical regime ($p = 0$). This is in contrast to the weakly driven non-degenerate $\Lambda$-system considered above, whose dynamics are {primarily} in the underdamped regime.

In the strong pumping regime, the discriminant (\ref{Disc_Poly}) of  the degenerate $\Lambda$-system becomes 
\begin{equation} \label{Discriminant_Degenerate}
    D = -108p^{2}\bar{n}^{2}\big[4+4p^{2}(5+2p^{2})\bar{n}+p^{2}(37+36p^{2})\bar{n}^{2}+6p^{2}(5+9p^{2})\bar{n}^{3}+9p^{2}(1+3p^{2})\bar{n}^{4}\big]
\end{equation}

We observe that the discriminant  is either zero or negative. Thus, the dynamics of the strongly driven degenerate $\Lambda$-system are restricted to either overdamped  ($p \neq 0$) or critical ($p = 0$) regimes. This is again different from the strongly driven non-degenerate $\Lambda$-system, which shows dynamics in all three regimes. 

\subsubsection{Eigenvalues and coherence lifetimes{in the underdamped regime [$\Delta/r > p^{2}/2$]} }
The eigenvalues in the weak pumping regime can be expressed as a polynomial function of $\bar{n} \ll 1$ (see above)
\begin{equation} \label{Eigval_WPL}
\lambda_{j}=\gamma \sum_{k=0}^{5}\tilde{z}_{jk} \bar{n}^{k} \enspace (j = 1 - 3).
\end{equation} 

Neglecting the terms with $k> 2$ we find
\begin{equation} \label{Eigval_WPL2}
\lambda_{j}=\gamma \big[\tilde{z}_{j0} + \tilde{z}_{j1} \bar{n} + \tilde{z}_{j2} \bar{n}^{2}\big] \enspace (j = 1 - 3).
\end{equation} 

Equation~(\ref{Eigval_WPL2}) is a general expression for the eigenvalues of the coefficient matrix (\ref{Amatrix_Lambda}) in the weak pumping limit. The above expression is valid for $\bar{n} \leq 0.01$ and it converges quickly with $\bar{n}$. The expansion coefficients $\tilde{z}_{jk}$ are listed in Table XIII.   

\begin{figure}[t]
\centering
\includegraphics[width=\linewidth]{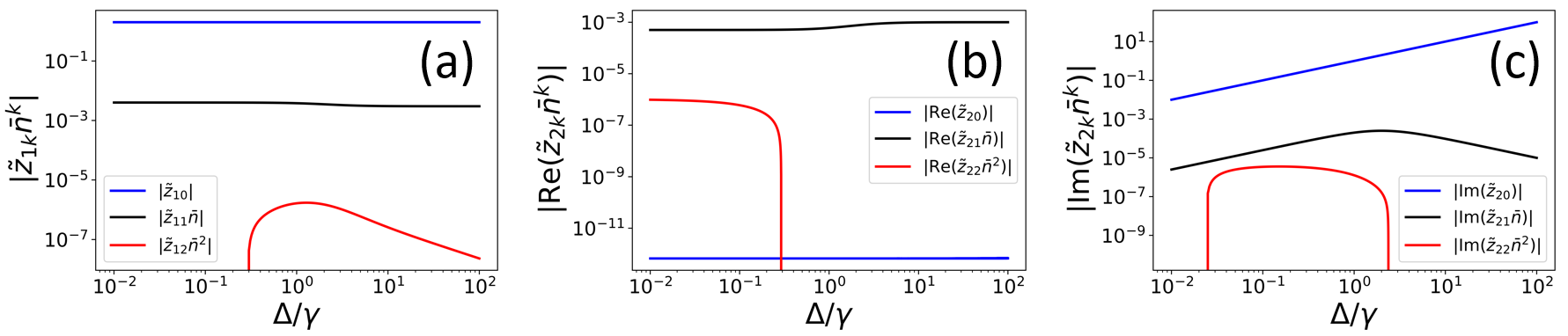}
\caption{Relative contributions of the different terms to the eigenvalues $\lambda_{1}$ (a), Real (b) and Imaginary (c) part of $\lambda_2$ plotted as a function of $\Delta/\gamma$ in the weak pumping limit with $\bar{n} = 10^{-3}$ and $p = 1$. Note that $\lambda_3$ is complex conjugate of $\lambda_2$.}
\label{fig: Zterms_WPL}
\end{figure}

Next we compare the relative contributions of the different $\tilde{z}_{jk} \bar{n}^{k}$ terms in Eq.~(\ref{Eigval_WPL2}) as a function of $\Delta/\gamma$ in Figure~\ref{fig: Zterms_WPL}. From Fig.~\ref{fig: Zterms_WPL}(a), we observe that the zeroth order term $\tilde{z}_{10}$ is the dominant term for $|\lambda_{1}|$ which can be written as
\begin{equation} \label{lambda1_WPL}
    \lambda_{1} = |\tilde{z}_{10}| \gamma,
\end{equation}
where the $p$-dependent coefficient $\tilde{z}_{10}$ is listed in Table~XIII.

Similarly, the real and imaginary parts of the eigenvalue $\lambda_{2}$ as a function of $\Delta/\gamma$ are plotted in Figs.~\ref{fig: Zterms_WPL}(b) and (c). We see that the imaginary part of $|\text{Im}(\lambda_{2})|$ depends significantly on the zeroth order term ($\text{Im}(\tilde{z}_{20})$) similar to $\lambda_{1}$. However, it is the first-order term $\text{Re}(\tilde{z}_{20})\bar{n}$ that is dominant in the real part of $|\text{Re}(\lambda_{2})|$. This approximation permits us to express the real and imaginary parts of $\lambda_{2}$ in the following form
\begin{align}
     \text{Re}(\lambda_{2}) &= \text{Re}(\tilde{z}_{21}) r, \nonumber \\
     \text{Im}(\lambda_{2}) &= \text{Im}(\tilde{z}_{20}) \gamma, \label{lambda2_WPL}
\end{align}
where the coefficients $\tilde{z}_{20}$ and $\tilde{z}_{21}$ are listed in Table~XIII. Here, we note that the eigenvalue $\lambda_{3}$ exhibits same features as $\lambda_{2}$ as they are complex conjugates of each other. \\

The eigenvalues in the weak pumping regime can be evaluated more simply by assuming the decoupling of the coherence term in Eq.~(\ref{BR_Eqns01_Lambda}). This is a valid assumption for the strong pumping regime ($\Delta/r \geq 1$) as long as $\Delta/\gamma > 1$. In the weak pumping limit with $\Delta/\gamma > 1$, the coefficient matrix (\ref{Amatrix_Lambda_UD}) after diagonalization results in the following three eigenvalues 
\begin{align}
    \lambda_{1} &= -2\gamma \label{EigVal1_WPL} \\
    \lambda_{2} &= -r + i \Delta \label{EigVal2_WPL} \\
    \lambda_{3} &= -r - i \Delta \label{EigVal3_WPL}
\end{align}

The eigenvalue spectrum $\lambda_{k}$ of the coefficient matrix $\textbf{A}$ 
%given by Eqs.~(\ref{EigVal1_WPL})-(\ref{EigVal3_WPL}) show that the eigenvalues in the weakly pumped limit 
consists of one real and two complex conjugate  eigenvalues. Thus, the dynamics of the $\Lambda$-system weakly driven by incoherent radiation is in the underdamped regime as discussed above.
It is important to note that we can not neglect the incoherent driving rate $r$ in the eigenvalues $\lambda_{2,3}$ in spite of its small magnitude in the weak pumping limit. \par 

If the ground state splitting is small in comparison to the spontaneous decay rate ($\Delta/\gamma < 1$), the eigenvalue (\ref{EigVal1_WPL}) is in complete agreement with the eigenvalue evaluated in (\ref{lambda1_WPL}). However, the eigenvalues (\ref{EigVal2_WPL})-(\ref{EigVal3_WPL}) deviate from the exact values. Next, we compute the eigenvalues ($\lambda_{2}$) using Eq.~(\ref{Eigval_WPL2}) for which the expansion coefficients $\tilde{z}_{2k}$ are evaluated with $(\alpha_{2},\beta_{2}) = (\omega^{2}, \omega)$. 

Substituting the coefficients $\tilde{z}_{2k}$, $\omega = (-1+i\sqrt{3})/2$, $\omega^{2} = (-1-i\sqrt{3})/2$ in Eq.~(\ref{lambda2_WPL}), we obtain 
\begin{align}
    \text{Re}(\lambda_{2}) &= r \bigg\lbrace-\frac{5}{3} +\frac{1}{6\tilde{K}}(8+6p^{2})+\frac{1}{18\tilde{K}}\Big(3\frac{\Delta^{2}}{\gamma^{2}}-4\Big)t_{1}+\frac{1}{18}\tilde{K}t_{1} \bigg\rbrace
     \label{lambda2_Real} \\
    \text{Im}(\lambda_{2}) &= -\sqrt{3}\gamma \bigg\lbrace\frac{1}{6\tilde{K}}\Big(3\frac{\Delta^{2}}{\gamma^{2}}-4\Big)+\frac{\tilde{K}}{6}\bigg)\bigg\rbrace \label{lambda2_Imag}
\end{align}

\begin{figure}[b]
\centering
\includegraphics[width=0.8\linewidth]{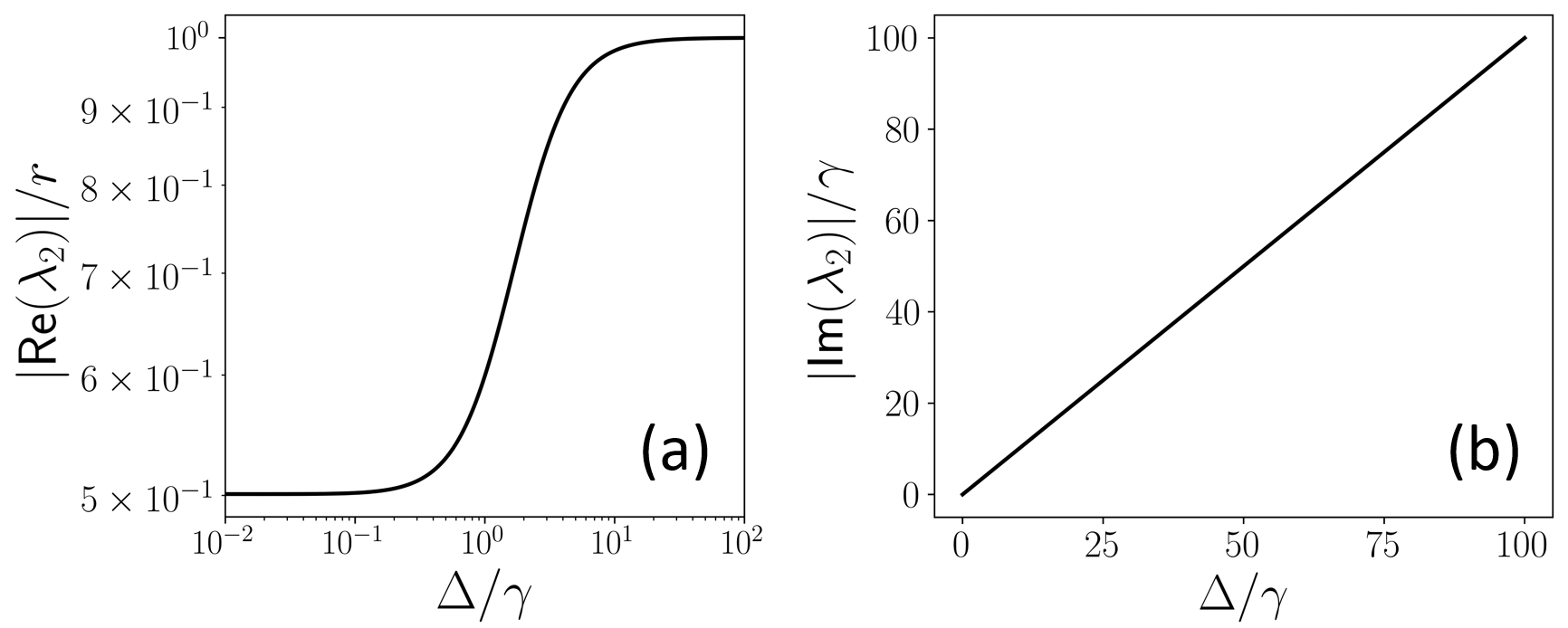}
\caption{The real (a) and imaginary (b) parts of the eigenvalue $\lambda_{2}$ in Eqs.~(\ref{lambda2_Real}) and (\ref{lambda2_Imag}) vs. $\Delta/\gamma$ for $\bar{n}=10^{-3}$ and $p=1$.}
\label{fig: Lambda2_vs_Delta_over_gamma}
\end{figure}

Figure~\ref{fig: Lambda2_vs_Delta_over_gamma} displays the absolute values of the real and imaginary parts of ($\lambda_{2}$) as a function of $\Delta/\gamma$ for $p = 1$. We observe from Fig.~\ref{fig: Lambda2_vs_Delta_over_gamma}(a) that the real part of $\lambda_{2}$ increases by a factor of two when going from the region of large to small ground state splitting. Furthermore, the real part of $\lambda_{2}$ attains asymptotic limits for large and small $\Delta/\gamma$. In contrast, the imaginary part of the eigenvalue $\lambda_{2}$ is a constant, equal to the ground state splitting ($\Delta$), for a wide range of $\Delta/\gamma$ ($10^{-2} \leq \Delta/\gamma \leq 10^{2}$) as shown in Fig.~\ref{fig: Lambda2_vs_Delta_over_gamma}(b). 

To evaluate the real part of $\lambda_{2}$ at the limiting values of $\Delta/\gamma$, we first consider the case of large $\Delta/\gamma \gg 1$. We begin by evaluating the various terms that define $\lambda_{2}$  
\begin{align}
    c_{0} &= 36 (\frac{\Delta}{\gamma})^{2}, \enspace
    d_{0} = 108 (\frac{\Delta}{\gamma})^{6}, \enspace 
    \tilde{K} = \sqrt{3} (\frac{\Delta}{\gamma}), \enspace 
    b_{0} = 3 (\frac{\Delta}{\gamma})^{2}, \enspace 
    b_{1} = -(8+6p^{2}), \enspace  
    t_{1} = \frac{2\sqrt{3}}{(\frac{\Delta}{\gamma})} \label{Coeffs_Large_Delta}
\end{align}

Using Eq.~(\ref{Coeffs_Large_Delta}) in Eq.~(\ref{lambda2_Real}), we find
\begin{align}
    \text{Re}(\lambda_{2}) &= r\bigg[-\frac{5}{3}-\frac{1}{2} \bigg\lbrace \frac{-(8+6p^{2})}{3\tilde{K}} -\frac{3(\frac{\Delta}{\gamma})^{2}t_{1}}{9\tilde{K}}\bigg\rbrace +\frac{1}{18} \sqrt{3}\Big(\frac{\Delta}{\gamma}\Big) t_{1} \bigg] \nonumber \\
    &= r\bigg[-\frac{5}{3}+\frac{1}{6\tilde{K}} \bigg\lbrace (8+6p^{2})+\Big(\frac{\Delta}{\gamma}\Big)^{2} \frac{2\sqrt{3}}{(\frac{\Delta}{\gamma})}\bigg\rbrace +\frac{1}{18} \sqrt{3}\Big(\frac{\Delta}{\gamma}\Big) \frac{2\sqrt{3}}{(\frac{\Delta}{\gamma})}\bigg]  \nonumber \\
    &= r\bigg[-\frac{5}{3}+\frac{1}{6 \sqrt{3} (\frac{\Delta}{\gamma})} \bigg\lbrace (8+6p^{2})+2\sqrt{3}\Big(\frac{\Delta}{\gamma}\Big)\bigg\rbrace +\frac{1}{3}\bigg]   \nonumber \\
    &= r\bigg[-\frac{4}{3}+\frac{1}{6 \sqrt{3} (\frac{\Delta}{\gamma})}  (8+6p^{2})+ \frac{1}{3}\bigg] 
    = r\bigg[-1+\frac{1}{6 \sqrt{3} (\frac{\Delta}{\gamma})}  (8+6p^{2})\bigg] = -r \label{Re_lambda2_large_Delta}
\end{align}
where we have neglected the last term $\frac{1}{6 \sqrt{3} (\frac{\Delta}{\gamma})}  (8+6p^{2})$ for large $\Delta/\gamma$. From Eq.~(\ref{Re_lambda2_large_Delta}), we observe that real part of the eigenvalue $\lambda_{2}$ is independent of the dipole alignment factor ($p$), and takes the asymptotic value of $r$ as seen in Fig.~\ref{fig: Lambda2_vs_Delta_over_gamma}(a).    

Next we calculate the real part of eigenvalue $\lambda_{2}$ when $\Delta/\gamma \ll 1$. We proceed by finding the terms relevant for $\lambda_{2}$ as in the previous case 
\begin{align}
    c_{0} &= 16, \enspace
    d_{0} = 0, \enspace 
    \tilde{K} = 2, \enspace 
    b_{0} = -4, \enspace 
    b_{1} = -(8+6p^{2}), \enspace  
    t_{1} = \frac{(12+9p^{2})}{4} \label{Coeffs_Small_Delta}
\end{align}

Using Eq.~(\ref{Coeffs_Small_Delta}) in Eq.~(\ref{lambda2_Real}), we find
\begin{align}
    \text{Re}(\lambda_{2}) &= r\bigg[-\frac{5}{3}-\frac{1}{2} \bigg\lbrace \frac{-(8+6p^{2})}{3\tilde{K}} -\frac{-4 \frac{(12+9p^{2})}{4}}{9\tilde{K}}\bigg\rbrace +\frac{1}{18}\tilde{K} \frac{(12+9p^{2})}{4} \bigg] \nonumber \\
    &= r\bigg[-\frac{5}{3}+\frac{1}{12} (8+6p^{2})+\frac{1}{36}(12+9p^{2}) -\frac{1}{36}(12+9p^{2})\bigg]  \nonumber \\
    &= r\bigg[-\frac{5}{3}+\frac{1}{12} (8+6p^{2})\bigg]  
    = r\bigg[\frac{(p^{2}-2)}{2}\bigg] \label{Re_lambda2_small_Delta}
\end{align}

The above expression shows that $\text{Re}(\lambda_{2})$ depends on the dipole alignment factor $p$. In the absence of Fano coherence ($p = 0$), $\text{Re}(\lambda_{2}) = - r$ and with full Fano coherence ($p = 1$), $\text{Re}(\lambda_{2}) = - r/2$ as demonstrated in Fig.~\ref{fig: Lambda2_vs_Delta_over_gamma}(a). This result shows that the coherence time, which is the inverse of the absolute value of the real part of the eigenvalue $\lambda_{2}$, is enhanced two-fold in the presence of Fano coherence when the ground state splitting is small ($\Delta \leq \gamma$). 

\subsubsection{Population and Coherence dynamics {in the underdamped regime [$\Delta/r > p^{2}/2$]}}

Here, we derive the analytical solutions of the BR equations for a weakly driven $\Lambda$-system. For $r \ll \gamma$ and $r \ll \Delta$, the excited-state population and coherences [see, e.g, Eq.~(\ref{GeneralSolnUD2})] can be further simplified as
\begin{align}
    \rho_{g_{1}g_{1}}(t) &= \frac{1}{2} \text{e}^{-2\gamma t} + \bigg(\frac{r+\gamma}{3r+2\gamma}\bigg)(1-\text{e}^{-2\gamma t}) = \frac{r+\gamma}{(3r+2\gamma)} + \frac{r}{2(3r+2\gamma)}\text{e}^{-2\gamma t} \label{BR_Pop_WPL2} \\
    \rho_{g_{1}g_{2}}^{R}(t) &= \frac{pr}{2[4\gamma^{2}+\Delta^{2}]} \bigg[2\gamma \text{e}^{-2\gamma t} - \big\lbrace 2\gamma \cos\Delta t + \Delta \sin\Delta t \big\rbrace \text{e}^{-rt} \bigg] \label{Real_Coh_UD_WPL2} \\
    \rho_{g_{1}g_{2}}^{I}(t) &= \frac{pr}{2[4\gamma^{2}+\Delta^{2}]} \bigg[\Delta \text{e}^{-2\gamma t} + \big\lbrace2\gamma \sin\Delta t - \Delta  \cos\Delta t \big\rbrace \text{e}^{-rt}\bigg] \label{Imag_Coh_UD_WPL2}
\end{align}

\begin{figure}[ht]
\centering
\includegraphics[width=\linewidth]{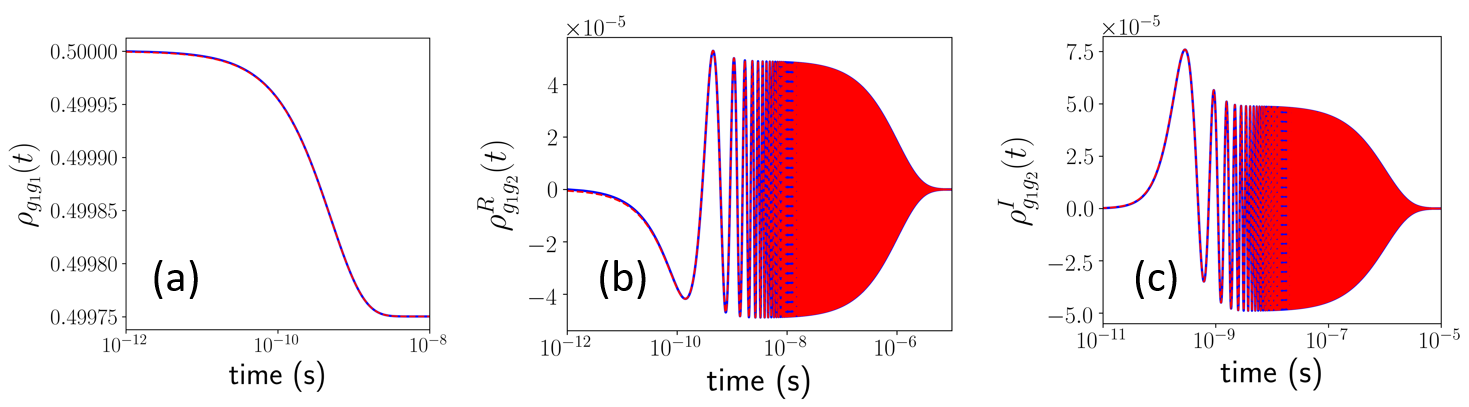}
\caption{Numerical [blue solid line] vs analytical [red dashed line] solutions of BR equations (\ref{BR_Eqns01_Lambda}) - (\ref{BR_Eqns02_Lambda}) in the underdamped regime. (a) Excited-state population [(a)] and coherence [(b), (c)] dynamics of the symmetric $\Lambda$-system irradiated by incoherent light in the weak pumping limit with $\bar{n}=10^{-3}$ and ${\Delta}/{\gamma}=10$.}
\label{fig: Lambda_System_Dynamics_UD_WPL01}
\end{figure}

\begin{figure}[ht]
\centering
\includegraphics[width=\linewidth]{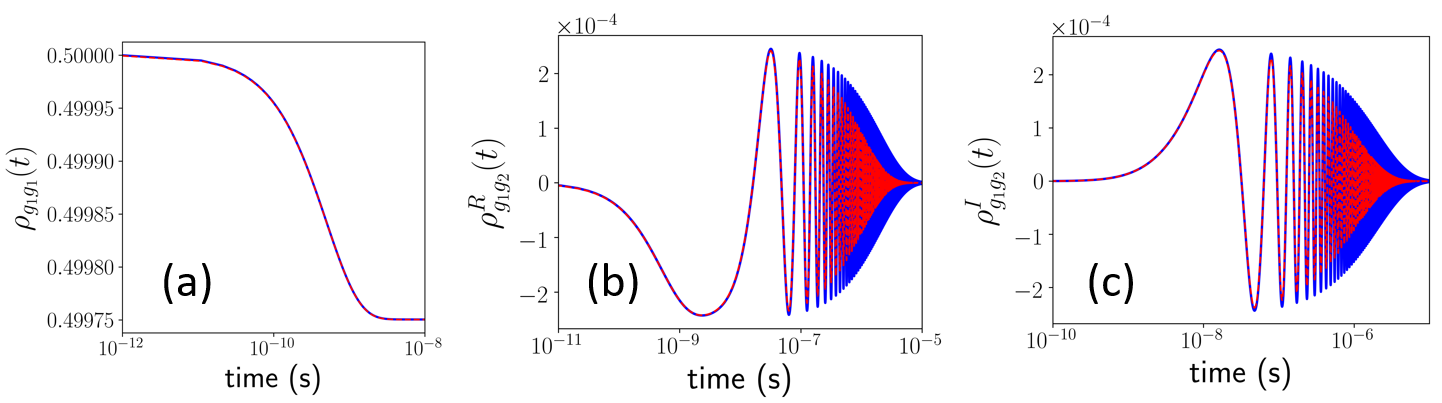}
\caption{Numerical [blue solid line] vs analytical [red dashed line] solutions of BR equations (\ref{BR_Eqns01_Lambda}) - (\ref{BR_Eqns02_Lambda}) in the underdamped regime. (a) Excited-state population [(a)] and coherence [(b), (c)] dynamics of the symmetric $\Lambda$-system irradiated by incoherent light in the weak pumping limit with $\bar{n}=10^{-3}$ and ${\Delta}/{\gamma}=10^{-1}$.}
\label{fig: Lambda_System_Dynamics_UD_WPL02}
\end{figure}

%The equations (\ref{BR_Pop_WPL}) - (\ref{Imag_Coh_UD_WPL}) present the analytic solutions for the coherence dynamics of the $\Lambda$-system weakly driven by incoherent radiation in the underdamped regime. 
Equations~(\ref{BR_Pop_WPL2}) - (\ref{Imag_Coh_UD_WPL2}) give the exact population and coherence dynamics of the weakly driven $\Lambda$-system when the ground state splitting is large in comparison to the spontaneous decay rate ($\Delta > \gamma$). Figure~\ref{fig: Lambda_System_Dynamics_UD_WPL01} shows a nearly perfect agreement of the analytical and numerical solutions for $\Delta > \gamma$. From the above equations, we note that there are two separate time scales $1/2\gamma$ and $1/r$ for the evolution of coherence. Figure~\ref{fig: Lambda_System_Dynamics_UD_WPL01}(b) - (c) exhibits the emergence of coherence from the coherence-free initial state which oscillates with a frequency of $\Delta$ that decays on the timescale $1/r$. 
%This implies that the coherence manifests long-lived behavior in the weak pumping limit.

However, the analytical and numerical solutions for $\Delta/\gamma \leq 1$ displayed in Fig.~\ref{fig: Lambda_System_Dynamics_UD_WPL02}   disagree at later times. This discrepancy can be understood from the behavior of $\lambda_{2,3}$ as a function of $\Delta/\gamma$ discussed in the previous section. While $\text{Im}(\lambda_{2,3}) = \Delta$ for a wide range of $\Delta/\gamma$, $\text{Re}(\lambda_{2,3})$ vary from $r$ to $r/2$ as $\Delta/\gamma$ decreases. This implies that the coherences decay on the time scale $2/r$ when $\Delta/\gamma < 1$. In this limit, the analytical solutions for coherence dynamics takes the form
\begin{align}
    \rho_{g_{1}g_{2}}^{R}(t) &= \frac{pr}{2[4\gamma^{2}+\Delta^{2}]} \bigg[2\gamma \text{e}^{-2\gamma t} - \big\lbrace 2\gamma \cos\Delta t + \Delta \sin\Delta t \big\rbrace {e}^{-\frac{r}{2}t} \bigg] \label{Real_Coh_UD_WPL3} \\
    \rho_{g_{1}g_{2}}^{I}(t) &= \frac{pr}{2[4\gamma^{2}+\Delta^{2}]} \bigg[\Delta \text{e}^{-2\gamma t} + \big\lbrace2\gamma \sin\Delta t - \Delta  \cos\Delta t \big\rbrace \text{e}^{-\frac{r}{2}t}\bigg] \label{Imag_Coh_UD_WPL3}
\end{align}

To interpolate between the two limits, we introduce a dimensionless function $Q(x)=Re(\lambda_2)/r$ plotted in  Fig.~\ref{fig: Lambda2_vs_Delta_over_gamma}(a), in terms of which
\begin{align}
    \rho_{g_{1}g_{2}}^{R}(t) &= \frac{pr}{2[4\gamma^{2}+\Delta^{2}]} \bigg[2\gamma \text{e}^{-2\gamma t} - \big\lbrace 2\gamma \cos\Delta t + \Delta \sin\Delta t \big\rbrace \text{e}^{-{r}{Q(\frac{\Delta}{\gamma}})t} \bigg] \label{Real_Coh_UD_Qfun} \\
    \rho_{g_{1}g_{2}}^{I}(t) &= \frac{pr}{2[4\gamma^{2}+\Delta^{2}]} \bigg[\Delta \text{e}^{-2\gamma t} + \big\lbrace2\gamma \sin\Delta t - \Delta  \cos\Delta t \big\rbrace \text{e}^{-{r}{Q(\frac{\Delta}{\gamma}})t} \bigg] \label{Imag_Coh_UD_Qfun}.
\end{align}

\color{blue}
We finally consider a weakly driven $\Lambda$-system with the ground states initially in an equal  coherent superposition. For $r \ll \gamma$ and $r \ll \Delta$, the coherent dynamics governed by Eq.~(\ref{GeneralSolnFinal2}) can be recast as    
\begin{align}
\begin{split}
    \rho_{g_{1}g_{2}}^{R}(t) &= \big[\rho^{R}_{g_{1} g_{2}}(0) \cos\Delta t + \rho^{I}_{g_{1} g_{2}}(0)  \sin\Delta t)\big] \text{e}^{-rt} + \frac{pr}{2[4\gamma^{2}+\Delta^{2}]} \bigg[2\gamma \text{e}^{-2\gamma t} - \Big\lbrace 2\gamma \cos \Delta t + \Delta \sin \Delta t \Big\rbrace \text{e}^{-rt} \bigg] 
\end{split} \nonumber \\
\begin{split}
    \rho_{g_{1}g_{2}}^{I}(t) &= \big[-\rho^{R}_{g_{1} g_{2}}(0) \sin\Delta t + \rho^{I}_{g_{1} g_{2}}(0) \cos\Delta t\big] \text{e}^{-rt} + \frac{pr}{2[4\gamma^{2}+\Delta^{2}]} \bigg[ \Delta \text{e}^{-2\gamma t} + \Big\lbrace 2\gamma) \sin \Delta t - \Delta  \cos \Delta t \Big\rbrace \text{e}^{-rt}\bigg] \label{GeneralSoln_WPL}
\end{split}
\end{align}

Equations (\ref{GeneralSoln_WPL}) accurately describe the coherent dynamics of a weakly driven $\Lambda$-system if the ground state splitting is greater than the spontaneous decay rate  ($\Delta > \gamma$). However, in the opposite limit ($\Delta \leq \gamma$) these analytical solutions  deviate from numerical solutions at later times. Following the same procedure as described above, we introduce a dimensionless function $Q(x)=Re(\lambda_2)/r$ that interpolates between the two limits of the eigenvalue $\lambda_{2}$ to find the general solutions in the form
\begin{align}
\begin{split}
    \rho_{g_{1}g_{2}}^{R}(t) &= \big[\rho^{R}_{g_{1} g_{2}}(0) \cos\Delta t + \rho^{I}_{g_{1} g_{2}}(0)  \sin\Delta t)\big] \text{e}^{-r Q(\frac{\Delta}{\gamma})t} + \frac{pr}{2[4\gamma^{2}+\Delta^{2}]} \bigg[2\gamma \text{e}^{-2\gamma t} - \Big\lbrace 2\gamma \cos \Delta t + \Delta \sin \Delta t \Big\rbrace \text{e}^{-r Q(\frac{\Delta}{\gamma}) t} \bigg] 
\end{split} \nonumber \\
\begin{split}
    \rho_{g_{1}g_{2}}^{I}(t) &= \big[-\rho^{R}_{g_{1} g_{2}}(0) \sin\Delta t + \rho^{I}_{g_{1} g_{2}}(0) \cos\Delta t\big] \text{e}^{-r Q(\frac{\Delta}{\gamma})t} + \frac{pr}{2[4\gamma^{2}+\Delta^{2}]} \bigg[ \Delta \text{e}^{-2\gamma t} + \Big\lbrace 2\gamma \sin \Delta t - \Delta  \cos \Delta t \Big\rbrace \text{e}^{-r Q(\frac{\Delta}{\gamma})t}\bigg] \label{GeneralSoln_WPL2}
\end{split}
\end{align}

This completes the derivation of Eqs.~(4) of the main text.
\color{black}

\subsection{von Neumann Entropy}

Here, we describe the procedure of computing the von Neumann entropy of the quantum states generated by incoherent driving of  the $\Lambda$-system. The von Neumann entropy of a quantum system described by the density operator $\rho$ is defined as
\begin{equation} \label{von_Neumann_Entropy}
    \mathcal{S}(t) = - \text{Tr}(\rho \text{ln}\rho)
\end{equation}
where $\text{Tr}$ is the trace operation.

By diagonalizing the density operator we can express it in the following form
\begin{equation} \label{rho_energy_eigenbasis}
    \rho = \sum_{i} \lambda_{i} \ket{i}\bra{i}
\end{equation}
where $\ket{i}$ is the eigenvector of $\rho$ with the eigenvalue $\lambda_{i}$. The von-Neumann entropy in the eigenbasis of  $\rho$ then becomes
% (\textcolor{red}{Suyesh, please clarify: It is the basis of eigenvectors of $\rho$.})
\begin{equation} \label{von_Neumann_Entropy_eigenbasis}
    \mathcal{S}(t) = - \sum_{i} \lambda_{i} \text{ln}{\lambda_{i}}
\end{equation}

The density operator of the $\Lambda$-system  driven by incoherent light in the energy eigenbasis is
\begin{equation} \label{density_op_Lambda}
    \rho = \begin{pmatrix}
         \rho_{ee} & 0 & 0 \\
         0 & \rho_{g_{1}g_{1}} & \rho_{g_{1}g_{2}} \\
         0 & \rho_{g_{2}g_{1}} & \rho_{g_{2}g_{2}}
    \end{pmatrix}
\end{equation}

\begin{figure}[t!]
\centering
\includegraphics[width=0.8\linewidth]{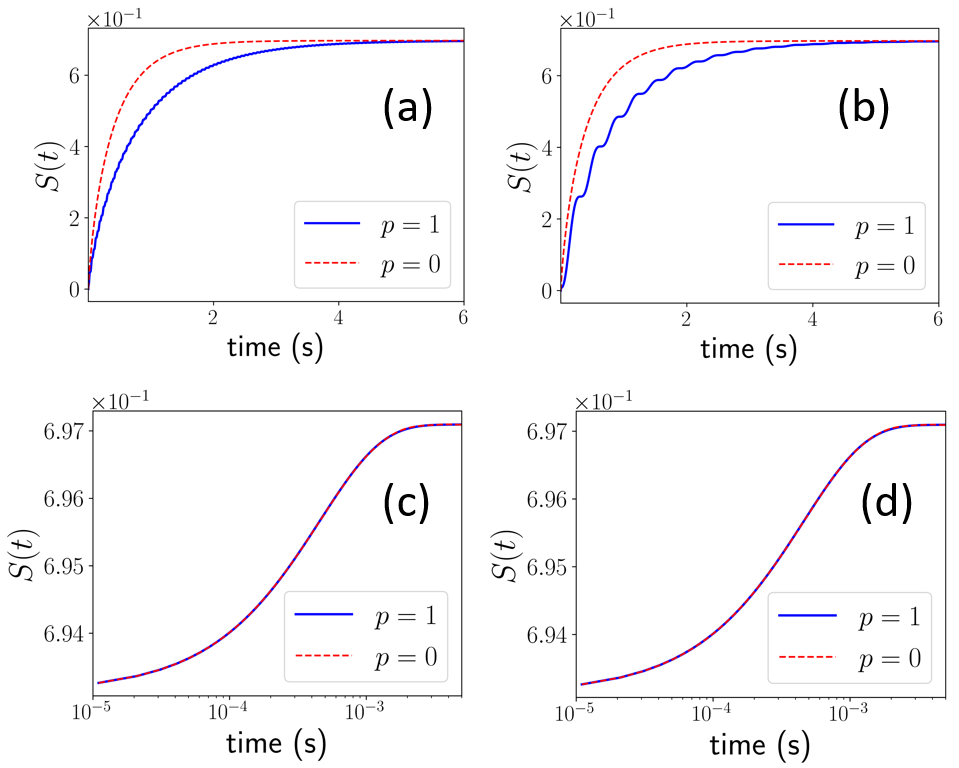}
\caption{Comparison of von Neumann Entropy defined in Eq.~(\ref{von_Neuman_Entropy_eigenbasis2}) with full Fano coherence [blue solid line] vs without Fano coherence [red dashed line] as a function of time in the weak pumping limit with $\bar{n} = 10^{-3}$. (a) Panels [(a)] and [(b)] display the evolution of the von Neumann entropy with $\Delta/\gamma = 10^{-1}$ and $10^{-2}$, and the ground states of the $\Lambda$-system initially in a coherent superposition. (b) Panel [(c)] von Neumann entropy is evaluated for $\Delta/\gamma = 10^{-1}$, and no initial coherence between the ground states. (c) Panel [(d)] same as panels[(b)]-[(c)] for $\Delta/\gamma = 10$. Note that time is given in the units of $r^{-1}$. }
\label{fig:von_Neumann_Entropy_WPL}
\end{figure}
The eigenvalues of $\rho$  are
\begin{align} \label{lambda_123}
    \lambda_{1} &= \rho_{ee} \nonumber \\
    \lambda_{2,3} &= \frac{1}{2} \Big[\big(\rho_{g_{1}g_{1}}+\rho_{g_{2}g_{2}}\big) \pm \sqrt{\big(\rho_{g_{1}g_{1}}-\rho_{g_{2}g_{2}}\big)^{2}+4\rho_{g_{1}g_{2}}\rho_{g_{2}g_{1}}}\Big] 
\end{align}
For the symmetric $\Lambda$-system, these  expressions simplify to
\begin{align} \label{lambda_123_Sym}
    \lambda_{1} &= \rho_{ee} \nonumber \\
    \lambda_{2,3} &= \rho_{g_{1}g_{1}} \pm |\rho_{g_{1}g_{2}}| 
\end{align}

Plugging Eqs.~(\ref{lambda_123_Sym}) into Eq.~(\ref{von_Neumann_Entropy_eigenbasis}), we obtain the von-Neumann entropy of the symmetric $\Lambda$-system 
\begin{equation} \label{von_Neuman_Entropy_eigenbasis2}
    \mathcal{S}(t) = -\Big[\big(\rho_{g_{1}g_{1}} + |\rho_{g_{1}g_{2}}|\big) \text{ln}\big(\rho_{g_{1}g_{1}} + |\rho_{g_{1}g_{2}}|\big) + \big(\rho_{g_{1}g_{1}} - |\rho_{g_{1}g_{2}}|\big) \text{ln}\big(\rho_{g_{1}g_{1}} - |\rho_{g_{1}g_{2}}|\big) + \rho_{ee} \text{ln}\rho_{ee}  \Big]
\end{equation}

\begin{figure}[ht]
\centering
\includegraphics[width=0.8\linewidth]{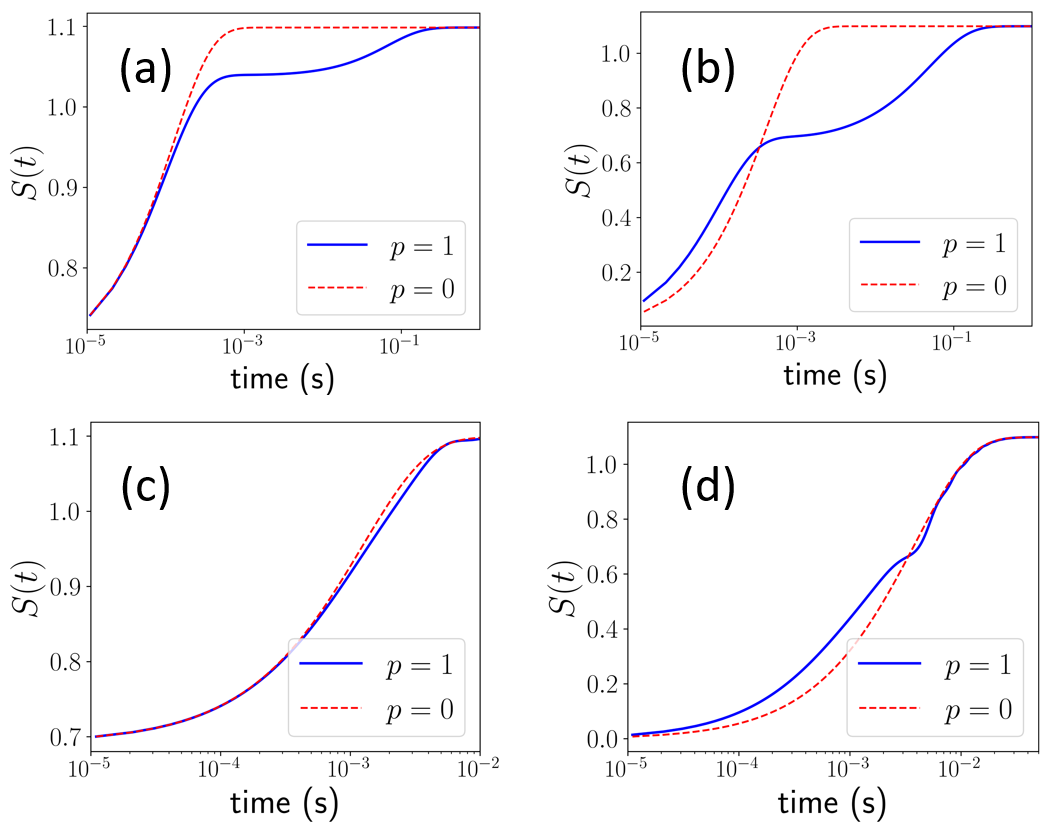}
\caption{von Neumann Entropy defined in Eq.~(\ref{von_Neuman_Entropy_eigenbasis2}) with full Fano coherence [blue solid line] vs without Fano coherence [red dashed line] as a function of time in the strong pumping limit. (a) Panels [(a)] and [(b)] display the evolution of von Neumann entropy for $\bar{n} = 10^{3}$, $\Delta/\gamma = 10^{2}$ with no initial coherence between the ground states, and the ground states initially in an equal coherent superposition. (b) Panels [(c)] and [(d)] show the evolution of von Neumann entropy for $\bar{n}=10^{2}$,  $\Delta/\gamma = 10^{3}$ with no initial coherence between the ground states, and the ground states initially in a coherent superposition. Note that time is given in the units of $\gamma^{-1}$. }
\label{fig:von_Neumann_Entropy_SPL}
\end{figure}

Substituting the time-dependent density matrix elements calculated above, we obtain the time evolution of von Neumann entropy in the weak and strong pumping limits plotted in Figs.~\ref{fig:von_Neumann_Entropy_WPL} and \ref{fig:von_Neumann_Entropy_SPL}, respectively.

\section{$\Lambda$-system driven by polarized incoherent light: The Bloch-Redfield equations and the steady-state solution}

\subsection{Bloch-Redfield equations for a $\Lambda$-system driven by polarized incoherent radiation}
%In Section 2.3, we explored the statistical properties of thermal radiation. The  radiation from a thermal source is isotropic and incoherent which means that the the fluctuation of the electric field of such light is random in all directions with no definite phase. Here, we will consider the interaction of matter with polarized incoherent radiation, where the electric field fluctuates randomly within a specific plane of polarization. We then present the derivation of the quantum master equation for a three-level $V$-system irradiated by polarized incoherent radiation. \par 

%We propose the metastable He (2$^3S_1$) as the physical $\Lambda$-system which is ideal due to the absence of the hyperfine structure. The electric dipole transitions between the excited state (electronic configuration: ${}^{3}P_{0}$) and the ground electronic states (electronic configuration: ${}^{3}S_{1}; \enspace m_{j}=\pm1$) forms a three-level $\Lambda$-system. \par 

Consider a beam of He$^*$ atoms irradiated by incoherent light, propagating along the ${{z}}$-direction, and linearly polarized in the ${{x}}$-direction. A uniform magnetic field  applied in the ${z}$-direction produces a tunable Zeeman shift between the ground states $\Delta = 2\mu_{B} B_{z}$, where $\mu_{B}$ is the Bohr magneton and $B_{z}$ is the magnitude of the applied magnetic field. Due to the polarized nature of the radiation field, the average number of photons depends on the direction of propagation $\hat{\bm{k}} = \textbf{k}/k$ \par 
\begin{equation}
    \langle n_{\textbf{k},\lambda} \rangle = \delta_{\hat{\textbf{k}},\hat{\textbf{z}}} \enspace \delta_{\bm{\varepsilon}_{\textbf{k}\lambda},\hat{\textbf{x}}} \enspace \bar{n}_{k},
\end{equation}
where $\bar{n}_{k}$ is the average photon number of the radiation field, $\textbf{k}$ is the wave-vector, $\lambda$ is the polarization mode index and $\bm{\varepsilon}_{\textbf{k}\lambda}$ represents the polarization vector for the field mode $(\textbf{k}, \lambda)$. \par 

The coupling between the polarized incoherent radiation field and the $\Lambda$ system is given by \cite{Dodin:18,Koyu:21}
\begin{equation}
   C =\sum_{ij} \sum_{kl} \sum_{\textbf{k}\lambda} g^{(i,j)}_{\textbf{k}\lambda} g^{(k,l)}_{\textbf{k}\lambda} \langle \hat{a}^{\dagger}_{\textbf{k}\lambda} \hat{a}_{\textbf{k}\lambda} \rangle=\bigg( \frac{\hbar\omega_{\textbf{k}}}{2\epsilon_{0}V} \bigg) \sum_{ij} \sum_{kl} \sum_{\textbf{k}\lambda} (\bm{\mu}_{ij}\cdot \bm{\varepsilon}_{\textbf{k}\lambda}) (\bm{\mu}_{kl} \cdot \bm{\varepsilon}_{\textbf{k}\lambda}) \langle n_{\textbf{k}\lambda} \rangle
\end{equation} \par 

The He$^*$ $\Lambda$-system under consideration consists of two quasi-degenerate ground states $\ket{g_{m}}$ and $\ket{g_{m'}}$ with a common excited state $\ket{e}$. We identify $i,k \rightarrow \ket{e}$, $j \rightarrow \ket{g_{m}}$ and $l \rightarrow \ket{g_{m'}}$ in the above equation to find the coupling term 
\begin{equation} \label{Coupling_Constant_Pol}
   C =\bigg( \frac{\hbar\omega_{\textbf{k}}}{2\epsilon_{0}V} \bigg) \sum_{\textbf{k}} (\bm{\mu}_{e,g_{m}}\cdot \bm{\varepsilon}_{\textbf{k}\lambda}) (\bm{\mu}_{e,g_{m'}} \cdot \bm{\varepsilon}_{\textbf{k}\lambda}) \langle n_{\textbf{k}\lambda} \rangle .
\end{equation} \par

While spontaneous emission from incoherently driven He$^*$ atoms occurs in all possible directions, stimulated emission is constrained to the direction of the beam of polarized light ($\hat{\bm{x}}$). 
%The coupling terms (\ref{Coupling_Constant_Pol}) when evaluated for the spontaneous and stimulated emissions display their contributions in the radiative process. 
% \textcolor{magenta}{[Suyesh, should we call this the {\it cross-coupling} term to distinguish it from the diagonal spontaneous emission term?] } 
Thus, the cross-coupling term in Eq.~(\ref{Coupling_Constant_Pol}) responsible for spontaneous emission is independent of whether the $\Lambda$-system is driven by polarized or isotropic incoherent light. This  term depends on the transition dipole alignment factor $p = \hat{\bm{\mu}}_{e,g_{m}} \cdot \hat{\bm{\mu}}_{e,g_{m'}}$ \cite{Koyu:21}, which can be  evaluated for the metastable He$^*$ $\Lambda$-system as $\hat{\bm{\mu}}_{e,g_{1}}= \langle ^{3}P_0|\bm{\mu}| ^{3}S_{m=1}\rangle = \frac{1}{\sqrt{2}} (\hat{x}+\hat{y})$ and $\hat{\bm{\mu}}_{e,g_{1}}= \langle ^{3}P_0|\bm{\mu}| ^{3}S_{m=-1}\rangle = \frac{1}{\sqrt{2}} (\hat{x}-\hat{y})$, where $\hat{x}$ and $\hat{y}$ are the unit vectors along the $X$ and $Y$ axes. Therefore, $\bm{\mu}_{e,g_{1}} \perp \bm{\mu}_{e,g_{-1}}$ and hence $p=0$ which means the cross-coupling term vanishes for spontaneous emission. 

To calculate the cross-coupling term due to stimulated emission, we substitute  $\bm{\varepsilon}_{\bm{k}\lambda} = \hat{\bm{x}}$ in Eq.~(\ref{Coupling_Constant_Pol}) to find
\begin{equation}
   C =\bigg( \frac{\hbar\omega_{\textbf{k}}}{2\epsilon_{0}V} \bigg) \sum_{\textbf{k}} \mu_{e,g_{m}} \mu_{e,g_{m'}} (\hat{\bm{\mu}}_{e,g_{m}}\cdot \hat{\bm{x}}) (\hat{\bm{\mu}}_{e,g_{m'}} \cdot \hat{\bm{x}})  \bar{n}_{k} 
\end{equation} 
 
Simplifying the scalar products  as  $\hat{\bm{\mu}}_{g,e_{\pm 1}}\cdot \hat{\bm{x}}  = \frac{1}{\sqrt{2}} (\hat{x} \pm \hat{y}) \cdot \hat{\bm{x}} = \frac{1}{\sqrt{2}}$, we compute the coupling term for the stimulated emission process as
\begin{equation}
   C =\bigg( \frac{\hbar\omega_{\textbf{k}}}{2\epsilon_{0}V} \bigg) \sum_{\textbf{k}} \frac{\mu_{e,g_{m}} \mu_{e,g_{m'}}}{2} \bar{n}_{k} 
\end{equation}

The coupling term is thus non-zero. The incoherent pumping rates for $m = m'$ are evaluated as
% by $x$-polarized incoherent radiation are attenuated by a factor of $16\pi/3$ in comparison to the isotropic case. The incoherent pumping rates when $m = m'$ are evaluated as
\begin{equation} \label{pumping_rates_attenuated}
    r_{\pm} = \frac{3}{16\pi}\gamma_{\pm} \bar{n}_{k} = \frac{3 }{16\pi}\gamma \bar{n}_{k},
\end{equation}
where $\gamma_{+} = \gamma_{-} = \gamma$ and $r_{+} = r_{-} = r$ for the symmetric $\Lambda$-system. Note that the polarized pumping are attenuated by the factor $16\pi/3$ compared to their isotropic counterparts due to the directed nature of the polarized beam, as in the case of the V-system explored before \cite{Koyu:21}.  \par  

Following a procedure similar to that described in the Appendix of Ref.~\cite{Koyu:21}, we obtain the BR equations for the $\Lambda$-system driven by polarized incoherent radiation
% by identifying the different coupling terms 
%on the Bloch-Redfield equations for the isotropic radiation case
% The incoherent pumping rate is modified as the radiation is available depending on the direction of the radiation. Thus the dipole transition alignment factor 'p' for the isotropic incoherent radiation case are different for isotropic spontaneous emission process and directional incoherent pumping process and behaves asymmetrically.
\begin{align}
\dot\rho_{g_{i}g_{i}} &= -r_{i} \rho_{g_{i}g_{i}} + (r_{i}+\gamma_{i}) \rho_{ee} - \sqrt{r_{1} r_{2}} \rho^R_{g_{1}g_{2}} \label{BR_Eqs_Pol1} \\
\begin{split}
\dot\rho_{g_{1}g_{2}} &= -\frac{1}{2}(r_{1}+r_{2}) \rho_{g_{1}g_{2}} - i\rho_{g_{1}g_{2}} \Delta + \sqrt{r_{1} r_{2}} \rho_{ee}-\frac{1}{2}\sqrt{r_{1} r_{2}}(\rho_{g_{1}g_{1}}+\rho_{g_{2}g_{2}})   \label{BR_Eqs_Pol2}
\end{split} 
\end{align}
where $\gamma_{i}$ is the spontaneous decay rate, which is the same as in the isotropic driving case, and the pumping rates $r_{i} = r$ are given by Eq.~(\ref{pumping_rates_attenuated}).
\color{blue}
We note that these equations are distinct from Eq. (1) of the main text. The latter is the Bloch-Redfield equation for the $\Lambda$-system driven by {\it isotropic} incoherent radiation, which  contains the interference term $p[\sqrt{r_1r_2} + \sqrt{\gamma_1\gamma_2}]\rho_{ee}$ that couples the ground-state coherence to the excited state population. 

In contrast, Eqs.~\eqref{BR_Eqs_Pol1} and \eqref{BR_Eqs_Pol2} are the BR equations for the metastable He$^*$  $\Lambda$-system driven by polarized incoherent radiation, as derived above. The transition dipole vectors between the ground states of the metastable He$^*$  $\Lambda$-system and the common excited state $|e\rangle$ are orthogonal, i.e, $p=0$. Thus, there is no cross-coupling in spontaneous emission, which implies that the $p\sqrt{\gamma_1\gamma_2}\rho_{ee}$  term in  the original BR equations \eqref{BR_Eqns02_Lambda}  vanishes. Note that the cross terms due to incoherent driving do not vanish because the orthogonal transition dipoles interact with the same x-polarized mode of the incoherent radiation, as discussed previously for the V-system \cite{Koyu:21}.
\color{black}

%Comparing Eqs.~(\ref{BR_Eqs_Pol2}) to the BR equations for isotropic excitation (see Sec. I), we note that $p = 0$ for spontaneous emission process and $p = 1$ for the stimulated emission process. Moreover we highlight the fact that the transition dipole alignment factor (p) is not defined for the stimulated emission process.  \textcolor{magenta}{[Suyesh, please check the above two sentences. They seem to contradict each other.}

\subsection{Steady State Solution}
We consider a symmetric $\Lambda$-system with $\gamma_{1}=\gamma_{2}=\gamma$ and $r_{1}=r_{2}=r$ for which the BR equations (\ref{BR_Eqs_Pol1}) - (\ref{BR_Eqs_Pol2}) may be written as
\begin{align}
    \Dot{\rho}_{g_{1}g_{1}} &= -(3r+2\gamma) \rho_{g_{1}g_{1}} - r \rho^{R}_{g_{1}g_{2}} + (r+\gamma)  \nonumber \\
    \Dot{\rho}^{R}_{g_{1}g_{2}} &= -3r \rho_{g_{1}g_{1}} - r \rho^{R}_{g_{1}g_{2}} + \Delta \rho^{I}_{g_{1}g_{2}}+r \nonumber \\
    \Dot{\rho}^{I}_{g_{1}g_{2}} &= -\Delta \rho^{R}_{g_{1}g_{2}} - r \rho^{I}_{g_{1}g_{2}} \label{QME2} 
\end{align} 
where we used $\rho_{ee}= 1-\rho_{g_{1}g_{1}}-\rho_{g_{2}g_{2}}$ to eliminate $\rho_{ee}$ from Eq.~(\ref{BR_Eqs_Pol2}). In matrix-vector form, the BR equations are expressed as
\begin{equation} \label{state_vector_DE}
    \Dot{\mathbf{x}}(t) = \mathbf{A}\mathbf{x}(t) + \mathbf{d},
\end{equation} 
where, as above, $\mathbf{x}(t)=[\rho_{g_{1}g_{1}}(t),\rho^{R}_{g_{1}g_{2}}(t),\rho^{I}_{g_{1}g_{2}}(t)]^{T}$ is the state vector in the Liouville representation, $\mathbf{d}=[r+\gamma,r,0]^{T}$ is the driving vector, and the matrix $\mathbf{A}$ in Eq. (\ref{state_vector_DE}) is 
\begin{equation}
    \mathbf{A} = 
\begin{bmatrix}
    -(3r+2\gamma) & -r & 0 \\
    -3r  & -r & \Delta \\
    0 & -\Delta  &  -r 
\end{bmatrix}
\end{equation} \par 

To explore the steady-state behavior of the $\Lambda$-system driven by polarized incoherent light, we set $\Dot{\mathbf{x}}(t) = 0$ and obtain the state vector in the steady state from Eq.~(\ref{state_vector_DE})
\begin{equation} \label{SSVec}
    \mathbf{x}_{s} = - \mathbf{A}^{-1} \mathbf{d}
\end{equation}  \par 

For the steady-state solution $\mathbf{x}_{s}$ to be unique, we must have $\text{det}(\mathbf{A}) \neq 0$ \cite{Koyu:21}. The determinant is given by $\text{det}(\mathbf{A}) = -[(3r+2\gamma)\Delta^{2} + 2r^{2} \gamma]$, which is non zero in general, and thus the steady-state vector (\ref{SSVec}) is unique. The inverse of the coefficient matrix $\mathbf{A}$ is evaluated as
\begin{equation} \label{Ainverse}
    \mathbf{A}^{-1} = \frac{1}{\text{det}(\mathbf{A})}
    \begin{bmatrix}
        \Delta^{2}+r^{2} & -r^{2} & -r\Delta \\
        -3r^{2}  & r(3r+2\gamma) & \Delta(3r+2\gamma) \\
        3r\Delta & -\Delta(3r+2\gamma) & 2r\gamma) \\
    \end{bmatrix}
\end{equation}

Finally, we substitute Eq.~(\ref{Ainverse}) and the expression for the driving vector into Eq.~(\ref{SSVec}) and compare the corresponding components to find the expressions for the ground-state population and the coherences in the steady state
\begin{align}
    \rho_{g_{1}g_{1}} &= \frac{(r+\gamma)\Delta^{2}+r^{2}\gamma}{(3r+2\gamma)\Delta^{2}+2r^{2}\gamma}, \nonumber \\
    \rho^{R}_{g_{1}g_{2}} &=- \frac{r^{2}\gamma}{(3r+2\gamma)\Delta^{2}+2r^{2}\gamma}, \nonumber \\
    \rho^{I}_{g_{1}g_{2}} &= -\frac{\Delta}{r} \rho^{R}_{g_{1}g_{2}}. \label{SS_Dyn}
\end{align} 

We note that the steady-state populations given by Eq. (\ref{SS_Dyn}) deviate from their canonical values in thermal equilibrium 
\begin{equation} \label{CanEqn_NoC}
    \rho^{(0)}_{g_{1}g_{1}} = \frac{r+\gamma}{(3r+2\gamma)}
\end{equation}
where the superscript '(0)' indicates the canonical steady state. We note that both the real and the imaginary coherences in Eq.~(\ref{SS_Dyn}) are non-zero in the steady-state, in contrast to the isotropic excitation case considered in Sec.~I. 

\bibliography{open_sys}

\begin{thebibliography}{60}%
\makeatletter
\providecommand \@ifxundefined [1]{%
 \@ifx{#1\undefined}
}%
\providecommand \@ifnum [1]{%
 \ifnum #1\expandafter \@firstoftwo
 \else \expandafter \@secondoftwo
 \fi
}%
\providecommand \@ifx [1]{%
 \ifx #1\expandafter \@firstoftwo
 \else \expandafter \@secondoftwo
 \fi
}%
\providecommand \natexlab [1]{#1}%
\providecommand \enquote  [1]{``#1''}%
\providecommand \bibnamefont  [1]{#1}%
\providecommand \bibfnamefont [1]{#1}%
\providecommand \citenamefont [1]{#1}%
\providecommand \href@noop [0]{\@secondoftwo}%
\providecommand \href [0]{\begingroup \@sanitize@url \@href}%
\providecommand \@href[1]{\@@startlink{#1}\@@href}%
\providecommand \@@href[1]{\endgroup#1\@@endlink}%
\providecommand \@sanitize@url [0]{\catcode `\\12\catcode `\$12\catcode
  `\&12\catcode `\#12\catcode `\^12\catcode `\_12\catcode `\%12\relax}%
\providecommand \@@startlink[1]{}%
\providecommand \@@endlink[0]{}%
\providecommand \url  [0]{\begingroup\@sanitize@url \@url }%
\providecommand \@url [1]{\endgroup\@href {#1}{\urlprefix }}%
\providecommand \urlprefix  [0]{URL }%
\providecommand \Eprint [0]{\href }%
\providecommand \doibase [0]{https://doi.org/}%
\providecommand \selectlanguage [0]{\@gobble}%
\providecommand \bibinfo  [0]{\@secondoftwo}%
\providecommand \bibfield  [0]{\@secondoftwo}%
\providecommand \translation [1]{[#1]}%
\providecommand \BibitemOpen [0]{}%
\providecommand \bibitemStop [0]{}%
\providecommand \bibitemNoStop [0]{.\EOS\space}%
\providecommand \EOS [0]{\spacefactor3000\relax}%
\providecommand \BibitemShut  [1]{\csname bibitem#1\endcsname}%
\let\auto@bib@innerbib\@empty
%</preamble>
\bibitem [{\citenamefont {Brumer}(2018)}]{Brumer:18}%
  \BibitemOpen
  \bibfield  {author} {\bibinfo {author} {\bibfnamefont {P.}~\bibnamefont
  {Brumer}},\ }\bibfield  {title} {\bibinfo {title} {Shedding (incoherent)
  light on quantum effects in light-induced biological processes},\ }\href
  {https://doi.org/10.1021/acs.jpclett.8b00874} {\bibfield  {journal} {\bibinfo
   {journal} {J. Phys. Chem. Lett.}\ }\textbf {\bibinfo {volume} {9}},\
  \bibinfo {pages} {2946} (\bibinfo {year} {2018})}\BibitemShut {NoStop}%
\bibitem [{\citenamefont {Cao}\ \emph {et~al.}(2020)\citenamefont {Cao},
  \citenamefont {Cogdell}, \citenamefont {Coker}, \citenamefont {Duan},
  \citenamefont {Hauer}, \citenamefont {Kleinekath{\"o}fer}, \citenamefont
  {Jansen}, \citenamefont {Man{\v c}al}, \citenamefont {Miller}, \citenamefont
  {Ogilvie}, \citenamefont {Prokhorenko}, \citenamefont {Renger}, \citenamefont
  {Tan}, \citenamefont {Tempelaar}, \citenamefont {Thorwart}, \citenamefont
  {Thyrhaug}, \citenamefont {Westenhoff},\ and\ \citenamefont
  {Zigmantas}}]{Cao:20}%
  \BibitemOpen
  \bibfield  {author} {\bibinfo {author} {\bibfnamefont {J.}~\bibnamefont
  {Cao}}, \bibinfo {author} {\bibfnamefont {R.~J.}\ \bibnamefont {Cogdell}},
  \bibinfo {author} {\bibfnamefont {D.~F.}\ \bibnamefont {Coker}}, \bibinfo
  {author} {\bibfnamefont {H.-G.}\ \bibnamefont {Duan}}, \bibinfo {author}
  {\bibfnamefont {J.}~\bibnamefont {Hauer}}, \bibinfo {author} {\bibfnamefont
  {U.}~\bibnamefont {Kleinekath{\"o}fer}}, \bibinfo {author} {\bibfnamefont
  {T.~L.~C.}\ \bibnamefont {Jansen}}, \bibinfo {author} {\bibfnamefont
  {T.}~\bibnamefont {Man{\v c}al}}, \bibinfo {author} {\bibfnamefont
  {R.~J.~D.}\ \bibnamefont {Miller}}, \bibinfo {author} {\bibfnamefont {J.~P.}\
  \bibnamefont {Ogilvie}}, \bibinfo {author} {\bibfnamefont {V.~I.}\
  \bibnamefont {Prokhorenko}}, \bibinfo {author} {\bibfnamefont
  {T.}~\bibnamefont {Renger}}, \bibinfo {author} {\bibfnamefont {H.-S.}\
  \bibnamefont {Tan}}, \bibinfo {author} {\bibfnamefont {R.}~\bibnamefont
  {Tempelaar}}, \bibinfo {author} {\bibfnamefont {M.}~\bibnamefont {Thorwart}},
  \bibinfo {author} {\bibfnamefont {E.}~\bibnamefont {Thyrhaug}}, \bibinfo
  {author} {\bibfnamefont {S.}~\bibnamefont {Westenhoff}},\ and\ \bibinfo
  {author} {\bibfnamefont {D.}~\bibnamefont {Zigmantas}},\ }\bibfield  {title}
  {\bibinfo {title} {Quantum biology revisited},\ }\href
  {https://doi.org/10.1126/sciadv.aaz4888} {\bibfield  {journal} {\bibinfo
  {journal} {Science Advances}\ }\textbf {\bibinfo {volume} {6}},\ \bibinfo
  {pages} {eaaz4888} (\bibinfo {year} {2020})}\BibitemShut {NoStop}%
\bibitem [{\citenamefont {Dodin}\ and\ \citenamefont
  {Brumer}(2021{\natexlab{a}})}]{Dodin:21}%
  \BibitemOpen
  \bibfield  {author} {\bibinfo {author} {\bibfnamefont {A.}~\bibnamefont
  {Dodin}}\ and\ \bibinfo {author} {\bibfnamefont {P.}~\bibnamefont {Brumer}},\
  }\bibfield  {title} {\bibinfo {title} {Noise-induced coherence in molecular
  processes},\ }\href {https://doi.org/10.1088/1361-6455/ac3e77} {\bibfield
  {journal} {\bibinfo  {journal} {J. Phys. B}\ }\textbf {\bibinfo {volume}
  {54}},\ \bibinfo {pages} {223001} (\bibinfo {year}
  {2021}{\natexlab{a}})}\BibitemShut {NoStop}%
\bibitem [{\citenamefont {Tscherbul}\ and\ \citenamefont
  {Brumer}(2014{\natexlab{a}})}]{Tscherbul:14}%
  \BibitemOpen
  \bibfield  {author} {\bibinfo {author} {\bibfnamefont {T.~V.}\ \bibnamefont
  {Tscherbul}}\ and\ \bibinfo {author} {\bibfnamefont {P.}~\bibnamefont
  {Brumer}},\ }\bibfield  {title} {\bibinfo {title} {Long-lived quasistationary
  coherences in a {$V$}-type system driven by incoherent light},\ }\href
  {https://doi.org/10.1103/PhysRevLett.113.113601} {\bibfield  {journal}
  {\bibinfo  {journal} {Phys. Rev. Lett.}\ }\textbf {\bibinfo {volume} {113}},\
  \bibinfo {pages} {113601} (\bibinfo {year} {2014}{\natexlab{a}})}\BibitemShut
  {NoStop}%
\bibitem [{\citenamefont {Ol{\v s}ina}\ \emph {et~al.}(2014)\citenamefont
  {Ol{\v s}ina}, \citenamefont {Dijkstra}, \citenamefont {Wang},\ and\
  \citenamefont {Cao}}]{Olsina:14}%
  \BibitemOpen
  \bibfield  {author} {\bibinfo {author} {\bibfnamefont {J.}~\bibnamefont
  {Ol{\v s}ina}}, \bibinfo {author} {\bibfnamefont {A.~G.}\ \bibnamefont
  {Dijkstra}}, \bibinfo {author} {\bibfnamefont {C.}~\bibnamefont {Wang}},\
  and\ \bibinfo {author} {\bibfnamefont {J.}~\bibnamefont {Cao}},\ }\bibfield
  {title} {\bibinfo {title} {Can natural sunlight induce coherent exciton
  dynamics?},\ }\href@noop {} {\bibfield  {journal} {\bibinfo  {journal}
  {arXiv:1408.5385}\ } (\bibinfo {year} {2014})}\BibitemShut {NoStop}%
\bibitem [{\citenamefont {Kosloff}\ and\ \citenamefont
  {Levy}(2014)}]{Kosloff:14}%
  \BibitemOpen
  \bibfield  {author} {\bibinfo {author} {\bibfnamefont {R.}~\bibnamefont
  {Kosloff}}\ and\ \bibinfo {author} {\bibfnamefont {A.}~\bibnamefont {Levy}},\
  }\bibfield  {title} {\bibinfo {title} {Quantum heat engines and
  refrigerators: Continuous devices},\ }\href
  {https://doi.org/10.1146/annurev-physchem-040513-103724} {\bibfield
  {journal} {\bibinfo  {journal} {Annu. Rev. Phys. Chem.}\ }\textbf {\bibinfo
  {volume} {65}},\ \bibinfo {pages} {365} (\bibinfo {year} {2014})}\BibitemShut
  {NoStop}%
\bibitem [{\citenamefont {Scully}\ \emph {et~al.}(2011)\citenamefont {Scully},
  \citenamefont {Chapin}, \citenamefont {Dorfman}, \citenamefont {Kim},\ and\
  \citenamefont {Svidzinsky}}]{Scully:11}%
  \BibitemOpen
  \bibfield  {author} {\bibinfo {author} {\bibfnamefont {M.~O.}\ \bibnamefont
  {Scully}}, \bibinfo {author} {\bibfnamefont {K.~R.}\ \bibnamefont {Chapin}},
  \bibinfo {author} {\bibfnamefont {K.~E.}\ \bibnamefont {Dorfman}}, \bibinfo
  {author} {\bibfnamefont {M.~B.}\ \bibnamefont {Kim}},\ and\ \bibinfo {author}
  {\bibfnamefont {A.}~\bibnamefont {Svidzinsky}},\ }\bibfield  {title}
  {\bibinfo {title} {Quantum heat engine power can be increased by
  noise-induced coherence},\ }\href {https://doi.org/10.1073/pnas.1110234108}
  {\bibfield  {journal} {\bibinfo  {journal} {Proc. Natl. Acad. Sci. USA}\
  }\textbf {\bibinfo {volume} {108}},\ \bibinfo {pages} {15097} (\bibinfo
  {year} {2011})}\BibitemShut {NoStop}%
\bibitem [{\citenamefont {Dorfman}\ \emph {et~al.}(2013)\citenamefont
  {Dorfman}, \citenamefont {Voronine}, \citenamefont {Mukamel},\ and\
  \citenamefont {Scully}}]{Dorfman:13}%
  \BibitemOpen
  \bibfield  {author} {\bibinfo {author} {\bibfnamefont {K.~E.}\ \bibnamefont
  {Dorfman}}, \bibinfo {author} {\bibfnamefont {D.~V.}\ \bibnamefont
  {Voronine}}, \bibinfo {author} {\bibfnamefont {S.}~\bibnamefont {Mukamel}},\
  and\ \bibinfo {author} {\bibfnamefont {M.~O.}\ \bibnamefont {Scully}},\
  }\bibfield  {title} {\bibinfo {title} {Photosynthetic reaction center as a
  quantum heat engine},\ }\href {https://doi.org/10.1073/pnas.1212666110}
  {\bibfield  {journal} {\bibinfo  {journal} {Proc. Natl. Acad. Sci. USA}\
  }\textbf {\bibinfo {volume} {110}},\ \bibinfo {pages} {2746} (\bibinfo {year}
  {2013})}\BibitemShut {NoStop}%
\bibitem [{\citenamefont {Beloy}\ \emph {et~al.}(2006)\citenamefont {Beloy},
  \citenamefont {Safronova},\ and\ \citenamefont {Derevianko}}]{Beloy:06}%
  \BibitemOpen
  \bibfield  {author} {\bibinfo {author} {\bibfnamefont {K.}~\bibnamefont
  {Beloy}}, \bibinfo {author} {\bibfnamefont {U.~I.}\ \bibnamefont
  {Safronova}},\ and\ \bibinfo {author} {\bibfnamefont {A.}~\bibnamefont
  {Derevianko}},\ }\bibfield  {title} {\bibinfo {title} {High-accuracy
  calculation of the blackbody radiation shift in the $^{133}\mathrm{Cs}$
  primary frequency standard},\ }\href
  {https://doi.org/10.1103/PhysRevLett.97.040801} {\bibfield  {journal}
  {\bibinfo  {journal} {Phys. Rev. Lett.}\ }\textbf {\bibinfo {volume} {97}},\
  \bibinfo {pages} {040801} (\bibinfo {year} {2006})}\BibitemShut {NoStop}%
\bibitem [{\citenamefont {Safronova}\ \emph {et~al.}(2012)\citenamefont
  {Safronova}, \citenamefont {Kozlov},\ and\ \citenamefont
  {Clark}}]{Safronova:12}%
  \BibitemOpen
  \bibfield  {author} {\bibinfo {author} {\bibfnamefont {M.~S.}\ \bibnamefont
  {Safronova}}, \bibinfo {author} {\bibfnamefont {M.~G.}\ \bibnamefont
  {Kozlov}},\ and\ \bibinfo {author} {\bibfnamefont {C.~W.}\ \bibnamefont
  {Clark}},\ }\bibfield  {title} {\bibinfo {title} {Blackbody radiation shifts
  in optical atomic clocks},\ }\href {https://doi.org/10.1109/TUFFC.2012.2213}
  {\bibfield  {journal} {\bibinfo  {journal} {IEEE Transactions on Ultrasonics,
  Ferroelectrics, and Frequency Control}\ }\textbf {\bibinfo {volume} {59}},\
  \bibinfo {pages} {439} (\bibinfo {year} {2012})}\BibitemShut {NoStop}%
\bibitem [{\citenamefont {Ovsiannikov}\ \emph {et~al.}(2011)\citenamefont
  {Ovsiannikov}, \citenamefont {Derevianko},\ and\ \citenamefont
  {Gibble}}]{Ovsiannikov:11}%
  \BibitemOpen
  \bibfield  {author} {\bibinfo {author} {\bibfnamefont {V.~D.}\ \bibnamefont
  {Ovsiannikov}}, \bibinfo {author} {\bibfnamefont {A.}~\bibnamefont
  {Derevianko}},\ and\ \bibinfo {author} {\bibfnamefont {K.}~\bibnamefont
  {Gibble}},\ }\bibfield  {title} {\bibinfo {title} {Rydberg spectroscopy in an
  optical lattice: Blackbody thermometry for atomic clocks},\ }\href
  {https://doi.org/10.1103/PhysRevLett.107.093003} {\bibfield  {journal}
  {\bibinfo  {journal} {Phys. Rev. Lett.}\ }\textbf {\bibinfo {volume} {107}},\
  \bibinfo {pages} {093003} (\bibinfo {year} {2011})}\BibitemShut {NoStop}%
\bibitem [{\citenamefont {Lisdat}\ \emph {et~al.}(2021)\citenamefont {Lisdat},
  \citenamefont {D\"orscher}, \citenamefont {Nosske},\ and\ \citenamefont
  {Sterr}}]{Lisdat:21}%
  \BibitemOpen
  \bibfield  {author} {\bibinfo {author} {\bibfnamefont {C.}~\bibnamefont
  {Lisdat}}, \bibinfo {author} {\bibfnamefont {S.}~\bibnamefont {D\"orscher}},
  \bibinfo {author} {\bibfnamefont {I.}~\bibnamefont {Nosske}},\ and\ \bibinfo
  {author} {\bibfnamefont {U.}~\bibnamefont {Sterr}},\ }\bibfield  {title}
  {\bibinfo {title} {Blackbody radiation shift in strontium lattice clocks
  revisited},\ }\href {https://doi.org/10.1103/PhysRevResearch.3.L042036}
  {\bibfield  {journal} {\bibinfo  {journal} {Phys. Rev. Research}\ }\textbf
  {\bibinfo {volume} {3}},\ \bibinfo {pages} {L042036} (\bibinfo {year}
  {2021})}\BibitemShut {NoStop}%
\bibitem [{\citenamefont {Kassal}\ \emph {et~al.}(2013)\citenamefont {Kassal},
  \citenamefont {Yuen-Zhou},\ and\ \citenamefont {Rahimi-Keshari}}]{Kassal:13}%
  \BibitemOpen
  \bibfield  {author} {\bibinfo {author} {\bibfnamefont {I.}~\bibnamefont
  {Kassal}}, \bibinfo {author} {\bibfnamefont {J.}~\bibnamefont {Yuen-Zhou}},\
  and\ \bibinfo {author} {\bibfnamefont {S.}~\bibnamefont {Rahimi-Keshari}},\
  }\bibfield  {title} {\bibinfo {title} {Does coherence enhance transport in
  photosynthesis?},\ }\href {https://doi.org/10.1021/jz301872b} {\bibfield
  {journal} {\bibinfo  {journal} {J. Phys. Chem. Lett.}\ }\textbf {\bibinfo
  {volume} {4}},\ \bibinfo {pages} {362} (\bibinfo {year} {2013})}\BibitemShut
  {NoStop}%
\bibitem [{\citenamefont {Le{\'o}n-Montiel}\ \emph {et~al.}(2014)\citenamefont
  {Le{\'o}n-Montiel}, \citenamefont {Kassal},\ and\ \citenamefont
  {Torres}}]{Leon-Montiel:14}%
  \BibitemOpen
  \bibfield  {author} {\bibinfo {author} {\bibfnamefont {R.~d.~J.}\
  \bibnamefont {Le{\'o}n-Montiel}}, \bibinfo {author} {\bibfnamefont
  {I.}~\bibnamefont {Kassal}},\ and\ \bibinfo {author} {\bibfnamefont {J.~P.}\
  \bibnamefont {Torres}},\ }\bibfield  {title} {\bibinfo {title} {Importance of
  excitation and trapping conditions in photosynthetic environment-assisted
  energy transport},\ }\href {https://doi.org/10.1021/jp505179h} {\bibfield
  {journal} {\bibinfo  {journal} {J. Phys. Chem. B}\ }\textbf {\bibinfo
  {volume} {118}},\ \bibinfo {pages} {10588} (\bibinfo {year}
  {2014})}\BibitemShut {NoStop}%
\bibitem [{\citenamefont {Tscherbul}\ and\ \citenamefont
  {Brumer}(2018)}]{Tscherbul:18}%
  \BibitemOpen
  \bibfield  {author} {\bibinfo {author} {\bibfnamefont {T.~V.}\ \bibnamefont
  {Tscherbul}}\ and\ \bibinfo {author} {\bibfnamefont {P.}~\bibnamefont
  {Brumer}},\ }\bibfield  {title} {\bibinfo {title} {Non-equilibrium stationary
  coherences in photosynthetic energy transfer under weak-field incoherent
  illumination},\ }\href {https://doi.org/10.1063/1.5028121} {\bibfield
  {journal} {\bibinfo  {journal} {J. Chem. Phys.}\ }\textbf {\bibinfo {volume}
  {148}},\ \bibinfo {pages} {124114} (\bibinfo {year} {2018})}\BibitemShut
  {NoStop}%
\bibitem [{\citenamefont {Yang}\ and\ \citenamefont {Cao}(2020)}]{Yang:20}%
  \BibitemOpen
  \bibfield  {author} {\bibinfo {author} {\bibfnamefont {P.-Y.}\ \bibnamefont
  {Yang}}\ and\ \bibinfo {author} {\bibfnamefont {J.}~\bibnamefont {Cao}},\
  }\bibfield  {title} {\bibinfo {title} {Steady-state analysis of
  light-harvesting energy transfer driven by incoherent light: From dimers to
  networks},\ }\href {https://doi.org/10.1021/acs.jpclett.0c01648} {\bibfield
  {journal} {\bibinfo  {journal} {J. Phys. Chem. Lett.}\ }\textbf {\bibinfo
  {volume} {11}},\ \bibinfo {pages} {7204} (\bibinfo {year}
  {2020})}\BibitemShut {NoStop}%
\bibitem [{\citenamefont {Chuang}\ and\ \citenamefont
  {Brumer}(2020)}]{Chuang:20}%
  \BibitemOpen
  \bibfield  {author} {\bibinfo {author} {\bibfnamefont {C.}~\bibnamefont
  {Chuang}}\ and\ \bibinfo {author} {\bibfnamefont {P.}~\bibnamefont
  {Brumer}},\ }\bibfield  {title} {\bibinfo {title} {{LH1-RC} light-harvesting
  photocycle under realistic light--matter conditions},\ }\href
  {https://doi.org/10.1063/5.0004490} {\bibfield  {journal} {\bibinfo
  {journal} {J. Chem. Phys.}\ }\textbf {\bibinfo {volume} {152}},\ \bibinfo
  {pages} {154101} (\bibinfo {year} {2020})}\BibitemShut {NoStop}%
\bibitem [{\citenamefont {Polli}\ \emph {et~al.}(2010)\citenamefont {Polli},
  \citenamefont {Alto{\`e}}, \citenamefont {Weingart}, \citenamefont
  {Spillane}, \citenamefont {Manzoni}, \citenamefont {Brida}, \citenamefont
  {Tomasello}, \citenamefont {Orlandi}, \citenamefont {Kukura}, \citenamefont
  {Mathies}, \citenamefont {Garavelli},\ and\ \citenamefont
  {Cerullo}}]{Polli:10}%
  \BibitemOpen
  \bibfield  {author} {\bibinfo {author} {\bibfnamefont {D.}~\bibnamefont
  {Polli}}, \bibinfo {author} {\bibfnamefont {P.}~\bibnamefont {Alto{\`e}}},
  \bibinfo {author} {\bibfnamefont {O.}~\bibnamefont {Weingart}}, \bibinfo
  {author} {\bibfnamefont {K.~M.}\ \bibnamefont {Spillane}}, \bibinfo {author}
  {\bibfnamefont {C.}~\bibnamefont {Manzoni}}, \bibinfo {author} {\bibfnamefont
  {D.}~\bibnamefont {Brida}}, \bibinfo {author} {\bibfnamefont
  {G.}~\bibnamefont {Tomasello}}, \bibinfo {author} {\bibfnamefont
  {G.}~\bibnamefont {Orlandi}}, \bibinfo {author} {\bibfnamefont
  {P.}~\bibnamefont {Kukura}}, \bibinfo {author} {\bibfnamefont {R.~A.}\
  \bibnamefont {Mathies}}, \bibinfo {author} {\bibfnamefont {M.}~\bibnamefont
  {Garavelli}},\ and\ \bibinfo {author} {\bibfnamefont {G.}~\bibnamefont
  {Cerullo}},\ }\bibfield  {title} {\bibinfo {title} {Conical intersection
  dynamics of the primary photoisomerization event in vision},\ }\href
  {https://doi.org/10.1038/nature09346} {\bibfield  {journal} {\bibinfo
  {journal} {Nature}\ }\textbf {\bibinfo {volume} {467}},\ \bibinfo {pages}
  {440} (\bibinfo {year} {2010})}\BibitemShut {NoStop}%
\bibitem [{\citenamefont {Schulten}\ and\ \citenamefont
  {Hayashi}(2014)}]{Schulten:14}%
  \BibitemOpen
  \bibfield  {author} {\bibinfo {author} {\bibfnamefont {K.}~\bibnamefont
  {Schulten}}\ and\ \bibinfo {author} {\bibfnamefont {S.}~\bibnamefont
  {Hayashi}},\ }\bibinfo {title} {Quantum biology of retinal},\ in\ \href
  {https://doi.org/DOI: 10.1017/CBO9780511863189.013} {\emph {\bibinfo
  {booktitle} {Quantum Effects in Biology}}},\ \bibinfo {editor} {edited by\
  \bibinfo {editor} {\bibfnamefont {G.~S.}\ \bibnamefont {Engel}}, \bibinfo
  {editor} {\bibfnamefont {M.~B.}\ \bibnamefont {Plenio}}, \bibinfo {editor}
  {\bibfnamefont {M.}~\bibnamefont {Mohseni}},\ and\ \bibinfo {editor}
  {\bibfnamefont {Y.}~\bibnamefont {Omar}}}\ (\bibinfo  {publisher} {Cambridge
  University Press},\ \bibinfo {address} {Cambridge},\ \bibinfo {year} {2014})\
  pp.\ \bibinfo {pages} {237--263}\BibitemShut {NoStop}%
\bibitem [{\citenamefont {Tscherbul}\ and\ \citenamefont
  {Brumer}(2014{\natexlab{b}})}]{Tscherbul:14b}%
  \BibitemOpen
  \bibfield  {author} {\bibinfo {author} {\bibfnamefont {T.~V.}\ \bibnamefont
  {Tscherbul}}\ and\ \bibinfo {author} {\bibfnamefont {P.}~\bibnamefont
  {Brumer}},\ }\bibfield  {title} {\bibinfo {title} {Excitation of biomolecules
  with incoherent light: Quantum yield for the photoisomerization of model
  retinal},\ }\href {https://doi.org/10.1021/jp501700t} {\bibfield  {journal}
  {\bibinfo  {journal} {J. Phys. Chem. A}\ }\textbf {\bibinfo {volume} {118}},\
  \bibinfo {pages} {3100} (\bibinfo {year} {2014}{\natexlab{b}})}\BibitemShut
  {NoStop}%
\bibitem [{\citenamefont {Tscherbul}\ and\ \citenamefont
  {Brumer}(2015{\natexlab{a}})}]{Tscherbul:15}%
  \BibitemOpen
  \bibfield  {author} {\bibinfo {author} {\bibfnamefont {T.~V.}\ \bibnamefont
  {Tscherbul}}\ and\ \bibinfo {author} {\bibfnamefont {P.}~\bibnamefont
  {Brumer}},\ }\bibfield  {title} {\bibinfo {title} {Quantum coherence effects
  in natural light-induced processes: cis--trans photoisomerization of model
  retinal under incoherent excitation},\ }\href
  {https://doi.org/10.1039/C5CP01388G} {\bibfield  {journal} {\bibinfo
  {journal} {Phys. Chem. Chem. Phys.}\ }\textbf {\bibinfo {volume} {17}},\
  \bibinfo {pages} {30904} (\bibinfo {year} {2015}{\natexlab{a}})}\BibitemShut
  {NoStop}%
\bibitem [{\citenamefont {Chuang}\ and\ \citenamefont
  {Brumer}(2022)}]{Chuang:22}%
  \BibitemOpen
  \bibfield  {author} {\bibinfo {author} {\bibfnamefont {C.}~\bibnamefont
  {Chuang}}\ and\ \bibinfo {author} {\bibfnamefont {P.}~\bibnamefont
  {Brumer}},\ }\bibfield  {title} {\bibinfo {title} {Steady state
  photoisomerization quantum yield of model rhodopsin: Insights from wavepacket
  dynamics?},\ }\href {https://doi.org/10.1021/acs.jpclett.2c01200} {\bibfield
  {journal} {\bibinfo  {journal} {J. Phys. Chem. Lett.}\ }\textbf {\bibinfo
  {volume} {13}},\ \bibinfo {pages} {4963} (\bibinfo {year}
  {2022})}\BibitemShut {NoStop}%
\bibitem [{\citenamefont {Cohen-Tannoudji}\ \emph {et~al.}(2004)\citenamefont
  {Cohen-Tannoudji}, \citenamefont {Dupont-Roc},\ and\ \citenamefont
  {Grynberg}}]{Cohen-Tannoudji:04}%
  \BibitemOpen
  \bibfield  {author} {\bibinfo {author} {\bibfnamefont {C.}~\bibnamefont
  {Cohen-Tannoudji}}, \bibinfo {author} {\bibfnamefont {J.}~\bibnamefont
  {Dupont-Roc}},\ and\ \bibinfo {author} {\bibfnamefont {G.}~\bibnamefont
  {Grynberg}},\ }\href@noop {} {\emph {\bibinfo {title} {{\it Atom-Photon
  Interactions: Basic Processes and Applications}}}}\ (\bibinfo  {publisher}
  {Wiley-VCH},\ \bibinfo {year} {2004})\BibitemShut {NoStop}%
\bibitem [{\citenamefont {Blum}(2011)}]{Blum:11}%
  \BibitemOpen
  \bibfield  {author} {\bibinfo {author} {\bibfnamefont {K.}~\bibnamefont
  {Blum}},\ }\href@noop {} {\emph {\bibinfo {title} {{\it Density Matrix Theory
  and Applications}}}},\ \bibinfo {number} {Chap. 8}\ (\bibinfo  {publisher}
  {Springer},\ \bibinfo {year} {2011})\BibitemShut {NoStop}%
\bibitem [{\citenamefont {Wilhelm}\ \emph {et~al.}(2007)\citenamefont
  {Wilhelm}, \citenamefont {Storcz}, \citenamefont {Hartmann},\ and\
  \citenamefont {Geller}}]{Wilhelm:07}%
  \BibitemOpen
  \bibfield  {author} {\bibinfo {author} {\bibfnamefont {F.~K.}\ \bibnamefont
  {Wilhelm}}, \bibinfo {author} {\bibfnamefont {M.~J.}\ \bibnamefont {Storcz}},
  \bibinfo {author} {\bibfnamefont {U.}~\bibnamefont {Hartmann}},\ and\
  \bibinfo {author} {\bibfnamefont {M.~R.}\ \bibnamefont {Geller}},\ }\bibfield
   {title} {\bibinfo {title} {Superconducting qubits ii: Decoherence},\ }in\
  \href@noop {} {\emph {\bibinfo {booktitle} {Manipulating Quantum Coherence in
  Solid State Systems}}},\ \bibinfo {editor} {edited by\ \bibinfo {editor}
  {\bibfnamefont {M.~E.}\ \bibnamefont {Flatt{\'e}}}\ and\ \bibinfo {editor}
  {\bibfnamefont {I.}~\bibnamefont {{\c T}ifrea}}}\ (\bibinfo  {publisher}
  {Springer Netherlands},\ \bibinfo {address} {Dordrecht},\ \bibinfo {year}
  {2007})\ pp.\ \bibinfo {pages} {195--232}\BibitemShut {NoStop}%
\bibitem [{\citenamefont {Jeske}\ \emph {et~al.}(2015)\citenamefont {Jeske},
  \citenamefont {Ing}, \citenamefont {Plenio}, \citenamefont {Huelga},\ and\
  \citenamefont {Cole}}]{Jeske:15}%
  \BibitemOpen
  \bibfield  {author} {\bibinfo {author} {\bibfnamefont {J.}~\bibnamefont
  {Jeske}}, \bibinfo {author} {\bibfnamefont {D.~J.}\ \bibnamefont {Ing}},
  \bibinfo {author} {\bibfnamefont {M.~B.}\ \bibnamefont {Plenio}}, \bibinfo
  {author} {\bibfnamefont {S.~F.}\ \bibnamefont {Huelga}},\ and\ \bibinfo
  {author} {\bibfnamefont {J.~H.}\ \bibnamefont {Cole}},\ }\bibfield  {title}
  {\bibinfo {title} {{Bloch-Redfield} equations for modeling light-harvesting
  complexes},\ }\href {https://doi.org/10.1063/1.4907370} {\bibfield  {journal}
  {\bibinfo  {journal} {J. Chem. Phys.}\ }\textbf {\bibinfo {volume} {142}},\
  \bibinfo {pages} {064104} (\bibinfo {year} {2015})}\BibitemShut {NoStop}%
\bibitem [{\citenamefont {Dodin}\ \emph
  {et~al.}(2016{\natexlab{a}})\citenamefont {Dodin}, \citenamefont
  {Tscherbul},\ and\ \citenamefont {Brumer}}]{Dodin:16}%
  \BibitemOpen
  \bibfield  {author} {\bibinfo {author} {\bibfnamefont {A.}~\bibnamefont
  {Dodin}}, \bibinfo {author} {\bibfnamefont {T.~V.}\ \bibnamefont
  {Tscherbul}},\ and\ \bibinfo {author} {\bibfnamefont {P.}~\bibnamefont
  {Brumer}},\ }\bibfield  {title} {\bibinfo {title} {Quantum dynamics of
  incoherently driven {V}-type systems: Analytic solutions beyond the secular
  approximation},\ }\href {https://doi.org/10.1063/1.4954243} {\bibfield
  {journal} {\bibinfo  {journal} {J. Chem. Phys.}\ }\textbf {\bibinfo {volume}
  {144}},\ \bibinfo {pages} {244108} (\bibinfo {year}
  {2016}{\natexlab{a}})}\BibitemShut {NoStop}%
\bibitem [{\citenamefont {Dodin}\ \emph {et~al.}(2018)\citenamefont {Dodin},
  \citenamefont {Tscherbul}, \citenamefont {Alicki}, \citenamefont {Vutha},\
  and\ \citenamefont {Brumer}}]{Dodin:18}%
  \BibitemOpen
  \bibfield  {author} {\bibinfo {author} {\bibfnamefont {A.}~\bibnamefont
  {Dodin}}, \bibinfo {author} {\bibfnamefont {T.~V.}\ \bibnamefont
  {Tscherbul}}, \bibinfo {author} {\bibfnamefont {R.}~\bibnamefont {Alicki}},
  \bibinfo {author} {\bibfnamefont {A.}~\bibnamefont {Vutha}},\ and\ \bibinfo
  {author} {\bibfnamefont {P.}~\bibnamefont {Brumer}},\ }\bibfield  {title}
  {\bibinfo {title} {Secular versus nonsecular {Redfield dynamics and Fano}
  coherences in incoherent excitation: {An} experimental proposal},\ }\href
  {https://doi.org/10.1103/PhysRevA.97.013421} {\bibfield  {journal} {\bibinfo
  {journal} {Phys. Rev. A}\ }\textbf {\bibinfo {volume} {97}},\ \bibinfo
  {pages} {013421} (\bibinfo {year} {2018})}\BibitemShut {NoStop}%
\bibitem [{\citenamefont {Koyu}\ and\ \citenamefont
  {Tscherbul}(2018)}]{Koyu:18}%
  \BibitemOpen
  \bibfield  {author} {\bibinfo {author} {\bibfnamefont {S.}~\bibnamefont
  {Koyu}}\ and\ \bibinfo {author} {\bibfnamefont {T.~V.}\ \bibnamefont
  {Tscherbul}},\ }\bibfield  {title} {\bibinfo {title} {Long-lived quantum
  coherences in a $\mathsf{V}$-type system strongly driven by a thermal
  environment},\ }\href {https://doi.org/10.1103/PhysRevA.98.023811} {\bibfield
   {journal} {\bibinfo  {journal} {Phys. Rev. A}\ }\textbf {\bibinfo {volume}
  {98}},\ \bibinfo {pages} {023811} (\bibinfo {year} {2018})}\BibitemShut
  {NoStop}%
\bibitem [{\citenamefont {Tscherbul}\ and\ \citenamefont
  {Brumer}(2015{\natexlab{b}})}]{Tscherbul:15b}%
  \BibitemOpen
  \bibfield  {author} {\bibinfo {author} {\bibfnamefont {T.~V.}\ \bibnamefont
  {Tscherbul}}\ and\ \bibinfo {author} {\bibfnamefont {P.}~\bibnamefont
  {Brumer}},\ }\bibfield  {title} {\bibinfo {title} {Partial secular
  {Bloch-Redfield} master equation for incoherent excitation of multilevel
  quantum systems},\ }\href {https://doi.org/10.1063/1.4908130} {\bibfield
  {journal} {\bibinfo  {journal} {J. Chem. Phys.}\ }\textbf {\bibinfo {volume}
  {142}},\ \bibinfo {pages} {104107} (\bibinfo {year}
  {2015}{\natexlab{b}})}\BibitemShut {NoStop}%
\bibitem [{\citenamefont {Eastham}\ \emph {et~al.}(2016)\citenamefont
  {Eastham}, \citenamefont {Kirton}, \citenamefont {Cammack}, \citenamefont
  {Lovett},\ and\ \citenamefont {Keeling}}]{Eastham:16}%
  \BibitemOpen
  \bibfield  {author} {\bibinfo {author} {\bibfnamefont {P.~R.}\ \bibnamefont
  {Eastham}}, \bibinfo {author} {\bibfnamefont {P.}~\bibnamefont {Kirton}},
  \bibinfo {author} {\bibfnamefont {H.~M.}\ \bibnamefont {Cammack}}, \bibinfo
  {author} {\bibfnamefont {B.~W.}\ \bibnamefont {Lovett}},\ and\ \bibinfo
  {author} {\bibfnamefont {J.}~\bibnamefont {Keeling}},\ }\bibfield  {title}
  {\bibinfo {title} {Bath-induced coherence and the secular approximation},\
  }\href {https://doi.org/10.1103/PhysRevA.94.012110} {\bibfield  {journal}
  {\bibinfo  {journal} {Phys. Rev. A}\ }\textbf {\bibinfo {volume} {94}},\
  \bibinfo {pages} {012110} (\bibinfo {year} {2016})}\BibitemShut {NoStop}%
\bibitem [{\citenamefont {Wang}\ \emph {et~al.}(2019)\citenamefont {Wang},
  \citenamefont {Wu},\ and\ \citenamefont {Wang}}]{Wang:19}%
  \BibitemOpen
  \bibfield  {author} {\bibinfo {author} {\bibfnamefont {Z.}~\bibnamefont
  {Wang}}, \bibinfo {author} {\bibfnamefont {W.}~\bibnamefont {Wu}},\ and\
  \bibinfo {author} {\bibfnamefont {J.}~\bibnamefont {Wang}},\ }\bibfield
  {title} {\bibinfo {title} {Steady-state entanglement and coherence of two
  coupled qubits in equilibrium and nonequilibrium environments},\ }\href
  {https://doi.org/10.1103/PhysRevA.99.042320} {\bibfield  {journal} {\bibinfo
  {journal} {Phys. Rev. A}\ }\textbf {\bibinfo {volume} {99}},\ \bibinfo
  {pages} {042320} (\bibinfo {year} {2019})}\BibitemShut {NoStop}%
\bibitem [{\citenamefont {Liao}\ and\ \citenamefont {Liang}(2021)}]{Liao:21}%
  \BibitemOpen
  \bibfield  {author} {\bibinfo {author} {\bibfnamefont {C.-Y.}\ \bibnamefont
  {Liao}}\ and\ \bibinfo {author} {\bibfnamefont {X.-T.}\ \bibnamefont
  {Liang}},\ }\bibfield  {title} {\bibinfo {title} {The {Lindblad and Redfield}
  forms derived from the {Born--Markov} master equation without secular
  approximation and their applications},\ }\href
  {https://doi.org/10.1088/1572-9494/abec65} {\bibfield  {journal} {\bibinfo
  {journal} {Comm. Theor. Phys.}\ }\textbf {\bibinfo {volume} {73}},\ \bibinfo
  {pages} {095101} (\bibinfo {year} {2021})}\BibitemShut {NoStop}%
\bibitem [{\citenamefont {Merkli}\ \emph {et~al.}(2015)\citenamefont {Merkli},
  \citenamefont {Song},\ and\ \citenamefont {Berman}}]{Merkli:15}%
  \BibitemOpen
  \bibfield  {author} {\bibinfo {author} {\bibfnamefont {M.}~\bibnamefont
  {Merkli}}, \bibinfo {author} {\bibfnamefont {H.}~\bibnamefont {Song}},\ and\
  \bibinfo {author} {\bibfnamefont {G.~P.}\ \bibnamefont {Berman}},\ }\bibfield
   {title} {\bibinfo {title} {Multiscale dynamics of open three-level quantum
  systems with two quasi-degenerate levels},\ }\href
  {https://doi.org/10.1088/1751-8113/48/27/275304} {\bibfield  {journal}
  {\bibinfo  {journal} {J. Phys. A}\ }\textbf {\bibinfo {volume} {48}},\
  \bibinfo {pages} {275304} (\bibinfo {year} {2015})}\BibitemShut {NoStop}%
\bibitem [{\citenamefont {Trushechkin}(2021)}]{Trushechkin:21}%
  \BibitemOpen
  \bibfield  {author} {\bibinfo {author} {\bibfnamefont {A.}~\bibnamefont
  {Trushechkin}},\ }\bibfield  {title} {\bibinfo {title} {Unified
  {Gorini-Kossakowski-Lindblad-Sudarshan} quantum master equation beyond the
  secular approximation},\ }\href {https://doi.org/10.1103/PhysRevA.103.062226}
  {\bibfield  {journal} {\bibinfo  {journal} {Phys. Rev. A}\ }\textbf {\bibinfo
  {volume} {103}},\ \bibinfo {pages} {062226} (\bibinfo {year}
  {2021})}\BibitemShut {NoStop}%
\bibitem [{\citenamefont {Fleischhauer}\ \emph {et~al.}(1992)\citenamefont
  {Fleischhauer}, \citenamefont {Keitel}, \citenamefont {Scully},\ and\
  \citenamefont {Su}}]{Fleischhauer:92}%
  \BibitemOpen
  \bibfield  {author} {\bibinfo {author} {\bibfnamefont {M.}~\bibnamefont
  {Fleischhauer}}, \bibinfo {author} {\bibfnamefont {C.~H.}\ \bibnamefont
  {Keitel}}, \bibinfo {author} {\bibfnamefont {M.~O.}\ \bibnamefont {Scully}},\
  and\ \bibinfo {author} {\bibfnamefont {C.}~\bibnamefont {Su}},\ }\bibfield
  {title} {\bibinfo {title} {Lasing without inversion and enhancement of the
  index of refraction via interference of incoherent pump processes},\ }\href
  {https://doi.org/https://doi.org/10.1016/0030-4018(92)90389-9} {\bibfield
  {journal} {\bibinfo  {journal} {Opt. Commun.}\ }\textbf {\bibinfo {volume}
  {87}},\ \bibinfo {pages} {109} (\bibinfo {year} {1992})}\BibitemShut
  {NoStop}%
\bibitem [{\citenamefont {Hegerfeldt}\ and\ \citenamefont
  {Plenio}(1993)}]{Hegerfeldt:93}%
  \BibitemOpen
  \bibfield  {author} {\bibinfo {author} {\bibfnamefont {G.~C.}\ \bibnamefont
  {Hegerfeldt}}\ and\ \bibinfo {author} {\bibfnamefont {M.~B.}\ \bibnamefont
  {Plenio}},\ }\bibfield  {title} {\bibinfo {title} {Coherence with incoherent
  light: A new type of quantum beat for a single atom},\ }\href
  {https://doi.org/10.1103/PhysRevA.47.2186} {\bibfield  {journal} {\bibinfo
  {journal} {Phys. Rev. A}\ }\textbf {\bibinfo {volume} {47}},\ \bibinfo
  {pages} {2186} (\bibinfo {year} {1993})}\BibitemShut {NoStop}%
\bibitem [{\citenamefont {Kozlov}\ \emph {et~al.}(2006)\citenamefont {Kozlov},
  \citenamefont {Rostovtsev},\ and\ \citenamefont {Scully}}]{Kozlov:06}%
  \BibitemOpen
  \bibfield  {author} {\bibinfo {author} {\bibfnamefont {V.~V.}\ \bibnamefont
  {Kozlov}}, \bibinfo {author} {\bibfnamefont {Y.}~\bibnamefont {Rostovtsev}},\
  and\ \bibinfo {author} {\bibfnamefont {M.~O.}\ \bibnamefont {Scully}},\
  }\bibfield  {title} {\bibinfo {title} {Inducing quantum coherence via decays
  and incoherent pumping with application to population trapping, lasing
  without inversion, and quenching of spontaneous emission},\ }\href
  {https://doi.org/10.1103/PhysRevA.74.063829} {\bibfield  {journal} {\bibinfo
  {journal} {Phys. Rev. A}\ }\textbf {\bibinfo {volume} {74}},\ \bibinfo
  {pages} {063829} (\bibinfo {year} {2006})}\BibitemShut {NoStop}%
\bibitem [{\citenamefont {Ou}\ \emph {et~al.}(2008)\citenamefont {Ou},
  \citenamefont {Liang},\ and\ \citenamefont {Li}}]{Ou:08}%
  \BibitemOpen
  \bibfield  {author} {\bibinfo {author} {\bibfnamefont {B.-Q.}\ \bibnamefont
  {Ou}}, \bibinfo {author} {\bibfnamefont {L.-M.}\ \bibnamefont {Liang}},\ and\
  \bibinfo {author} {\bibfnamefont {C.-Z.}\ \bibnamefont {Li}},\ }\bibfield
  {title} {\bibinfo {title} {Coherence induced by incoherent pumping field and
  decay process in three-level {$\Lambda$}-type atomic system},\ }\href
  {https://doi.org/https://doi.org/10.1016/j.optcom.2008.06.037} {\bibfield
  {journal} {\bibinfo  {journal} {Opt. Commun.}\ }\textbf {\bibinfo {volume}
  {281}},\ \bibinfo {pages} {4940} (\bibinfo {year} {2008})}\BibitemShut
  {NoStop}%
\bibitem [{\citenamefont {Koyu}\ \emph {et~al.}(2021)\citenamefont {Koyu},
  \citenamefont {Dodin}, \citenamefont {Brumer},\ and\ \citenamefont
  {Tscherbul}}]{Koyu:21}%
  \BibitemOpen
  \bibfield  {author} {\bibinfo {author} {\bibfnamefont {S.}~\bibnamefont
  {Koyu}}, \bibinfo {author} {\bibfnamefont {A.}~\bibnamefont {Dodin}},
  \bibinfo {author} {\bibfnamefont {P.}~\bibnamefont {Brumer}},\ and\ \bibinfo
  {author} {\bibfnamefont {T.~V.}\ \bibnamefont {Tscherbul}},\ }\bibfield
  {title} {\bibinfo {title} {Steady-state {Fano coherences} in a {V}-type
  system driven by polarized incoherent light},\ }\href
  {https://doi.org/10.1103/PhysRevResearch.3.013295} {\bibfield  {journal}
  {\bibinfo  {journal} {Phys. Rev. Research}\ }\textbf {\bibinfo {volume}
  {3}},\ \bibinfo {pages} {013295} (\bibinfo {year} {2021})}\BibitemShut
  {NoStop}%
\bibitem [{\citenamefont {Ficek}\ and\ \citenamefont {Swain}(2005)}]{Ficek:05}%
  \BibitemOpen
  \bibfield  {author} {\bibinfo {author} {\bibfnamefont {Z.}~\bibnamefont
  {Ficek}}\ and\ \bibinfo {author} {\bibfnamefont {S.}~\bibnamefont {Swain}},\
  }\href@noop {} {\emph {\bibinfo {title} {{\it Quantum Interference and
  Coherence: Theory and Experiments}}}}\ (\bibinfo  {publisher}
  {Springer-Verlag, New York},\ \bibinfo {year} {2005})\BibitemShut {NoStop}%
\bibitem [{\citenamefont {Patnaik}\ and\ \citenamefont
  {Agarwal}(1999)}]{Patnaik:99}%
  \BibitemOpen
  \bibfield  {author} {\bibinfo {author} {\bibfnamefont {A.~K.}\ \bibnamefont
  {Patnaik}}\ and\ \bibinfo {author} {\bibfnamefont {G.~S.}\ \bibnamefont
  {Agarwal}},\ }\bibfield  {title} {\bibinfo {title} {Cavity-induced coherence
  effects in spontaneous emissions from preselection of polarization},\ }\href
  {https://doi.org/10.1103/PhysRevA.59.3015} {\bibfield  {journal} {\bibinfo
  {journal} {Phys. Rev. A}\ }\textbf {\bibinfo {volume} {59}},\ \bibinfo
  {pages} {3015} (\bibinfo {year} {1999})}\BibitemShut {NoStop}%
\bibitem [{\citenamefont {Kapale}\ \emph {et~al.}(2003)\citenamefont {Kapale},
  \citenamefont {Scully}, \citenamefont {Zhu},\ and\ \citenamefont
  {Zubairy}}]{Kapale:03}%
  \BibitemOpen
  \bibfield  {author} {\bibinfo {author} {\bibfnamefont {K.~T.}\ \bibnamefont
  {Kapale}}, \bibinfo {author} {\bibfnamefont {M.~O.}\ \bibnamefont {Scully}},
  \bibinfo {author} {\bibfnamefont {S.-Y.}\ \bibnamefont {Zhu}},\ and\ \bibinfo
  {author} {\bibfnamefont {M.~S.}\ \bibnamefont {Zubairy}},\ }\bibfield
  {title} {\bibinfo {title} {Quenching of spontaneous emission through
  interference of incoherent pump processes},\ }\href
  {https://doi.org/10.1103/PhysRevA.67.023804} {\bibfield  {journal} {\bibinfo
  {journal} {Phys. Rev. A}\ }\textbf {\bibinfo {volume} {67}},\ \bibinfo
  {pages} {023804} (\bibinfo {year} {2003})}\BibitemShut {NoStop}%
\bibitem [{\citenamefont {Jung}\ and\ \citenamefont {Brumer}(2020)}]{Jung:20}%
  \BibitemOpen
  \bibfield  {author} {\bibinfo {author} {\bibfnamefont {K.~A.}\ \bibnamefont
  {Jung}}\ and\ \bibinfo {author} {\bibfnamefont {P.}~\bibnamefont {Brumer}},\
  }\bibfield  {title} {\bibinfo {title} {Energy transfer under natural
  incoherent light: Effects of asymmetry on efficiency},\ }\href
  {https://doi.org/10.1063/5.0020576} {\bibfield  {journal} {\bibinfo
  {journal} {J. Chem. Phys.}\ }\textbf {\bibinfo {volume} {153}},\ \bibinfo
  {pages} {114102} (\bibinfo {year} {2020})}\BibitemShut {NoStop}%
\bibitem [{\citenamefont {Jankovi{\'c}}\ and\ \citenamefont {Man{\v
  c}al}(2020)}]{Jankovic:20}%
  \BibitemOpen
  \bibfield  {author} {\bibinfo {author} {\bibfnamefont {V.}~\bibnamefont
  {Jankovi{\'c}}}\ and\ \bibinfo {author} {\bibfnamefont {T.}~\bibnamefont
  {Man{\v c}al}},\ }\bibfield  {title} {\bibinfo {title} {Nonequilibrium
  steady-state picture of incoherent light-induced excitation harvesting},\
  }\href {https://doi.org/10.1063/5.0029918} {\bibfield  {journal} {\bibinfo
  {journal} {J. Chem. Phys.}\ }\textbf {\bibinfo {volume} {153}},\ \bibinfo
  {pages} {244110} (\bibinfo {year} {2020})}\BibitemShut {NoStop}%
\bibitem [{\citenamefont {Tomasi}\ \emph {et~al.}(2021)\citenamefont {Tomasi},
  \citenamefont {Rouse}, \citenamefont {Gauger}, \citenamefont {Lovett},\ and\
  \citenamefont {Kassal}}]{Tomasi:21}%
  \BibitemOpen
  \bibfield  {author} {\bibinfo {author} {\bibfnamefont {S.}~\bibnamefont
  {Tomasi}}, \bibinfo {author} {\bibfnamefont {D.~M.}\ \bibnamefont {Rouse}},
  \bibinfo {author} {\bibfnamefont {E.~M.}\ \bibnamefont {Gauger}}, \bibinfo
  {author} {\bibfnamefont {B.~W.}\ \bibnamefont {Lovett}},\ and\ \bibinfo
  {author} {\bibfnamefont {I.}~\bibnamefont {Kassal}},\ }\bibfield  {title}
  {\bibinfo {title} {Environmentally improved coherent light harvesting},\
  }\href {https://doi.org/10.1021/acs.jpclett.1c01303} {\bibfield  {journal}
  {\bibinfo  {journal} {J. Phys. Chem. Lett.}\ }\textbf {\bibinfo {volume}
  {12}},\ \bibinfo {pages} {6143} (\bibinfo {year} {2021})}\BibitemShut
  {NoStop}%
\bibitem [{\citenamefont {Latune}\ \emph {et~al.}(2020)\citenamefont {Latune},
  \citenamefont {Sinayskiy},\ and\ \citenamefont {Petruccione}}]{Latune:20}%
  \BibitemOpen
  \bibfield  {author} {\bibinfo {author} {\bibfnamefont {C.~L.}\ \bibnamefont
  {Latune}}, \bibinfo {author} {\bibfnamefont {I.}~\bibnamefont {Sinayskiy}},\
  and\ \bibinfo {author} {\bibfnamefont {F.}~\bibnamefont {Petruccione}},\
  }\bibfield  {title} {\bibinfo {title} {Negative contributions to entropy
  production induced by quantum coherences},\ }\href
  {https://doi.org/10.1103/PhysRevA.102.042220} {\bibfield  {journal} {\bibinfo
   {journal} {Phys. Rev. A}\ }\textbf {\bibinfo {volume} {102}},\ \bibinfo
  {pages} {042220} (\bibinfo {year} {2020})}\BibitemShut {NoStop}%
\bibitem [{\citenamefont {Agarwal}\ and\ \citenamefont
  {Menon}(2001)}]{Agarwal:01}%
  \BibitemOpen
  \bibfield  {author} {\bibinfo {author} {\bibfnamefont {G.~S.}\ \bibnamefont
  {Agarwal}}\ and\ \bibinfo {author} {\bibfnamefont {S.}~\bibnamefont
  {Menon}},\ }\bibfield  {title} {\bibinfo {title} {Quantum interferences and
  the question of thermodynamic equilibrium},\ }\href
  {https://doi.org/10.1103/PhysRevA.63.023818} {\bibfield  {journal} {\bibinfo
  {journal} {Phys. Rev. A}\ }\textbf {\bibinfo {volume} {63}},\ \bibinfo
  {pages} {023818} (\bibinfo {year} {2001})}\BibitemShut {NoStop}%
\bibitem [{\citenamefont {Han}\ \emph {et~al.}(2021)\citenamefont {Han},
  \citenamefont {Lee}, \citenamefont {Sinha}, \citenamefont {Fatemi},\ and\
  \citenamefont {Rolston}}]{Han:21}%
  \BibitemOpen
  \bibfield  {author} {\bibinfo {author} {\bibfnamefont {H.~S.}\ \bibnamefont
  {Han}}, \bibinfo {author} {\bibfnamefont {A.}~\bibnamefont {Lee}}, \bibinfo
  {author} {\bibfnamefont {K.}~\bibnamefont {Sinha}}, \bibinfo {author}
  {\bibfnamefont {F.~K.}\ \bibnamefont {Fatemi}},\ and\ \bibinfo {author}
  {\bibfnamefont {S.~L.}\ \bibnamefont {Rolston}},\ }\bibfield  {title}
  {\bibinfo {title} {Observation of vacuum-induced collective quantum beats},\
  }\href {https://doi.org/10.1103/PhysRevLett.127.073604} {\bibfield  {journal}
  {\bibinfo  {journal} {Phys. Rev. Lett.}\ }\textbf {\bibinfo {volume} {127}},\
  \bibinfo {pages} {073604} (\bibinfo {year} {2021})}\BibitemShut {NoStop}%
\bibitem [{\citenamefont {Fleischhauer}\ \emph {et~al.}(2005)\citenamefont
  {Fleischhauer}, \citenamefont {Imamoglu},\ and\ \citenamefont
  {Marangos}}]{Fleischhauer:05}%
  \BibitemOpen
  \bibfield  {author} {\bibinfo {author} {\bibfnamefont {M.}~\bibnamefont
  {Fleischhauer}}, \bibinfo {author} {\bibfnamefont {A.}~\bibnamefont
  {Imamoglu}},\ and\ \bibinfo {author} {\bibfnamefont {J.~P.}\ \bibnamefont
  {Marangos}},\ }\bibfield  {title} {\bibinfo {title} {Electromagnetically
  induced transparency: Optics in coherent media},\ }\href
  {https://doi.org/10.1103/RevModPhys.77.633} {\bibfield  {journal} {\bibinfo
  {journal} {Rev. Mod. Phys.}\ }\textbf {\bibinfo {volume} {77}},\ \bibinfo
  {pages} {633} (\bibinfo {year} {2005})}\BibitemShut {NoStop}%
\bibitem [{\citenamefont {Altenm{\"u}ller}(1995)}]{Altenmuller:95}%
  \BibitemOpen
  \bibfield  {author} {\bibinfo {author} {\bibfnamefont {T.~P.}\ \bibnamefont
  {Altenm{\"u}ller}},\ }\bibfield  {title} {\bibinfo {title} {Are there quantum
  beats from vacuum-induced coherence?},\ }\href
  {https://doi.org/10.1007/BF01437684} {\bibfield  {journal} {\bibinfo
  {journal} {Z. Phys. D}\ }\textbf {\bibinfo {volume} {34}},\ \bibinfo {pages}
  {157} (\bibinfo {year} {1995})}\BibitemShut {NoStop}%
\bibitem [{\citenamefont {Hoki}\ and\ \citenamefont {Brumer}(2011)}]{Hoki:11}%
  \BibitemOpen
  \bibfield  {author} {\bibinfo {author} {\bibfnamefont {K.}~\bibnamefont
  {Hoki}}\ and\ \bibinfo {author} {\bibfnamefont {P.}~\bibnamefont {Brumer}},\
  }\bibfield  {title} {\bibinfo {title} {Excitation of biomolecules by coherent
  vs. incoherent light: {Model} rhodopsin photoisomerization},\ }\href
  {https://doi.org/https://doi.org/10.1016/j.proche.2011.08.019} {\bibfield
  {journal} {\bibinfo  {journal} {Procedia Chem.}\ }\textbf {\bibinfo {volume}
  {3}},\ \bibinfo {pages} {122} (\bibinfo {year} {2011})}\BibitemShut {NoStop}%
\bibitem [{\citenamefont {Breuer}\ and\ \citenamefont
  {Petruccione}(2006)}]{Breuer:06}%
  \BibitemOpen
  \bibfield  {author} {\bibinfo {author} {\bibfnamefont {H.-P.}\ \bibnamefont
  {Breuer}}\ and\ \bibinfo {author} {\bibfnamefont {F.}~\bibnamefont
  {Petruccione}},\ }\href@noop {} {\emph {\bibinfo {title} {{\it The Theory of
  Open Quantum Systems}}}}\ (\bibinfo  {publisher} {Clarendon Press, Oxford},\
  \bibinfo {year} {2006})\BibitemShut {NoStop}%
\bibitem [{\citenamefont {McCauley}\ \emph {et~al.}(2020)\citenamefont
  {McCauley}, \citenamefont {Cruikshank}, \citenamefont {Bondar},\ and\
  \citenamefont {Jacobs}}]{McCauley:20}%
  \BibitemOpen
  \bibfield  {author} {\bibinfo {author} {\bibfnamefont {G.}~\bibnamefont
  {McCauley}}, \bibinfo {author} {\bibfnamefont {B.}~\bibnamefont
  {Cruikshank}}, \bibinfo {author} {\bibfnamefont {D.~I.}\ \bibnamefont
  {Bondar}},\ and\ \bibinfo {author} {\bibfnamefont {K.}~\bibnamefont
  {Jacobs}},\ }\bibfield  {title} {\bibinfo {title} {Accurate {Lindblad-form}
  master equation for weakly damped quantum systems across all regimes},\
  }\href {https://doi.org/10.1038/s41534-020-00299-6} {\bibfield  {journal}
  {\bibinfo  {journal} {npj Quantum Information}\ }\textbf {\bibinfo {volume}
  {6}},\ \bibinfo {pages} {74} (\bibinfo {year} {2020})}\BibitemShut {NoStop}%
\bibitem [{SM()}]{SM}%
  \BibitemOpen
  \href@noop {} {}\bibinfo {note} {See Supplementary Material at [URL] for a
  derivation of analytical solutions of the BR equations.}\BibitemShut {Stop}%
\bibitem [{\citenamefont {Dodin}\ \emph
  {et~al.}(2016{\natexlab{b}})\citenamefont {Dodin}, \citenamefont
  {Tscherbul},\ and\ \citenamefont {Brumer}}]{Dodin:16b}%
  \BibitemOpen
  \bibfield  {author} {\bibinfo {author} {\bibfnamefont {A.}~\bibnamefont
  {Dodin}}, \bibinfo {author} {\bibfnamefont {T.~V.}\ \bibnamefont
  {Tscherbul}},\ and\ \bibinfo {author} {\bibfnamefont {P.}~\bibnamefont
  {Brumer}},\ }\bibfield  {title} {\bibinfo {title} {Coherent dynamics of
  v-type systems driven by time-dependent incoherent radiation},\ }\href
  {https://doi.org/10.1063/1.4972140} {\bibfield  {journal} {\bibinfo
  {journal} {J. Chem. Phys.}\ }\textbf {\bibinfo {volume} {145}},\ \bibinfo
  {pages} {244313} (\bibinfo {year} {2016}{\natexlab{b}})}\BibitemShut
  {NoStop}%
\bibitem [{\citenamefont {Dodin}\ and\ \citenamefont
  {Brumer}(2021{\natexlab{b}})}]{Dodin:21b}%
  \BibitemOpen
  \bibfield  {author} {\bibinfo {author} {\bibfnamefont {A.}~\bibnamefont
  {Dodin}}\ and\ \bibinfo {author} {\bibfnamefont {P.}~\bibnamefont {Brumer}},\
  }\bibfield  {title} {\bibinfo {title} {Generalized adiabatic theorems:
  Quantum systems driven by modulated time-varying fields},\ }\href
  {https://doi.org/10.1103/PRXQuantum.2.030302} {\bibfield  {journal} {\bibinfo
   {journal} {PRX Quantum}\ }\textbf {\bibinfo {volume} {2}},\ \bibinfo {pages}
  {030302} (\bibinfo {year} {2021}{\natexlab{b}})}\BibitemShut {NoStop}%
\bibitem [{\citenamefont {Vassen}\ \emph {et~al.}(2012)\citenamefont {Vassen},
  \citenamefont {Cohen-Tannoudji}, \citenamefont {Leduc}, \citenamefont
  {Boiron}, \citenamefont {Westbrook}, \citenamefont {Truscott}, \citenamefont
  {Baldwin}, \citenamefont {Birkl}, \citenamefont {Cancio},\ and\ \citenamefont
  {Trippenbach}}]{Vassen:12}%
  \BibitemOpen
  \bibfield  {author} {\bibinfo {author} {\bibfnamefont {W.}~\bibnamefont
  {Vassen}}, \bibinfo {author} {\bibfnamefont {C.}~\bibnamefont
  {Cohen-Tannoudji}}, \bibinfo {author} {\bibfnamefont {M.}~\bibnamefont
  {Leduc}}, \bibinfo {author} {\bibfnamefont {D.}~\bibnamefont {Boiron}},
  \bibinfo {author} {\bibfnamefont {C.~I.}\ \bibnamefont {Westbrook}}, \bibinfo
  {author} {\bibfnamefont {A.}~\bibnamefont {Truscott}}, \bibinfo {author}
  {\bibfnamefont {K.}~\bibnamefont {Baldwin}}, \bibinfo {author} {\bibfnamefont
  {G.}~\bibnamefont {Birkl}}, \bibinfo {author} {\bibfnamefont
  {P.}~\bibnamefont {Cancio}},\ and\ \bibinfo {author} {\bibfnamefont
  {M.}~\bibnamefont {Trippenbach}},\ }\bibfield  {title} {\bibinfo {title}
  {Cold and trapped metastable noble gases},\ }\href
  {https://doi.org/10.1103/RevModPhys.84.175} {\bibfield  {journal} {\bibinfo
  {journal} {Rev. Mod. Phys.}\ }\textbf {\bibinfo {volume} {84}},\ \bibinfo
  {pages} {175} (\bibinfo {year} {2012})}\BibitemShut {NoStop}%
\bibitem [{\citenamefont {Vassen}\ \emph {et~al.}(2016)\citenamefont {Vassen},
  \citenamefont {Notermans}, \citenamefont {Rengelink},\ and\ \citenamefont
  {van~der Beek}}]{Vassen:16}%
  \BibitemOpen
  \bibfield  {author} {\bibinfo {author} {\bibfnamefont {W.}~\bibnamefont
  {Vassen}}, \bibinfo {author} {\bibfnamefont {R.~P. M. J.~W.}\ \bibnamefont
  {Notermans}}, \bibinfo {author} {\bibfnamefont {R.~J.}\ \bibnamefont
  {Rengelink}},\ and\ \bibinfo {author} {\bibfnamefont {R.~F. H.~J.}\
  \bibnamefont {van~der Beek}},\ }\bibfield  {title} {\bibinfo {title}
  {Ultracold metastable helium: Ramsey fringes and atom interferometry},\
  }\href {https://doi.org/10.1007/s00340-016-6563-0} {\bibfield  {journal}
  {\bibinfo  {journal} {Appl. Phys. B}\ }\textbf {\bibinfo {volume} {122}},\
  \bibinfo {pages} {289} (\bibinfo {year} {2016})}\BibitemShut {NoStop}%
\bibitem [{\citenamefont {Degen}\ \emph {et~al.}(2017)\citenamefont {Degen},
  \citenamefont {Reinhard},\ and\ \citenamefont {Cappellaro}}]{Degen:17}%
  \BibitemOpen
  \bibfield  {author} {\bibinfo {author} {\bibfnamefont {C.~L.}\ \bibnamefont
  {Degen}}, \bibinfo {author} {\bibfnamefont {F.}~\bibnamefont {Reinhard}},\
  and\ \bibinfo {author} {\bibfnamefont {P.}~\bibnamefont {Cappellaro}},\
  }\bibfield  {title} {\bibinfo {title} {Quantum sensing},\ }\href
  {https://doi.org/10.1103/RevModPhys.89.035002} {\bibfield  {journal}
  {\bibinfo  {journal} {Rev. Mod. Phys.}\ }\textbf {\bibinfo {volume} {89}},\
  \bibinfo {pages} {035002} (\bibinfo {year} {2017})}\BibitemShut {NoStop}%
\end{thebibliography}

%\newpage

\section{Tables}
The following Tables list the coefficients parametrizing the analytical solutions of the BR equations derived in Sec.~I.\\
\begin{table}[htb]
\begin{tabular}{ |p{3.0cm}||p{10.0cm}| |p{2.5cm}|}
 \hline
 \multicolumn{3}{|c|}{Table I: The expansion coefficients $c_{k}$, $d_{k}$ and $b_{k}$} \\
 \hline
$c_{k}$& $d_{k}$ & $b_{k}$\\
 \hline
$c_{0} = 36\frac{\Delta^{2}}{\gamma^{2}}+16$ & $d_{0} = 4\Big(3\frac{\Delta^{2}}{\gamma^{2}}-4\Big)^{3}+ \Big(36\frac{\Delta^{2}}{\gamma^{2}}+16\Big)^{2}$ & $K= \sqrt[3]{\frac{c_{3}}{54}+\frac{\sqrt{d_{6}}}{54}}$  \\

$c_{1}=36\frac{\Delta^{2}}{\gamma^{2}}+48+36p^{2}$ &$d_{1} = -12(8+6p^{2})\Big(3\frac{\Delta^{2}}{\gamma^{2}}-4 \Big)^{2} + 2\Big(36\frac{\Delta^{2}}{\gamma^{2}}+16\Big)\Big(36\frac{\Delta^{2}}{\gamma^{2}}+48+36p^{2}\Big)$ & $b_{1} = \frac{c_{2}+\sqrt{d_{6}}u_{1}}{(c_{3}+\sqrt{d_{6}})}$ \\

$c_{2}=48+90p^{2}$&$d_{2} = 12(8+6p^{2})^{2}\Big(3\frac{\Delta^{2}}{\gamma^{2}}-4\Big) - 12(4+9p^{2})\Big(3\frac{\Delta^{2}}{\gamma^{2}}-4\Big)^{2} + \Big(36\frac{\Delta^{2}}{\gamma^{2}}+48+36p^{2}\Big)^{2} + 2(48+90p^{2})\Big(36\frac{\Delta^{2}}{\gamma^{2}}+16\Big)$ & $b_{2} = \frac{c_{1}+\sqrt{d_{6}}u_{2}}{(c_{3}+\sqrt{d_{6}})}$ \\

$c_{3}=16+54p^{2}$ &$d_{3} = 24(8+6p^{2})(4+9p^{2})\Big(3\frac{\Delta^{2}}{\gamma^{2}}-4\Big) - 4(8+6p^{2})^{3} + 2(16+54p^{2})\Big(36\frac{\Delta^{2}}{\gamma^{2}}+16\Big) +2(48+90p^{2})\Big(36\frac{\Delta^{2}}{\gamma^{2}}+48+36p^{2}\Big)$ & $b_{3} = \frac{c_{0}+\sqrt{d_{6}}u_{3}}{(c_{3}+\sqrt{d_{6}})}$ \\

0& $d_{4} = 12(4+9p^{2})^{2}\Big(3\frac{\Delta^{2}}{\gamma^{2}}-4\Big)-12(8+6p^{2})^{2}(4+9p^{2}) + (48+90p^{2})^{2} + 2(16+54p^{2})\Big(36\frac{\Delta^{2}}{\gamma^{2}}+48+36p^{2}\Big)$ & $b_{4} = \frac{\sqrt{d_{6}}u_{4}}{(c_{3}+\sqrt{d_{6}})}$ \\

0& $d_{5} = -12(8+6p^{2})(4+9p^{2})^{2} + 2(48+90p^{2})(16+54p^{2})$ & $b_{5} = \frac{\sqrt{d_{6}}u_{5}}{(c_{3}+\sqrt{d_{6}})}$ \\

0&$d_{6} = -4(4+9p^{2})^{3} + (16+54p^{2})^{2}$ & $b_{6} = \frac{\sqrt{d_{6}}u_{6}}{(c_{3}+\sqrt{d_{6}})}$ \\
 \hline
\end{tabular}\\
\end{table}

\begin{table}[htb]
\begin{tabular}{ |p{10cm}||p{6cm}| }
 \hline
 \multicolumn{2}{|c|}{Table II: The expansion coefficients $u_{k}$ and $v_{k}$ in the expressions for $\sqrt{\mathcal{D}}$ and $\Lambda(x)$} \\
 \hline
$u_{k}$ & $v_{k}$\\
 \hline
$u_{1}=\frac{1}{2} \frac{d_{5}}{d_{6}}$& $v_{1}=\frac{b_{1}}{3}$\\

$u_{2}=\frac{1}{2}\frac{d_{4}}{d_{6}}-\frac{1}{8}(\frac{d_{5}}{d_{6}})^{2}$& $v_{2}=\frac{b_{2}}{3}-\frac{b_{1}^{2}}{9}$\\

$u_{3}=\frac{1}{2}\frac{d_{3}}{d_{6}}-\frac{1}{8}2\frac{d_{4}d_{5}}{d_{6}^{2}}+\frac{3}{48}(\frac{d_{5}}{d_{6}})^{3}$& $v_{3}=\frac{b_{3}}{3}-\frac{2b_{1}b_{2}}{9}+\frac{5b_{1}^{3}}{81}$\\

$u_{4}=\frac{1}{2}\frac{d_{2}}{d_{6}}-\frac{1}{8}({2\frac{d_{3}d_{5}}{d_{6}^{2}}+(\frac{d_{4}}{d_{6}})^{2}})+\frac{3}{48}(3\frac{d_{4}d_{5}^{2}}{d_{6}^{3}})-\frac{15}{384}(\frac{d_{5}}{d_{6}})^{4}$&$v_{4}=\frac{b_{4}}{3}-\frac{(2b_{1}b_{3}+b_{2}^{2})}{9}+\frac{5(3b_{1}^{2}b_{2})}{81}$\\

$u_{5}=\frac{1}{2}\frac{d_{1}}{d_{6}}-\frac{1}{8}(2\frac{d_{2}d_{5}}{d_{6}^{2}}+2\frac{d_{3}d_{4}}{d_{6}^{2}})+\frac{3}{48}(3\frac{d_{3}d_{5}^{2}}{d_{6}^{3}}+3\frac{d_{4}^{2}d_{5}}{d_{6}^{3}})-\frac{15}{384}(4\frac{d_{4}d_{5}^{3}}{d_{6}^{4}})$&$v_{5}=\frac{b_{5}}{3}-\frac{(2b_{1}b_{4}+2b_{2}b_{3})}{9}+\frac{5(3b_{1}^{2}b_{3}+3b_{1}b_{2}^{2})}{81}$\\

$u_{6}=\frac{1}{2}\frac{d_{0}}{d_{6}}-\frac{1}{8}(2\frac{d_{1}d_{5}}{d_{6}^{2}}+2\frac{d_{2}d_{4}}{d_{6}^{2}}+(\frac{d_{3}}{d_{6}})^{3})+\frac{3}{48}(3\frac{d_{2}d_{5}^{2}}{d_{6}^{3}}+6\frac{d_{3}d_{4}d_{5}}{d_{6}^{3}}+(\frac{d_{4}}{d_{6}})^{3})-\frac{15}{384}(4\frac{d_{3}d_{5}^{3}}{d_{6}^{4}}+6\frac{d_{4}^{2}d_{5}^{2}}{d_{6}^{4}})$&$v_{6}=\frac{b_{6}}{3}-\frac{(2b_{1}b_{5}+2b_{2}b_{4}+b_{3}^{2})}{9}+\frac{5(3b_{1}^{2}b_{4}+6b_{1}b_{2}b_{3}+b_{2}^{3})}{81}$\\
 \hline
\end{tabular}
\end{table}

\begin{table}
\begin{tabular}{ |p{7cm}||p{9cm}| }
 \hline
 \multicolumn{2}{|c|}{Table III: The expansion coefficients $\mathcal{W}_{k}$ and $z_{jk}$ in the expression of $\frac{1}{\mathcal{T}}$ in Eq. (50) and $\lambda_{j}$ in Eq. (53)} \\
 \hline
$\mathcal{W}_{k}$ & $z_{jk}$\\
 \hline
0  & $z_{j0}= -\frac{5}{3} - \frac{\alpha_{j}}{9K} (4+9p^{2}) - \beta_{j} K$ \\

$\mathcal{W}_{1}=-v_{1}$& $z_{j1}= -\frac{2}{3} - \frac{\alpha_{j}}{9K} [(8+6p^{2})+(4+9p^{2})\mathcal{W}_{1}] - \beta_{j} K v_{1}$ \\

$\mathcal{W}_{2}=-v_{2}+v_{1}^{2}$&$z_{j2}= \frac{\alpha_{j}}{9K}[(3\frac{\Delta^{2}}{\gamma^{2}}-4)-(8+6p^{2})\mathcal{W}_{1}-(4+9p^{2})\mathcal{W}_{2})] - \beta_{j} K v_{2}$ \\

$\mathcal{W}_{3}=-v_{3}+2v_{1}v_{2}-v_{1}^{3}$& $z_{j3}= \frac{\alpha_{j}}{9K}[(3\frac{\Delta^{2}}{\gamma^{2}}-4)\mathcal{W}_{1}-(8+6p^{2})\mathcal{W}_{2}-(4+9p^{2})\mathcal{W}_{3})] - \beta_{j} K v_{3}$ \\

$\mathcal{W}_{4}=-v_{4}+(2v_{1}v_{3}+v_{2}^{2})-3v_{1}^{2}v_{2}+v_{1}^{4}$& $z_{j4}= \frac{\alpha_{j}}{9K}[(3\frac{\Delta^{2}}{\gamma^{2}}-4)\mathcal{W}_{2}-(8+6p^{2})\mathcal{W}_{3}-(4+9p^{2})\mathcal{W}_{4})] - \beta_{j} K v_{4}$ \\

$\mathcal{W}_{5}=-v_{5}+(2v_{1}v_{4}+2v_{2}v_{3})-(3v_{1}^{2}v_{3}+3v_{1}v_{2}^{2})+4v_{1}^{3}v_{2}-v_{1}^{5}$& $z_{j5}= \frac{\alpha_{j}}{9K}[(3\frac{\Delta^{2}}{\gamma^{2}}-4)\mathcal{W}_{3}-(8+6p^{2})\mathcal{W}_{4}-(4+9p^{2})\mathcal{W}_{5})] - \beta_{j} K v_{5}$ \\

$\mathcal{W}_{6}=-v_{6}+(2v_{1}v_{5}+2v_{2}v_{4}+v_{3}^{2})-(3v_{1}^{2}v_{4}+6v_{1}v_{2}v_{3}+v_{3}^{2})+(4v_{1}^{3}v_{3}+6v_{1}^{2}v_{2}^{2})+v_{1}^{6}$& $z_{j6}= \frac{\alpha_{j}}{9K}[(3\frac{\Delta^{2}}{\gamma^{2}}-4)\mathcal{W}_{4}-(8+6p^{2})\mathcal{W}_{5}-(4+9p^{2})\mathcal{W}_{6})] - \beta_{j} K v_{6}$\\

0 & $z_{j7}= \frac{\alpha_{j}}{9K}[(3\frac{\Delta^{2}}{\gamma^{2}}-4)\mathcal{W}_{5}-(8+6p^{2})\mathcal{W}_{6})]$\\

0 & $z_{j8}= \frac{\alpha_{j}}{9K}[(3\frac{\Delta^{2}}{\gamma^{2}}-4)\mathcal{W}_{6}]$ \\
 \hline
\end{tabular}
\end{table}

\begin{table}
\begin{tabular}{ |p{9.5cm}| |p{7.5cm}| }
 \hline
 \multicolumn{2}{|c|}{Table IV: Coefficients $z_{j2}$ and $v_{2}$}\\
 \hline
$z_{j2}$ & $v_{2}$\\
 \hline
$f_{j1}(p)=\frac{\alpha_{j}}{3K}+[\frac{\alpha_{j}}{9K}(4+9p^{2})-\beta_{j} K]s_{1}(p)$ & $s_{1}(p)=\frac{l_{1}(p)}{3}$\\

$f_{j2}(p)=-\frac{\alpha_{j}}{9K} [4-(8+6p^{2})v_{1}+(4+9p^{2})v_{1}^{2}]+[\frac{\alpha_{j}}{9K}(4+9p^{2})-\beta_{j} K]s_{2}(p)$ & $s_{2}(p)=\frac{l_{2}(p)}{3}-[\frac{1}{27 K^{3}}\lbrace(8+15p^{2})+p^{2}\frac{(5+9p^{2})}{(1+3p^{2})}\sqrt{3(1+3p^{2})}i\rbrace]^{2}$\\
 \hline
\end{tabular}\\
\end{table}

\begin{table}
\begin{tabular}{ |p{6.25cm}| |p{5.25cm}| |p{5.0cm}| }
 \hline
 \multicolumn{3}{|c|}{Table V: Coefficients $d_{4}$, $u_{2}$ and $b_{2}$ in the expansion of $z_{j2}$ }\\
 \hline
$b_{2}$ &  $u_{2}$ & $d_{4}$ \\
 \hline
$l_{1}(p)=\frac{2}{3K^{3}}+\frac{p^{2}}{3K^{3}}\sqrt{3(1+3p^{2})}g_{1}(p)i$ & $g_{1}(p)=-\frac{h_{1}(p)}{1944p^{4}(1+3p^{2})}$ & $h_{1}(p)=108(16+60p^{2}+27p^{4})$\\

$l_{2}(p)=\frac{2(4+3p^{2})}{9K^{3}}+\frac{p^{2}}{3K^{3}}\sqrt{3(1+3p^{2})}g_{2}(p)i$ & $g_{2}(p)=-\frac{h_{2}(p)}{1944p^{4}(1+3p^{2})}-\frac{1}{18}(\frac{5+9p^{2}}{1+3p^{2}})^{2}$ & $h_{2}(p)=-108p^{2}(37+36p^{2})$\\
 \hline
\end{tabular}\\
\end{table}

\begin{table}
\begin{tabular}{ |p{4cm}| |p{3.5cm}| |p{2.5cm}| |p{6cm}|}
 \hline
 \multicolumn{4}{|c|}{Table VI: Coefficients for the eigenvectors $V_{j}$, $j$ = 1, 2 in Eq.~(87) and (88)}\\
 \hline
$L_{jk}; j = 1,2,3$& $k_{jm}; j=1,2$ & $a_{jm}; j=1,2$ & $b_{jm}; j =1,2$\\
 \hline
$L_{j0}=z_{j0}^{2}+4z_{j0}+3(1-p^{2})$ & 0 & $a_{j0}=1$ & $b_{j0}= 3 + z_{j0}$\\

$L_{j1}=2z_{j0}z_{j1}+2z_{j0}+4z_{j1}+4(1-p^{2})$ & $k_{j1}=-\frac{L_{j3}}{L_{j2}}$ &$a_{j1}= 1 + k_{j1}$ & $b_{j1}= (1+z_{j1})+(3+z_{j0})k_{j1}$\\

$L_{j2}=z_{j1}^{2}+2z_{j0}z_{j2}+2z_{j1}+4z_{j2}+(1-p^{2})$ & $k_{j2}=-\frac{L_{j4}}{L_{j2}}+(\frac{L_{j3}}{L_{j2}})^{2}$ & $a_{j2}= k_{j1} + k_{j2}$ & $b_{j2}=(1+z_{j1})k_{j1}+(3+z_{j0})k_{j2}+z_{j2}$\\

$L_{j3}=2z_{j1}z_{j2}+2z_{j2}$ & $k_{j3}=\frac{2 L_{j3} L_{j4}}{L_{j2}^{2}}-(\frac{L_{j3}}{L_{j2}})^{3}$ & $a_{j3}= k_{j2} + k_{j3}$ & $b_{j3}=(1+z_{j1})k_{j2}+(3+z_{j0})k_{j3}+z_{j2}k_{j1}$\\

$L_{j4}=z_{j2}^{2}$ & $k_{j4}=(\frac{L_{j4}}{L_{j2}})^{2}- \frac{3 L_{j3}^{2} L_{j4}}{L_{j2}^{3}}$ & $a_{j4}= k_{j3} + k_{j4}$ & $b_{j4}=(1+z_{j1})k_{j3}+(3+z_{j0})k_{j4}+z_{j2}k_{j2}$\\

0 & $k_{j5}=-\frac{3 L_{j3} L_{j4}^{2}}{L_{j2}^{3}}$ & $a_{j5}= k_{j4} + k_{j5}$ & $b_{j5}=(1+z_{j1})k_{j4}+(3+z_{j0})k_{j5}+z_{j2}k_{j3}$\\

0 & $k_{j6}=-(\frac{L_{j4}}{L_{j2}})^{2}$ & $a_{j6} = k_{j5} + k_{j6}$ & $b_{j6}=(1+z_{j1})k_{j5}+(3+z_{j0})k_{j6}+z_{j2}k_{j4}$\\

0 & 0 & $a_{j7}=  k_{j6}$ & $b_{j7}=(1+z_{j1})k_{j6}+(3+z_{j0})k_{j7}+z_{j2}k_{j5}$\\

0 & 0 & 0 & $b_{j8}= z_{j2}k_{j6}$\\
 \hline
\end{tabular}\\
\end{table}

\begin{table}
\begin{tabular}{ |p{9cm}| |p{2.5cm}| |p{5cm}| }
 \hline
 \multicolumn{3}{|c|}{Table VII: Coefficients for the eigenvector $V_{j}$, $j$ = 3 in Eq. (99)}\\
 \hline
$k_{jm}; j=3$ & $a_{jm}; j=3$ & $b_{jm}; j =3$\\
 \hline
0 & $a_{30}=1$ & $b_{30}= 3 + z_{30}$\\

$k_{31}=-\frac{L_{31}}{L_{30}}$ &$a_{31}= 1 + k_{31}$ & $b_{31}= (1+z_{31})+(3+z_{30})k_{31}$\\

$k_{32}=-\frac{L_{32}}{L_{30}}+(\frac{L_{31}}{L_{30}})^{2}$ & $a_{32}= k_{31} + k_{32}$ & $b_{32}=(1+z_{31})k_{31}+(3+z_{30})k_{32}+z_{32}$\\

$k_{33}=-\frac{L_{33}}{L_{30}}+\frac{2 L_{31} L_{32}}{L_{30}^{2}}-(\frac{L_{31}}{L_{30}})^{3}$ & $a_{33}= k_{32} + k_{33}$ & $b_{33}=(1+z_{31})k_{32}+(3+z_{30})k_{33}+z_{32}k_{31}$\\

$k_{34}=-\frac{L_{34}}{L_{30}}+(\frac{2 L_{31} L_{33}}{L_{30}^{2}}+(\frac{L_{32}}{L_{30}})^{2})-(\frac{3 L_{31}^{2} L_{32}}{L_{30}^{3}}+\frac{3 L_{31} L_{32}^{2}}{L_{30}^{3}})+(\frac{L_{31}}{L_{30}})^{4}$ & $a_{34}= k_{33} + k_{34}$ & $b_{34}=(1+z_{31})k_{33}+(3+z_{30})k_{34}+z_{32}k_{32}$\\

$k_{35}=(\frac{2 L_{31} L_{34}}{L_{30}^{2}}+\frac{2 L_{32} L_{33}}{L_{30}^{2}})-\frac{3 L_{31}^{2} L_{33}}{L_{30}^{3}}+(\frac{4 L_{31}^{3}L_{32}}{L_{30}^{4}}-(\frac{L_{31}}{L_{30}})^{5})$ & $a_{35}= k_{34} + k_{35}$ & $b_{35}=(1+z_{31})k_{34}+(3+z_{30})k_{35}+z_{32}k_{33}$\\
 \hline
\end{tabular}\\
\end{table}

\begin{table}
\begin{tabular}{ |p{9cm}||p{7cm}| }
 \hline
 \multicolumn{2}{|c|}{Table VIII: Coefficients for $L_{j2}; j = 1, 2$ and the determinant of the eigenvector matrix $\bf{M}$}\\
 \hline
$L_{j2};\enspace j = 1,\enspace 2$ & det($\bf{M}$)\\
 \hline
$F_{j1}(p)=(4+2z_{j0}(p))f_{j1}(p)$ & $T_{1}(p)= p(-\frac{m_{1}}{F_{11}}-\frac{m_{5}}{F_{21}})$\\

$F_{j2}(p)=(4+2z_{j0}(p))f_{j2}(p)+(z_{j1}(p)^{2}+2z_{j1}(p)+(1-p^{2}))$ & $T_{2}(p)= p(-\frac{m_{2}}{F_{11}}+\frac{f_{11}}{F_{11}}\frac{1}{(2z_{10}+4)}-\frac{m_{6}}{F_{21}}-\frac{m_{11}}{L_{30}})$\\
 \hline
\end{tabular}
\end{table}

\begin{table}
\begin{tabular}{ |p{7.5cm}| |p{4.5cm}| |p{4.0cm}| }
 \hline
 \multicolumn{3}{|c|}{Table IX: Components $V_{ij}; i, j$ = 1, 2, 3 of the eigenvector matrix $\bf{M}$ in Eq. (101)}\\
 \hline
$V_{1i}$ & $V_{2i}$ & $V_{3i}$\\
 \hline
$V_{11} = p[-\frac{1}{F_{11}}+\frac{f_{11}}{F_{11}}\frac{1}{(2z_{10}+4)}\frac{1}{\bar{n}^{2}}(\frac{\Delta}{\gamma})^{2}]\bar{n}(\frac{\Delta}{\gamma})^{-1}$ & $V_{21} = -\frac{p}{F_{21}}\bar{n}(\frac{\Delta}{\gamma})^{-1}$ & $V_{31}= -\frac{p}{L_{30}} (\frac{1}{\bar{n}})(\frac{\Delta}{\gamma})$\\

$V_{12} = [\frac{(3+z_{10})}{F_{11}}+(\frac{f_{11}}{F_{11}})\frac{(1+z_{10})}{(4+2z_{10})}(\frac{1}{\bar{n}})^{2}(\frac{\Delta}{\gamma})^{2}]\bar{n}(\frac{\Delta}{\gamma})^{-1}$ & $V_{22}= \frac{(3+z_{20})}{F_{21}}\bar{n}(\frac{\Delta}{\gamma})^{-1}$ & $V_{32}= \frac{(3+z_{30})}{L_{30}}(\frac{1}{\bar{n}})(\frac{\Delta}{\gamma})$\\

$V_{13}= 1$ & $V_{23}= 1$ & $V_{33}= 1$\\
 \hline
\end{tabular}\\
\end{table}

\begin{table}
\begin{tabular}{ |p{5.5cm}| |p{5.0cm}| |p{6.0cm}| }
 \hline
 \multicolumn{3}{|c|}{Table X: Cofactors $T_{ij}; i, j$ = 1, 2, 3 of the eigenvector matrix $\bf{M}$ in Eq. (101)}\\
 \hline
$T_{1i}$ & $T_{2i}$ & $T_{3i}$\\
 \hline
$T_{11} = [m_{1}(p)+m_{2}(p)\frac{1}{\bar{n}^{2}}(\frac{\Delta}{\gamma})^{2}]\bar{n}(\frac{\Delta}{\gamma})^{-1}$ & $T_{21} = [m_{5}(p)+m_{6}(p)\frac{1}{\bar{n}^{2}}(\frac{\Delta}{\gamma})^{2}]\bar{n}(\frac{\Delta}{\gamma})^{-1}$ & $T_{31}= [m_{11}(p)+m_{12}(p)\frac{1}{\bar{n}^{2}}(\frac{\Delta}{\gamma})^{2}]\bar{n}(\frac{\Delta}{\gamma})^{-1}$\\

$T_{12} = [m_{3}(p)+m_{4}(p)\frac{1}{\bar{n}^{2}}(\frac{\Delta}{\gamma})^{2}]\bar{n}(\frac{\Delta}{\gamma})^{-1}$ & $T_{22}= [m_{7}(p)+m_{8}(p)\frac{1}{\bar{n}^{2}}(\frac{\Delta}{\gamma})^{2}]\bar{n}(\frac{\Delta}{\gamma})^{-1}$ & $T_{32}= [m_{13}(p)+m_{14}(p)\frac{1}{\bar{n}^{2}}(\frac{\Delta}{\gamma})^{2}]\bar{n}(\frac{\Delta}{\gamma})^{-1}$\\

$T_{13}= \frac{p}{F_{21}L_{30}}(z_{20}-z_{30})$ & $T_{23}= [m_{9}(p)+m_{10}(p)\frac{1}{\bar{n}^{2}}(\frac{\Delta}{\gamma})^{2}]$ & $T_{33}= [m_{15}(p)+m_{16}(p)\frac{1}{\bar{n}^{2}}(\frac{\Delta}{\gamma})^{2}]\bar{n}^{2}(\frac{\Delta}{\gamma})^{-2}$\\
 \hline
\end{tabular}\\
\end{table}

\begin{table}
\begin{tabular}{ |p{3cm}| |p{4.5cm}| |p{4.5cm}| |p{4cm}|}
 \hline
 \multicolumn{4}{|c|}{Table XI: Coefficients $m_{i}(p)$ for the cofactors $T_{ij}$ of the eigenvector matrix $\bf{M}$}\\
 \hline
$m_{1}(p)=\frac{(3+z_{20})}{F21}$ & $m_{5}(p)=-\frac{(3+z_{10})}{F_{11}}$ & $m_{9}(p)=\frac{p}{F_{11}L_{30}}(z_{10}+z_{30}+6)$ & $m_{13}(p)=p(\frac{1}{F_{11}}-\frac{1}{F_{21}})$\\

$m_{2}(p)=-\frac{(3+z_{30})}{F31}$ & $m_{6}(p)=-(\frac{f_{11}}{F_{11}}\frac{(1+z_{10})}{(4+2z_{10})}-\frac{(3+z_{30})}{L_{30}})$ & $m_{10}(p)= p(\frac{f_{11}}{F_{11}L_{30}})\frac{(z_{10}-z_{30}-2)}{(2z_{10}+4)}$ & $m_{14}(p)= -(\frac{f_{11}}{F_{11}}\frac{1}{(2z_{10}+4)})$\\

$m_{3}(p)=\frac{p}{F21}$ & $m_{7}(p)=-\frac{p}{F_{11}}$ & $m_{11}(p)=(\frac{(3+z_{10})}{F_{11}}-\frac{(3+z_{20})}{F_{21}})$ & $m_{15}(p)=p(\frac{(z_{10-z_{20}})}{F_{11}F_{21}})$\\

$m_{4}(p)=-\frac{p}{L_{30}}$ & $m_{8}(p)= p (\frac{f_{11}}{F_{11}}\frac{1}{(2z_{10}+4)}+\frac{1}{L_{30}})$ & $m_{12}(p)= (\frac{f_{11}}{F_{11}})\frac{(1+z_{10})}{(4+2z_{10})}$ & $m_{16}(p)= p(\frac{f_{11}}{F_{11}F_{21}}\frac{(z_{10}+z_{20}+4)}{(2z_{10}+4)})$\\
 \hline
\end{tabular}\\
\end{table}

\begin{table}
\begin{tabular}{ |p{6.5cm}| |p{6.5cm}| |p{3cm}| }
 \hline
 \multicolumn{3}{|c|}{Table XII: Coefficients $A_{i}(p)$, $B_{i}(p)$ and $C_{i}(p)$ for the population and coherence terms in Eqs. (129) to (131) }\\
 \hline
$A_{i}$ & $B_{i}$ & $C_{i}$\\
 \hline
$A_{1}= -\frac{p}{F_{11}}(m_{1}+pm_{3})$ & $B_{1}= \frac{(3+z_{10})}{F_{11}}(m_{1}+pm_{3})$ & $C_{1}= (m_{1}+pm_{3})$\\

$A_{2}= p(\frac{f_{11}}{F_{11}}\frac{1}{(2z_{10}+4)}(m_{1}+pm_{3})-\frac{1}{F_{11}}(m_{2}+pm_{4}))$ & $B_{2}= (\frac{f_{11}}{F_{11}}\frac{1}{(2z_{10}+4)}(m_{1}+pm_{3})+\frac{(3+z_{10})}{F_{11}}(m_{2}+pm_{4}))$ & $C_{2}= (m_{2}+pm_{4})$\\

$A_{3}= -\frac{p}{F_{21}}(m_{5}+pm_{7})$ & $B_{3}= \frac{(3+z_{20})}{F_{21}}(m_{5}+pm_{7})$ & $C_{3}= (m_{5}+pm_{7})$\\

$A_{4}= -\frac{p}{F_{21}}(m_{6}+pm_{8})$ & $B_{4}= \frac{(3+z_{20})}{F_{21}}(m_{6}+pm_{8})$ & $C_{4}= (m_{6}+pm_{8})$\\

$A_{5}= -\frac{p}{L_{30}}(m_{11}+pm_{13})$ & $B_{5}= \frac{(3+z_{30})}{L_{30}}(m_{11}+pm_{13})$ & $C_{5}= (m_{11}+pm_{13})$\\

$A_{6}= -\frac{p}{L_{30}}(m_{12}+pm_{14})$ & $B_{6}= \frac{(3+z_{30})}{L_{30}}(m_{12}+pm_{14})$ & $C_{6}= (m_{12}+pm_{14})$\\
 \hline
\end{tabular}\\
\end{table}

\begin{table}
\begin{tabular}{ |p{3.0cm}| |p{9.5cm}| }
 \hline
 \multicolumn{2}{|c|}{Table XIII: The expansion coefficients $\tilde{K}$, $t_{k}$ and $\tilde{z}_{jk}(p)$ for the eigenvalues in weak pumping limit} \\
 \hline
$t_{k}$ & $\tilde{z}_{jk}$ \\
 \hline
$\tilde{K}= \sqrt[3]{\frac{c_{0}}{2}+\frac{\sqrt{d_{0}}}{2}}$ & $\tilde{z}_{j0}= -\frac{2}{3} + \frac{\alpha_{j}}{3\tilde{K}}\big(3\frac{\Delta^{2}}{\gamma^{2}}-4\big) - \frac{1}{3} \beta_{j} \tilde{K}$  \\

$t_{1}= \frac{c_{1}+\frac{d_{1}}{2\sqrt{d_{0}}}}{c_{0}+\sqrt{d_{0}}}$ & $\tilde{z}_{j1}= -\frac{5}{3} - \frac{\alpha_{j}}{3\tilde{K}}\Big[(8+6p^{2})+\frac{1}{3}\big(3\frac{\Delta^{2}}{\gamma^{2}}-4\big)t_{1}\Big] - \frac{1}{3} \beta_{j} \tilde{K}t_{1}$   \\

$t_{2}= \frac{c_{2}+\frac{d_{2}}{2\sqrt{d_{0}}}}{c_{0}+\sqrt{d_{0}}}$ & $\tilde{z}_{j2}= - \frac{\alpha_{j}}{3\tilde{K}}\Big[(4+9p^{2})-\frac{1}{3}(8+6p^{2})t_{1}+\frac{1}{3}\big(3\frac{\Delta^{2}}{\gamma^{2}}-4\big)t_{2}\Big] - \frac{1}{9} \beta_{j} \tilde{K}t_{2}$ \\

$t_{3}= \frac{c_{3}+\frac{d_{3}}{2\sqrt{d_{0}}}}{c_{0}+\sqrt{d_{0}}}$ & $\tilde{z}_{j3}=  \frac{\alpha_{j}}{9\tilde{K}}\Big[(4+9p^{2})t_{1}+(8+6p^{2})t_{2}-\big(3\frac{\Delta^{2}}{\gamma^{2}}-4\big)t_{3}\Big] - \frac{1}{9} \beta_{j}\tilde{K} t_{3}$ \\

0 & $\tilde{z}_{j4}=  \frac{\alpha_{j}}{9\tilde{K}}\Big[(4+9p^{2})t_{2}+(8+6p^{2})t_{3}\Big]$ \\

0 & $\tilde{z}_{j5}=  \frac{\alpha_{j}}{9\tilde{K}}(4+9p^{2})t_{3}$ \\

 \hline
\end{tabular}\\
\end{table}

 \end{document}